\documentclass[prd,aps,preprint,tightenlines,groupedaddress,nofootinbib,showpacs,floatfix]{revtex4}
\usepackage{float}
\usepackage{mathrsfs}
\usepackage{amsfonts}
\usepackage{array}
\usepackage{epsfig}
\usepackage{amsmath}    % need for subequations
\usepackage{amssymb}   % defines \lesssim, etc
\usepackage{graphicx}   % need for figures
\usepackage{verbatim}   % useful for program listings
\usepackage{color}      % use if color is used in text
\usepackage{subfigure}  % use for side-by-side figures

\begin{document}
\begin{flushright}
MSUHEP-18-006
\end{flushright}

\title{Updating and Optimizing Error PDFs in the Hessian Approach}

\author{Carl Schmidt}
\email{schmidt@pa.msu.edu}
\author{Jon Pumplin}
\email{pumplin@pa.msu.edu}
\author{ C.--P. Yuan}
\email{yuan@pa.msu.edu}
\affiliation{
Department of Physics and Astronomy, Michigan State University,\\
 East Lansing, MI 48824 U.S.A. }

\begin{abstract}

We discuss how to apply the Hessian method (i) to predict the impact of a new data set (or sets) on the parton distribution functions (PDFs) and their errors, by producing an updated best-fit PDF and error PDF sets, such as the CTEQ-TEA PDFs; (ii) to predict directly the effect of a new data set on the PDF errors of any other set of observables, without the need to recalculate using the new error PDFs; and (iii) to transform the original set into a reduced set of error PDFs which is optimized for a specific set of observables to reproduce the PDF-induced uncertainties to any specified precision. We present a software package, {\tt ePump} (error PDF Updating Method Package), that can be used to update or optimize a set of PDFs, including the best-fit PDF set and Hessian eigenvector pairs of PDF sets (i.e., error PDFs), and to update any other set of observables. We demonstrate the potential of the program by presenting selected phenomenological applications relevant to the Large Hadron Collider.  Special care is given to discuss the assumptions made and the limitations of this theoretical framework compared to a treatment by the full global-analysis program.

\end{abstract}



\maketitle

\section{Introduction}\label{sec:intro}

Predictions for high-energy cross sections at the Large Hadron Collider (LHC) and other colliders require the use of Parton Distribution Functions (PDFs), which
supply the long-distance hadronic contribution.
The PDFs cannot be calculated from first principles to sufficient accuracy; therefore they are generally extracted from a global analysis of high-energy scattering data, including both data from Deep Inelastic Scattering (DIS) at lepton-hadron colliders and data from hadron-hadron colliders~\cite{Dulat:2015mca,Harland-Lang:2014zoa,Ball:2014uwa,Abramowicz:2015mha,Alekhin:2013nda,Accardi:2016qay}.  In addition to a best-fit set of
PDFs, modern PDF global-analysis groups also provide a number of additional sets that can be used to estimate the uncertainty in the predictions due to the uncertainties in the PDFs themselves.

The two most commonly-used methods for obtaining the PDF uncertainties are the Monte Carlo method~\cite{Giele:1998gw,Giele:2001mr} and the Hessian method~\cite{Pumplin:2001ct}.  In the Monte Carlo method,
a statistical ensemble of PDF sets is provided, which is assumed to approximate the probability distribution of possible PDFs, as constrained from the global analysis of the data.  The advantage of this method is that it can, in principle, reproduce the probability distribution without approximations. A disadvantage is that it
may require a sizable number of sets (around 1000) in the ensemble to get an accurate estimate of the uncertainty.  In the Hessian method a smaller number of ``error sets'' (about 50) is used to obtain an estimate of the error.  These error sets correspond to the plus and minus eigenvector directions in the space of PDF parameters,
which are used to approximate the $\chi^2$-function near its global minimum. The Hessian method relies on a quadratic approximation for the parameter dependence of the $\chi^2$ minimization function and a linear approximation for the parameter dependence of the observable in question.  In practice the Hessian method works quite well for most observables.

An understanding of uncertainties due to PDFs is crucial to precision studies of the standard model, as well as to searches for new physics beyond the standard model.  In turn, new measurements of standard model processes can be used to constrain the uncertainties on the PDFs.  The most complete method for obtaining constraints from the new data on the PDFs would be to add the new data into the global-analysis package and to do a full re-analysis on the PDFs.  However, this is impractical for most users of PDFs for several reasons.  First, it requires the complete knowledge of both the experimental measurements and the
theoretical predictions as a function of the PDFs, for all of the data sets in the global analysis.  Second, given the complexity of the theoretical predictions at Next-to-Next-to-Leading Order (NNLO) and the fact that the observables must be calculated numerous times in order to probe the terrain of the $\chi^2$-function, the minimization and error analysis of the PDFs is very computationally intensive.

For these reasons a technique for estimating the impact of new data on the PDFs, without performing a full global analysis, is extremely useful.
In the context of the Monte Carlo PDFs, the PDF reweighting method has become commonplace.  This involves applying a weight factor, which is dependent on the
new data and the theory predictions, to each of the PDFs in the ensemble~\cite{Giele:1998gw, Ball:2010gb,Ball:2011gg} when performing ensemble averages. Because the weight factor for some of of the PDFs in the ensemble may be small, the effective number of PDFs in the ensemble is reduced. Therefore, the number of initial PDFs in the ensemble must be increased to get sufficient statistics in the reweighted averages.
The CT14MC PDF set is such an example, in which a large number of
Monte Carlo replicas have been generated from the Hessian PDFs in order to use this reweighting procedure~\cite{Hou:2016sho}

It is also possible to estimate the impact of new data directly using Hessian PDFs, as has been shown by Paukkunen and Zurita~\cite{Paukkunen:2014zia}, who expanded on ideas presented in Ref.~\cite{Paukkunen:2013grz}.
A version of this method has also been included in the xFitter package~\cite{Camarda:2015zba}, where it is called Hessian profiling.
The advantage of this  Hessian updating method over the Monte Carlo reweighting method is that it directly works with the (small set of) Hessian PDFs and it is  a simpler
and much faster way to estimate the effects of the new data.  This method directly calculates the minimum of the $\chi^2$ function within the Hessian approximation, and was shown by Paukkunen and Zurita to be equivalent to the Monte Carlo reweighting method, if the Giele-Keller weights~\cite{Giele:1998gw} (appropriately scaled to include the tolerance criterion) are used in that method.
In this paper, we extend the method proposed in Ref.~\cite{Paukkunen:2014zia} and develop a software package called {\tt ePump} (error PDF Updating Method
Package) to be used to updated any Hessian PDF sets obtained from an earlier global analysis. We will use the CTEQ-TEA (CT) PDFs as an example. Namely, we shall demonstrate how to use ePump to update CT14 PDFs when new experimental data are included in the global fit. We emphasize, however, that ePump may be used to update any PDF set containing Hessian error PDFs in LHAPDF format~\cite{lhapdf}, such as those supplied by CT, MMHT, and others, and it is flexible enough to accommodate different non-global tolerance criterions.

Another useful and related, but distinct operation on the Hessian error PDF sets is to obtain a reduced set of error PDFs optimized for a particular experimental analysis.
In a detailed experimental analysis, it is quite common to investigate uncertainties on the kinematic acceptance and the systematic errors induced by the uncertainties in PDFs.
The large number of Monte Carlo PDFs, and even the smaller number of Hessian PDFs can make this type of analysis very difficult due to the large
computational time needed for the theoretical calculation to simulate the experimental data with different cuts and bins.  For this purpose a smaller reduced set of Hessian error PDFs
that cover the majority of the PDF dependence of the relevant observables is necessary to make this detailed experimental analysis possible.
In Refs.~\cite{Gao:2013bia,Carrazza:2015aoa,Carrazza:2016htc} several methods were developed to obtain a reduced set of Hessian PDFs from a Monte Carlo or Hessian PDF set.
In this paper, we adopt a very different method which is based on ideas similar to that used in the data set diagonalization
method developed by Pumplin~\cite{Pumplin:2009nm}. The advantage of our method is that it takes a set of Hessian error PDFs and constructs an equivalent set
of error PDFs that exactly reproduces the Hessian symmetric PDF uncertainties, but in addition each new eigenvector pair has an eigenvalue that quantitatively describes its contribution to the PDF uncertainty of the data set. The new optimized error PDF pairs are ordered by
their eigenvalues, and so it is easy to choose a reduced set that covers the PDF uncertainty
for the data set to any desired accuracy.

The remainder of this paper is organized in the following manner. In Sec.~II, we discuss the method of updating Hessian PDFs (including the central and error PDFs) and predictions on any physical observables (including its PDF uncertainties) from new experimental data.
Careful attention is given to how to include non-global tolerances in the updating of both the central and error PDFs.
 In this section we also discuss limitations of the method and present some realistic applications. In Sec.~III, we discuss the method of optimizing the Hessian error PDFs for a given experimental data and give some sample applications of the PDF optimization procedure. A brief description of the ePump code is given in Sec.~IV. Sec.~V contains our conclusion.

\section{Updating Hessian Error PDFs and Observable Predictions From New Experimental Data}\label{sec:Updating}

\subsection{Review of the Hessian Method}\label{sec:ReviewHessian}

In order to set notation and present the necessary background, we first give the results of the Hessian Method for obtaining the contribution of the PDF
uncertainty to the theoretical uncertainty of a general observable.  The PDFs\footnote{Note that we suppress the flavor index of the PDFs.  All PDF indices in
these sections will correspond to eigenvector directions.} $f(x,Q_0;{\bf z})$, defined at the initial scale $Q_0$, are parametrized by $N$ parameters $\{z_i;\ i=1,N\}$, which we write as a vector $\bf z$.   The determination of the PDFs is obtained using a $\chi^2$-function, which quantifies the discrepancy between the theoretical predictions and the experimental measurements of a global set of experiments, including the experimental errors.  The best-fit PDFs are then obtained by minimizing $\chi^2$ as a function of the parameters.

By evaluating the $\chi^2$ function around its minimum, it is possible to choose and scale the parameters in terms of the Hessian eigenvalue directions, such that the $\chi^2$ function can be written, up to quadratic order in ${\bf z}$, as
\begin{eqnarray}
\Delta\chi^2({\bf z})&=&T^2\sum_{i=1}^Nz_i^2\,,\label{eq:chisquare}
\end{eqnarray}
where $T$ is the tolerance parameter.\footnote{If the data errors are precisely Gaussian and internally consistent,
 then one should use a value of $T=1$ at the 68\% CL (corresponding to $T=1.645$ at 90\% CL).
 However, to accommodate experimental inconsistencies between the data sets, as well as uncertainties arising from the choice of non-perturbative parametrization,
 the value of $T$ is often increased.  The CTEQ-TEA group has used the value $T=10$ at the 90\% CL in their analyses to date.}
 The uncertainty in the PDF parameters is then set by the requirement $\Delta\chi^2\le T^2$ at some prescribed confidence level (CL), yielding
\begin{eqnarray}
\sum_{i=1}^Nz_i^2\le1\label{eq:hypersphere}
\end{eqnarray}
at the same CL.
With this choice of parameters, the best-fit PDFs are given by
$f^0(x,Q_0)\equiv f(x,Q_0;\mathrm{\bf 0})$, and the $2N$ error PDFs (2 for each eigenvector direction) are defined by $f^{\pm j}(x,Q_0)\equiv f(x,Q_0;\pm{\bf e}^{j})$,
where $({\bf e}^{j})_i=\delta^j_i$.

The usefulness of the Hessian error PDFs is in determining the PDF uncertainty of a theoretical calculation of some observable $X$.
The theoretical prediction for $X$ can be written, via the PDFs, as a function of the PDF parameters, $X({\bf z})$. Expanding this function in a Taylor expansion around the best fit gives
\begin{eqnarray}
X({\bf z})&=&X(\mathrm{\bf 0})+\sum_{j=1}^N\frac{\partial X}{\partial z_{j}}\Bigg|_{\bf 0}z_j+\frac{1}{2}\sum_{i,j=1}^N\frac{\partial^2 X}{\partial z_{i}\partial z_{j}}\Bigg|_{\bf 0}z_iz_j+\cdots\ .\label{eq:observable}
\end{eqnarray}
We can calculate $X(\mathrm{\bf 0})\equiv X(f^0)$ using the best-fit PDFs, and the first derivatives can be calculated numerically using the
$+$ and $-$ error PDFs in each eigenvector direction $j$ as
\begin{eqnarray}
\frac{\partial X}{\partial z_{j}}\Bigg|_{\bf 0}\,\approx\,\Delta X^{j}&=&
\frac{X(f^{+j})-X(f^{-j})}{2}\,.\label{eq:dx}
\end{eqnarray}
The error PDFs are not sufficient to calculate the full set of second derivatives, although the diagonal ones can be obtained as
\begin{eqnarray}
\frac{\partial^2 X}{\partial z_{j}^2}\Bigg|_{\bf 0}&\approx&
X(f^{+j})+X(f^{-j})-2X(f^0)\,.\label{eq:R}
\end{eqnarray}

The theoretical uncertainty in $X$ due to the PDFs at the specified CL is determined by the maximum and minimum values of $X$,
subject to the constraint $\sum z_i^2\le1$.  In the quadratic approximation of Eq.~(\ref{eq:chisquare}) and keeping only the linear term in
Eq.~(\ref{eq:observable}), we obtain the extrema at
\begin{eqnarray}
z_i&=&\frac{\Delta X^i}{\sqrt{\sum_{j=1}^N\bigl(\Delta X^j\bigr)^2}}.\label{eq:zextreme}
\end{eqnarray}
Thus, the upper and lower limits on the observable at the specified CL  are given by
\begin{eqnarray}
X^\pm&=&X(\mathrm{\bf 0})\pm \Delta X\,,\label{eq:limits}
\end{eqnarray}
where
\begin{eqnarray}
\Delta X&=&\sqrt{\sum_{j=1}^N\Bigl(\Delta X^j\Bigr)^2}\,.\label{eq:uncertainty}
\end{eqnarray}
This is the Symmetric Master Equation for the Hessian PDF uncertainty on the observable $X$.   The Hessian approximation is contained in the
quadratic assumption of Eq.~(\ref{eq:chisquare}) and the linear limit of Eq.~(\ref{eq:observable}), which are then supplemented by the tolerance criterion
$\Delta\chi^2\le T^2$ to obtain the symmetric uncertainties for the observable.  An Asymmetric Master Equation, which estimates asymmetric positive and negative uncertainties at the same confidence level is given in Refs.~\cite{Nadolsky:2001yg,Lai:2010vv}.

\subsection{Updating of Hessian error PDFs}\label{sec:UpdatePDFs}

Our first goal is to assess the impact of new data on the PDFs and their uncertainties.  Short of performing a full global analysis that includes the new
data, one desires a simpler and much faster way to estimate the effects of the new data.  A method that utilized the Hessian eigenvector PDFs was suggested by Paukkunen and Zurita~\cite{Paukkunen:2014zia}.  In this section we discuss the details of this method, which have been implemented in the software package {\tt ePump} (error PDF Updating Method Package).

Suppose that we have measurements for $N_X$ observables with experimental values given by $X^E_\alpha$, as well as the inverse covariance matrix $C^{-1}_{\alpha\beta}$ for the correlated experimental errors in the measurements.  The inclusion of these new data in the global $\chi^2$ function becomes
\begin{eqnarray}
\Delta\chi^2({\bf z})_{\rm new}&=&T^2\sum_{i=1}^Nz_i^2\,+\,\sum_{\alpha,\beta=1}^{N_X}\left(X_\alpha({\bf z})-X_{\alpha}^E\right)C^{-1}_{\alpha\beta}\left(X_\beta({\bf z})-X_{\beta}^E\right)\,.\label{eq:chi2new}
\end{eqnarray}
Using the generalization of Eqs.~(\ref{eq:observable}) and (\ref{eq:dx}) to expand $X_\alpha({\bf z})$ to linear order in ${\bf z}$, gives
\begin{eqnarray}
\Delta\chi^2({\bf z})_{\rm new}&=&
\sum_{\alpha,\beta=1}^{N_X}\left(X_\alpha(\mathrm{\bf 0})-X^E_\alpha\right)C^{-1}_{\alpha\beta}\left(X_\beta(\mathrm{\bf 0})-X^E_\beta\right)\nonumber\\
 &&+\,T^2\left[\sum_{i=1}^Nz_i^2\,+\,\sum_{i,j=1}^Nz_iM^{ij}z_j\,-\,2\sum_{i=1}^Nz_iA^i\right]\,,\label{eq:chi2newPZ}
\end{eqnarray}
where
\begin{eqnarray}
A^i&=&
\frac{1}{T^2}\sum_{\alpha,\beta=1}^{N_X}\left(X^E_\alpha-X_\alpha(\mathrm{\bf 0})\right)\,C^{-1}_{\alpha\beta}\,\Delta{X}^{i}_{\beta}\,,\nonumber\\
M^{ij} &=&\frac{1}{T^2}\sum_{\alpha,\beta=1}^{N_X}\Delta{X}^{i}_{\alpha}\,C^{-1}_{\alpha\beta}\,\Delta{X}^{j}_{\beta}\ .\label{eq:AandM}
\end{eqnarray}

With the additional data added to the $\chi^2$ function, the new best-fit parameters are
\begin{eqnarray}
z_i^0&=&\sum_{j=1}^N(\delta+M)^{-1}_{ij}A^j\, .\label{eq:bestfit}
\end{eqnarray}
The normalized eigenvectors $U_i^{(r)}$ and eigenvalues $\lambda^{(r)}$ of the matrix $M_{ij}$ satisfy
\begin{eqnarray}
\sum_{j=1}^N M_{ij}U^{(r)}_j&=&\lambda^{(r)} U_i^{(r)}\,,\nonumber\\
\sum_{i=1}^NU_i^{(r)}U_i^{(s)}&=&\delta_{rs}\,.
\end{eqnarray}
Note that the eigenvalues satisfy $\lambda^{(r)}\ge 0$, and if $N_X<N$ then there will be at least $N-N_X$ eigenvectors with zero eigenvalue, corresponding to directions orthogonal to the new data set.
We can now simplify the equation for $\Delta\chi^2$ by introducing new coordinates ${\bf c}\equiv(c_1,\dots,c_N)$
defined by
\begin{eqnarray}
z_i=z_i^0+\sum_{r=1}^N\frac{1}{\sqrt{1+\lambda^{(r)}}}\,c_r\,U_i^{(r)}\,,
\end{eqnarray}
to obtain
\begin{eqnarray}
\Delta\chi^2({\bf z})_{\rm new}&=&\Delta\chi^2({\bf z}^0)_{\rm new}\,+\,T^2\sum_{r=1}^Nc_r^2\,,\label{eq:chi2newPZdiagonalized}
\end{eqnarray}
where
\begin{eqnarray}
\Delta\chi^2({\bf z}^0)_{\rm new}&=&\sum_{\alpha,\beta=1}^{N_X}\left(X_\alpha(\mathrm{\bf 0})-X^E_\alpha\right)C^{-1}_{\alpha\beta}\left(X_\beta(\mathrm{\bf 0})-X^E_\beta\right)\,-\,T^2\sum_{i,j=1}^NA^i(\delta+M)_{ij}^{-1}A^j\,.\label{eq:reducedchi2}
\end{eqnarray}
Assuming the same tolerance $T$ as for the original $\chi^2$ function, gives
\begin{eqnarray}
\sum_{r=1}^Nc_r^2\le1\label{eq:hypersphere2}
\end{eqnarray}
at the same CL.

Expanding the PDFS $f(x,Q_0;{\bf z})$ to linear  order in $\bf z$, we can approximate the new best-fit PDFs, which include the impact of the new data by
\begin{eqnarray}
f^0(x,Q_0)_{\rm new}&=&f^0(x,Q_0)+\sum_{i=1}^Nz_i^0\,\Delta{f}^i(x,Q_0)\, ,\label{eq:bestfitPDF}
\end{eqnarray}
where
\begin{eqnarray}
\Delta f^i(x,Q_0)&=&\frac{f^{+i}(x,Q_0)-f^{-i}(x,Q_0)}{2}\, .\label{eq:DF}
\end{eqnarray}
In a similar manner, we can obtain updated error PDFs, by setting the new coordinates to ${\bf c}^{\pm(r)}=\pm{\bf e}^r$.  We define the new error PDFs by
\begin{eqnarray}
f^{\pm(r)}(x,Q_0)&=&f^0(x,Q_0)_{\rm new}+\frac{1}{\sqrt{1+\lambda^{(r)}}}  \sum_{i=1}^NU_i^{(r)}\,\left(f^{\pm i}(x,Q_0)-f^0(x,Q_0)\right)\, .\label{eq:newev}
\end{eqnarray}

Note that $\left(f^{\pm i}-f^0\right)=\pm\Delta f^i$ up to linear order in ${\bf z}$, so the definition of the error PDFs given in Eq.~(\ref{eq:newev}) is just one of several
equally valid choices  at linear order. On the other hand, due to nonlinearities, $(f^{+i}-f^0)\ne-(f^{-i}-f^0)$ in general, so that with the choice used here, the overall sign of the eigenvectors
$U_i^{(r)}$ becomes relevant.  We therefore supplement this definition with the choice of sign: $\sum_{i=1}^NU^{(r)}_i\ge0$.
Not that, even in the limit of $C^{-1}_{\alpha\beta}\rightarrow0$, the updated asymmetric errors will
not recover exactly the asymmetric PDF uncertainties of the original PDF set, due to the nontrivial rotation of the eigenvector basis.
However, we have found that the definition of the error PDFs given here, supplemented by the sign convention for the eigenvectors, typically gives better agreement with the original asymmetric PDF uncertainties than other choices in this limit.
In particular, in the combined limit $C^{-1}_{\alpha\beta}\rightarrow0$ and $U_i^{(r)}\rightarrow\pm\delta_i^r$ this sign choice recovers exactly the original error PDFs.

On the other hand, the symmetric uncertainties for the PDFs, as given by Eq.~(\ref{eq:uncertainty}) will coincide with the original symmetric uncertainties in the limit $C^{-1}_{\alpha\beta}\rightarrow0$.  In this limit we find
\begin{eqnarray}
\Delta f^{(r)}&=&\frac{f^{+(r)}(x,Q_0)-f^{-(r)}(x,Q_0)}{2}\ \rightarrow\ \sum_{i=1}^N U_i^{(r)}\Delta f^i \, ,\label{eq:DFr}
\end{eqnarray}
so that
\begin{eqnarray}
\Delta f&=&\sqrt{\sum_{r=1}^N\Bigl(\Delta f^{(r)}\Bigr)^2}\ \rightarrow\ \sqrt{\sum_{i=1}^N\Bigl(\Delta f^{i}\Bigr)^2}\,,\label{eq:PDFuncertainty}
\end{eqnarray}
due to the unitarity of the matrix $U_i^{(r)}$.
For this reason we advocate the use of symmetric errors when using ePump to assess the impact of new data on the PDFs, since the symmetric errors are not affected by the rotation of the eigenvector basis and therefore display exclusively the impact of the new experimental data and errors.

\subsection{Tier-2 Penalties and Dynamical tolerances}\label{sec:Weight}

The discussion up to this point assumes that the $\chi^2$ function is a smooth analytic function of the parameters, and that the confidence levels defined by the Hessian eigenvector PDFs are specified by a single global tolerance value given by $\Delta\chi^2\le T^2$.  However, in practice the confidence levels of the error PDFs are often determined by separate constraints from individual experiments, rather than by a single global tolerance.  The CTEQ-TEA group incorporates these separate constraints by adding a Tier-2 penalty to the $\chi^2$ function when determining the error PDFs, such that the tolerance requirement becomes modified
to
\begin{eqnarray}
\Delta\chi^2+\mbox{Tier-2}&\le&T^2\,.\label{eq:tier2}
\end{eqnarray}
The specific details of the Tier-2 penalty, which was introduced in Ref.~\cite{Lai:2010vv} and was discussed in detail in Ref.~\cite{Dulat:2013hea}, are not too important here, other than the fact that it increases rapidly as a function of the parameters if the disagreement between any individual experiment and theory becomes large.  Furthermore, we note that only the standard $\chi^2$ function without the Tier-2 penalty is used in determining the best-fit PDFs.

We can include the effect of these Tier-2 penalties by identifying $\Delta\chi^2=(T_i^\pm)^2\le T^2$ at the parameter values that correspond to each of the positive and negative Hessian eigenvalue PDFs at the given confidence level.  These dynamical tolerances, $T_i^\pm$, are determined either by the separate constraints from individual experiments (MMHT) or through the $\Delta\chi^2+$Tier-2 penalty (CTEQ-TEA), which limits the deviation in the particular eigenvalue direction.  Again, we note that in the CTEQ-TEA analysis, the dynamical tolerances are only used in determining the uncertainties and are not included in the $\chi^2$ function that is used to determine the best-fit PDFs.  Therefore, all of the results from section \ref{sec:UpdatePDFs} for obtaining the new best-fit parameters are unchanged, except for the numerical approximations for the coefficients in the Taylor expansion, Eq.~(\ref{eq:observable}), obtained from the error PDFs.

The relevant effect of the dynamical tolerances is that the Hessian error PDFs are now evaluated at the parameter values ${\bf z}^{\pm j}=\pm(T_j^\pm/T){\bf e}^j$.  Using this, we obtain
\begin{eqnarray}
\frac{\partial X}{\partial z_{j}}\Bigg|_{\bf 0}\,\approx\,\widehat{\Delta X}^{j}&=&
\frac{T_j^-}{T_j^++T_j^-}\left(\frac{X_\alpha(f^{+j})-X_\alpha(f^0)}{T_j^+/T}\right)\nonumber\\
&&+\frac{T_j^+}{T_j^++T_j^-}\left(\frac{X_\alpha(f^0)-X_\alpha(f^{-j})}{T_j^-/T}\right)
\,,\label{eq:dxalphaDyn}
\end{eqnarray}
and
\begin{eqnarray}
\frac{\partial^2 X}{\partial z_{j}^2}\Bigg|_{\bf 0}&\approx&
\frac{2T}{T_j^++T_j^-}\left(\frac{X_\alpha(f^{+j})-X_\alpha(f^0)}{T_j^+/T}-\frac{X_\alpha(f^0)-X_\alpha(f^{-j})}{T_j^-/T}\right)\,.\label{eq:RalphaDyn}
\end{eqnarray}
Note that the approximation for the first derivative, Eq.~(\ref{eq:dxalphaDyn}), is just the weighted average of the approximations using the positive and negative eigenvector directions.
With this modification, all of the results for the updated best-fit predictions still hold, with the replacement $\Delta X^j_\alpha\rightarrow\widehat{\Delta X}^j_\alpha$
everywhere.  We also make the replacement $\Delta f^i(x,Q_0)\rightarrow\widehat{\Delta f}^i(x,Q_0)$ to obtain the best-fit PDFs
\begin{eqnarray}
f^0(x,Q_0)_{\rm new}&=&f^0(x,Q_0)+\sum_{i=1}^Nz_i^0\,\widehat{\Delta{f}}^i(x,Q_0)\, .\label{eq:bestfitPDFdt}
\end{eqnarray}
Note that there is actually no dependence on the global tolerance $T$ in any of these formulae, since it can be removed simply by re-scaling the parameters $\bf{z}$.

The generalization of the best-fit results to include the dynamical tolerances is relatively straightforward, as we have seen.
However, the
calculation of the Hessian eigenvector PDFs requires that we first address some practical aspects of the updating method.
In the original global analysis the dynamical tolerances were obtained by the requirement
$\Delta\chi^2=(T^\pm_i)^2$ along each of the positive and negative eigenvalue directions.
The new eigenvector PDFs, however, will also depend on the dynamical tolerances, $T^\pm_{(r)}$, along the updated eigenvector directions.  To obtain these, one would need to calculate the constraints from each of the individual data sets included in the global analysis on each of the new updated eigenvector PDFs.  This is beyond the applicability of the present method, which assumes that all information of the global $\chi^2$ function is contained in the original error PDFs, except for the contribution of the new data set.
Even if one assumes that the criterion for constraining the eigenvector directions is the same for the original and updated error PDFs, it is not possible to map directly from the $T^\pm_i$ to the $T^\pm_{(r)}$ without doing a more detailed calculation involving all of the global analysis data sets, since the original and updated eigenvector directions are not parallel.

If the new dynamical tolerances were known, one could then calculate the corresponding eigenvector parameters,
\begin{eqnarray}
{z}_i^{\pm (r)}=z_i^0\pm\frac{1}{\sqrt{1+\lambda^{(r)}}}\left(\frac{T^\pm_{(r)}}{T}\right)U_i^{(r)}\,.\label{eq:zErrorDyn}
\end{eqnarray}
and obtain the updated error PDFs using the linear expansion as before.  In their absence we proceed by noting that we should recover
the original error PDFs in the combined limits, $C^{-1}_{\alpha\beta}\rightarrow0$ and $U_i^{(r)}\rightarrow\pm\delta_i^r$.
Thus, a reasonable assumption is to replace $(T^\pm_{(r)}/T)$ with $(T^\pm_{i}/T)$ in Eq.~(\ref{eq:zErrorDyn}), and approximate the numerical derivatives using the positive or negative eigenvector PDFs, correspondingly.
Roughly, this assumes that the effect of the new dynamical tolerances is not substantially different from the effect of the old ones
(and in particular, that the main contribution of the new data to the reduction of PDF errors comes through the standard $\chi^2$ treatment and not through additional constraints, such as the Tier-2 penalty).
Note that with the above replacement, the factors of $(T^\pm_{i}/T)$ in Eq.~(\ref{eq:zErrorDyn}) cancel against the same factors
in the numerical derivatives, so that we still use Eq.~(\ref{eq:newev}) for the updated error PDFs in the presence of dynamical tolerances.  The only difference is that now the eigenvalues $\lambda^{(r)}$ and rotation matrices $U_i^{(r)}$ depend on the
original dynamical tolerances (since the matrix $M^{ij}$ depends on the original dynamical tolerances).
Again, we advocate the use of symmetric errors when evaluating the effects of new data, since they are less sensitive to nonlinear
effects arising from the rotation of the eigenvectors directions.

\subsection{Direct Updating of Observable Predictions and Uncertainties}\label{sec:UpdateObservables}

Given the updated central and Hessian error PDFs, obtained as described in the last section, it is possible now to calculate updated predictions
and uncertainties for any other set of observables $Y_\alpha$ (including the original observables $X_\alpha$
that were used for updating the PDFs).  This would require running the code to calculate the observables using the updated central and each of the updated error PDFs.  However, frequently the predictions with the original central and error PDFs are already known, so it would be more time-efficient to use these to calculate
the updated predictions and uncertainties directly.  It is very easy to obtain the updated results using the approach presented in the last section.

Suppose we have already calculated $Y_\alpha(\mathrm{\bf 0})\equiv Y_\alpha(f^0)$ and $Y_\alpha(f^{\pm j})$.  Then using the linear expansion, we obtain
\begin{eqnarray}
Y_{\alpha}(f^0_{\rm new})&=&Y_{\alpha}(f^0)+\sum_{i=1}^Nz_i^0\,\widehat{\Delta Y}_\alpha^i\, ,\label{eq:centralY}
\end{eqnarray}
where $z_i^0$ was given in Eq.~(\ref{eq:bestfit}) and $\widehat{\Delta Y}_\alpha$ is defined in Eq.~(\ref{eq:dxalphaDyn}).
For the uncertainty, we use the convention described at the end of the last section and note that
\begin{eqnarray}
\Delta Y_\alpha^{(r)}&=&\frac{Y_\alpha(f^{+(r)})-Y_\alpha(f^{-(r)})}{2}\nonumber\\
&=&\frac{1}{\sqrt{1+\lambda^{(r)}}}\sum_{i=1}^NU_i^{(r)}\Delta Y_\alpha^{i}\,,\label{eq:dyr}
\end{eqnarray}
which is valid in the linear approximation.
Thus, we can obtain the updated symmetric uncertainties for the observable $Y_\alpha$ as
\begin{eqnarray}
\left(\Delta Y_\alpha\right)^2_{\rm new}&=&\sum_{r=1}^N\Bigl(\Delta Y_\alpha^{(r)}\Bigr)^2\nonumber\\
&=&\sum_{r=1}^N\frac{1}{1+\lambda^{(r)}}\sum_{i,j=1}^N\left(\Delta Y_\alpha^iU_i^{(r)}\right)\left(\Delta Y_\alpha^jU_j^{(r)}\right)
\nonumber\\
&=&\sum_{i,j=1}^N\Delta Y_\alpha^i\,(\delta+M)_{ij}^{-1}\,\Delta Y_\alpha^j\,,\label{eq:uncertaintyY}
\end{eqnarray}
where we have used the completeness of the eigenvectors $U_i^{(r)}$ in the last equation, and the matrix $M_{ij}$ was given in Eq.~(\ref{eq:AandM}).

Equations (\ref{eq:centralY}) and (\ref{eq:uncertaintyY}) give the results for the updated central prediction and updated symmetric uncertainty for the observable $Y_\alpha$
in terms of calculations that use only the original central and Hessian error PDFs.
These results can also be generalized to the
correlation cosine between two observables $Y_\alpha$ and $Y_\beta$,
which can be experimental observables or PDF distributions at specific $x$ and $Q$ values.
The original correlation cosine between the observables is given by~\cite{Nadolsky:2008zw}
\begin{eqnarray}
\cos(\theta_{\alpha\beta})
&=&\frac{\sum_{i=1}^N\Delta Y_\alpha^i\Delta Y_\beta^i}{\left|\Delta Y_{\alpha}\right|\,\left|\Delta Y_{\beta}\right|}\,.\label{eq:corrcosine}
\end{eqnarray}
It represents the degree of correlation between the two observables as determined by the PDFs, with $\cos(\theta_{\alpha\beta})=1$ completely correlated,
 $\cos(\theta_{\alpha\beta})=-1$ completely anticorrelated, and $\cos(\theta_{\alpha\beta})=0$ no correlation.
Using the same steps as for Eq.~(\ref{eq:uncertaintyY}), we obtain the updated correlation between the two observables to be
\begin{eqnarray}
\cos(\theta_{\alpha\beta})_{\rm new}
&=&\frac{\sum_{i,j=1}^N\Delta Y_\alpha^i\,(\delta+M)_{ij}^{-1}\,\Delta Y_\beta^j}{\left|\Delta Y_{\alpha}\right|_{\rm new}\,\left|\Delta Y_{\beta}\right|_{\rm new}}\,.\label{eq:corrcosinenew}
\end{eqnarray}

\subsection{Limitations on the use of the error PDF updating method}\label{sec:limitations}

The method used in ePump is very powerful and can very quickly give results on the impact of new data on the PDFs.  However, it cannot be used to
replace a full PDF global analysis of experimental data.  In this section we discuss the limitations of this method and consider ways to assess where it might
fail.  There are basically four categories of assumptions that go into the method, and we will discuss each of them below.

The first assumption that goes into the Hessian updating method is the quadratic dependence of the chi-square function on the parameters for the original data.  Although this assumption is expected to be good in most of the eigenvalue directions, it is possible that some directions that are not well-probed by the original data
can have noticeable deviations from quadratic dependence.  Unfortunately, this cannot be assessed within ePump, since it would require a detailed analysis of the original global chi-square function from the original global analysis.  However, a reasonable expectation is that the further the parameters deviate from the original best fit, the more likely that non-quadratic behavior of the chi-square function becomes important.  A measure of the deviation of the parameters from the original best fit is
\begin{eqnarray}
\Delta\chi^2({\bf z}^0)_{\rm original}&=&T^2\sum_{i=1}^N\left(z^0_i\right)^2\,,\label{eq:chi2originaldata}
\end{eqnarray}
which is the change in chi-square for the original data set,  evaluated at the updated best-fit parameters.  This was called the ``penalty term'' in Ref.~\cite{Paukkunen:2014zia}.
A comparable measure in the presence of dynamical tolerances would be the fractional distance-squared between the new and old best-fits relative to the confidence-level boundary in the space of parameters, which can be written
\begin{eqnarray}
(d^0)^2&=&\sum_{i=1}^N\left(\frac{T}{T_i}z^0_i\right)^2\,,\label{eq:distance1}
\end{eqnarray}
where we have let $T_i$ be the average of $T^\pm_i$.
In the calculation of the PDF uncertainties in the Hessian method, it is assumed that the chi-square function is quadratic in the parameters for $d^2\le1$.  In addition,
 independent of the quadratic approximation, a value $(d^0)^2$ larger than one would indicate
that either there is tension between the new data and the original data, or else the uncertainties in the original global analysis were
under-estimated.  This latter interpretation might occur if the new data probe a region of $x$ and $Q^2$ for the PDFs for which they are undetermined, so
that the original error estimate itself would not be well-determined.

The second assumption that goes into the Hessian updating method is the linear dependence of the observables $X_\alpha$ on the parameters.
This can be tested somewhat by comparing results using the linear approximation for the dependence of the observables on the parameters, as presented above, with a calculation which also includes the diagonal quadratic contribution to the observables.  We present an efficient method for the calculation including these terms in appendix A.  Note that since we can only calculate the diagonal quadratic terms using the error PDFs, this nonlinear prediction of the updated PDFs
is not guaranteed to be more correct than the strictly linear approximation.  However, if the differences between the two predictions are small, one may expect that linear approximation for the observables is probably good.

Related to this second assumption, one might be interested to know if the original best-fit is outside the uncertainty bounds of the 
updated fit, even if the updated best fit is within the original uncertainties.  
This might occur if the new data strongly constrain the PDFs in a
region of the parameter space that was essentially unconstrained in the original fit.  In analogy to Eq.~(\ref{eq:distance1}), we can define the
fractional distance-squared between the old and new best-fits relative to the new confidence boundary:
\begin{eqnarray}
(\tilde{d}^0)^2&=&\sum_{r=1}^N\left(\frac{T}{T_{(r)}}c^0_r\right)^2\,,\label{eq:distance2}
\end{eqnarray}
where
\begin{eqnarray}
c_r^0&=&-\sqrt{1+\lambda^{(r)}}\sum_{i=1}^Nz_i^0U_i^{(r)}\,\label{eq:newparam}
\end{eqnarray}
is the location of the original best-fit written in the updated coordinates.  If we use the same replacement
$(T_{(r)}/T)\rightarrow(T_{i}/T)$ as we did in determining the updated eigenvector PDFs, we can approximate this by 
\begin{eqnarray}
(\tilde{d}^0)^2&=&\sum_{i,j=1}^N\left(\frac{T}{T_{i}}z^0_i\right)\left(\delta +M\right)_{ij}\left(\frac{T}{T_{j}}z^0_j\right)\,.\label{eq:distance2b}
\end{eqnarray}
Note that $\tilde{d}^0\ge d^0$, because $\lambda^{(r)}\ge0$ or equivalently $M_{ij}$ is non-negative.  This is related to the fact
that in the Hessian approximation the uncertainties cannot increase when new data are added, although this can be possible
if nonlinear effects not included in the approximation are important.
Also note that if this measure is larger than one, it does not mean that the new data are in tension with the old data, or even that any of the assumptions in the error PDF updating method have necessarily broken down.  However, if $\tilde{d}^0\gg1\gg d^0$, it implies that the grid of points evaluated in the parameter space (defined by the original error PDFs) is large compared to the scale of the variations in the contribution of the new data to $\chi^2$.  In this case the modeling of the $X_\alpha({\bf z})$ near the new best fit may be more likely to be affected by nonlinearities, which could produce results that differ from the true global fit.

The third assumption that goes into the Hessian updating method is the linear dependence of the PDFs, or any other observable whose prediction is to be
updated by the impact of the new data. We note that this assumption is the same as that used in any determination of uncertainties using the Hessian method.
In particular, the predictions of the updated best-fit PDFs should be good in the regions where they have some constraints from the original data, but in regions where they are essentially unconstrained, such as at very large or small $x$, one might expect the results of ePump to be less reliable.

Finally, and probably most seriously, the error PDF updating method is tied to all of the systematic assumptions that went into the original global analysis.
This includes the choice of parametrization and number of parameters in the PDFs; any constraints on PDFs, such as $s=\bar{s}$ or constraints on sea-quark
distributions; choice of global tolerance and treatment of dynamical tolerances ({\it i.e.,} Tier-2 penalty in the case of the CT distributions); and any other systematic choices (such as the mass of the heavy quarks, the value of $\alpha_s$, etc.) that went into the original global analysis.  Therefore, new insights that may change some of the original assumptions, or the inclusion of new data
that may require additional flexibility in the parametrization (either through additional parameters or relaxation of constraints between parameters) cannot be
probed using the present method.

\subsection{Examples of Updating using ePump}\label{sec:ExampleU}

In this section we show some examples of updating the PDFs and observables using ePump.  We first begin by checking the results of updating with ePump versus refitting the PDFs with a full global analysis.  Then we consider an example of how ePump might be used in
practice to analyze the impact of a new data set by using ePump to update a current standard PDF set with the inclusion of the new data.

As a check on the consistency of the ePump results with the full global analysis, we have carried out the following studies.
Firstly, we have checked that the ePump analysis can reproduce the full CT14HERA2 best-fit to a very good accuracy.
This was done by taking the complete original CT14HERA2 data sets as the ``new" input data set and using ePump
to update the CT14HERA2 PDFs. As expected the updated best-fit PDFs are essentially unchanged, and the total $\chi^2$ of the best-fit found in the ePump analysis is only higher by
0.8 units as compared to the minimum total $\chi^2$ of the original CT14HERA2 best-fit, which is 3594 (for the total of 3302 data points). We have also checked that the updated predictions for physical observables from the ePump updating code agree perfectly with our global fitting code for all the data points included in our CT14HERA2 fit. The uncertainties on physical observables induced by the PDF errors predicted by the ePump code
also agree with our global fit after taking into account the factor of $1/\sqrt{2} \sim 0.707$ reduction in the  ePump result because the CT14HERA2 data set was effectively counted twice in this exercise.  Although the fact that the central value of the best-fit parameters
does not change in this exercise makes the updating in this case less complex, it does demonstrate that ePump can easily handle a large
data set and that the uncertainties for the full global analysis are reasonably given in the Hessian approximation.
More detailed discussions about this comparison will be presented in a separate paper~\cite{paper-2}.

\begin{figure}[t]
\includegraphics[width=0.43\textwidth]{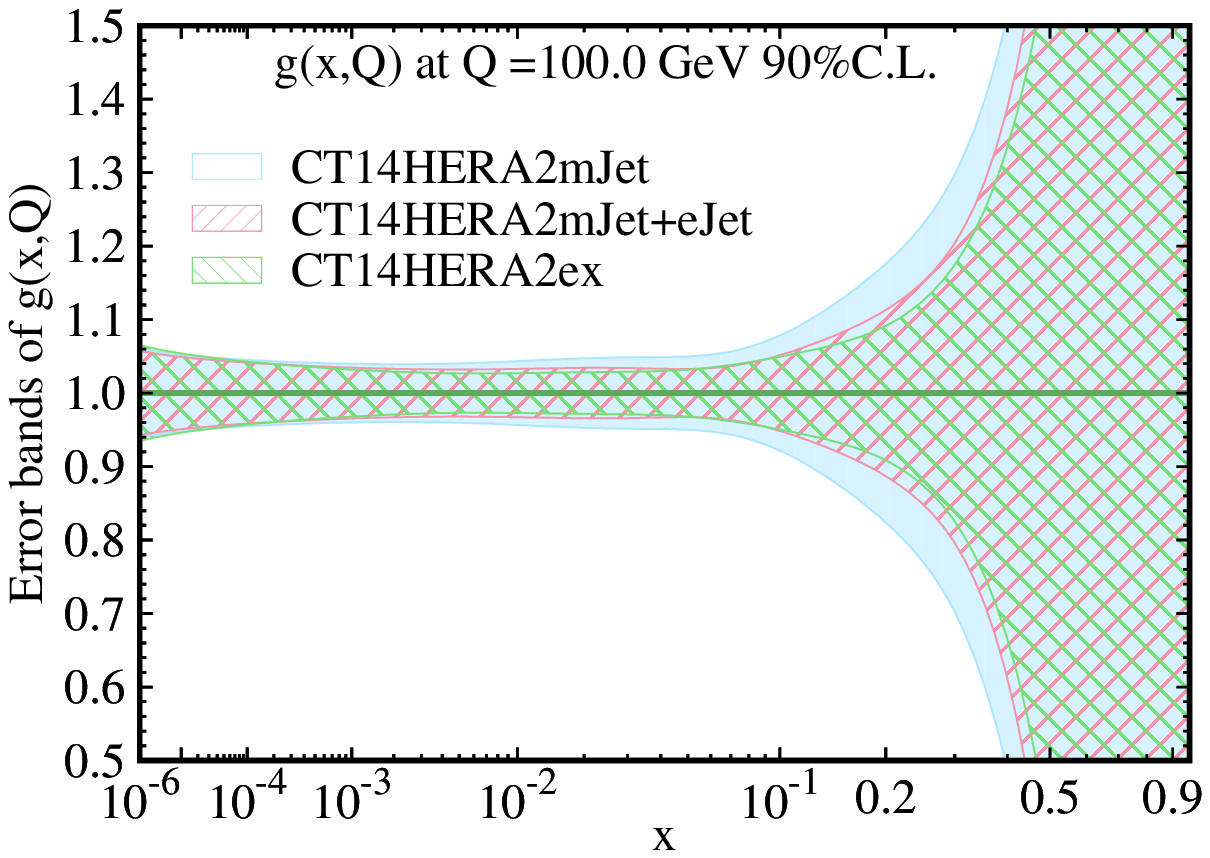}
\includegraphics[width=0.43\textwidth]{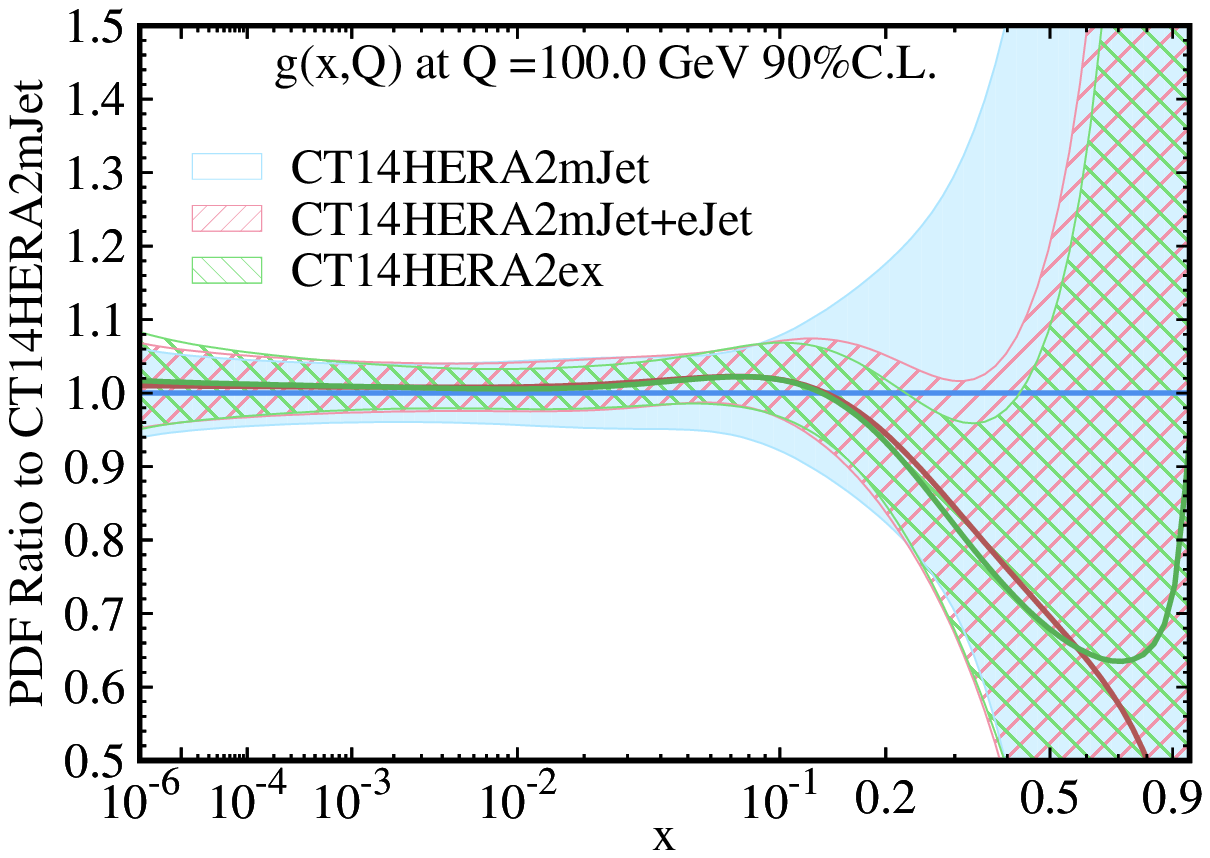}
\caption{Comparison of the gluon-PDF for three analyses: CT14HERA2mJet is the standard CT14HERA2 analysis but with the
four inclusive jet data sets removed, CT14HERA2mJet$+$eJet is the CT14HERA2mJet set updated by ePump to include the missing jet data sets, and CT14HERA2ex is the full CT14HERA2 analysis (but with the two gluon extreme sets excluded in computing the uncertainty bands).
}\label{fig:ct14hera2mjet}
\end{figure}

Secondly, we have compared the results of adding new data with ePump versus adding them with the complete global fitting code.
Specifically, we prepared a base set of PDFs from a global fit to the full CT14HERA2 data set with all the jet data removed, called the CT14HERA2mJet PDF set.  This includes a central set (00) and error PDF sets 01 through 54, so that there are $N_i=27$ eigenvector directions.  This number of error PDFs is the same as that in the CT14HERA2 PDF set, provided we exclude the two extreme sets from CT14HERA2, which were added to enlarge the gluon-PDF errors in the small $x$ region\footnote{As compared to the original CT14NNLO PDF set, the CT14HERA2 PDF set contains one more shape parameter to describe the strange-PDF. Hence, it was not necessary to introduce two additional extreme sets to enlarge the strange-PDF uncertainty in the small $x$ region (in addition to the two gluon extreme sets), as was done in the CT14NNLO PDFs.}.
We then compared the results of adding back the four sets of inclusive jet data (CDF \cite{Aaltonen:2008eq}, D\O~ \cite{Abazov:2008ae},
ATLAS \cite{Aad:2011fc}, and CMS \cite{Chatrchyan:2012bja}) using ePump (labeled CT14HERA2$+$eJet) with the results of adding them back using the full global analysis ({\it i.e.,} the published CT14HERA2 global fit, with the two extreme $g$-PDF sets removed, labeled CT14HERA2ex).
In Fig.~\ref{fig:ct14hera2mjet}, we show the comparison of the $g$-PDFs at the scale $Q=100$ GeV for the CT14HERA2mJet, CT14HERA2ex, and the CT14HERA2$+$eJet fits.
The left curves and error bands are normalized to their respective central fits, while the right curves and error bands are each normalized to the CT14HERA2mJet central PDF.
These plots clearly show that the jet data included in the CT14HERA2 fit further constrain the $g$-PDF for $x$ between a few times $10^{-4}$ to about 0.5.
In addition, they show that ePump reproduces the result of the original CT14HERA2 fit for $x$ 
in the same range.  Outside of that range, particularly at very large $x$, the PDFs are nearly unconstrained relative to the best fit,
so that nonlinear dependence on the parameters becomes significant, which explains the apparent difference between the result of ePump updating and true global fit.  We note that the distance in parameter-space between the original and new best-fits, relative to the original confidence-level boundary, (given by Eq.~(\ref{eq:distance1})) is
$d^0=0.6$, indicating that the new data is consistent within the uncertainties of the CT14HERA2mJet fit.  However, relative to the new confidence-level boundary (Eq.~(\ref{eq:distance2})) the distance is $\tilde{d}^0=1.3$, indicating that the original parameter values are 
outside the confidence-level boundaries of the updated fit.  This might lead to some concern that the approximations used in evaluating the contributions to $\chi^2$ of the new data might break down in this analysis.  To check on this we have also performed the update with ePump while including the diagonal quadratic terms, as discussed in Appendix A, and found that the gluon PDFs are essentially indistinguishable with the linear result except at very large $x\gtrsim0.6$.  Of course, the ePump result is further vindicated by the direct comparison with the full CT14HERA2 global fit shown in Fig.~\ref{fig:ct14hera2mjet}.

\begin{figure}[t]
\includegraphics[width=0.43\textwidth]{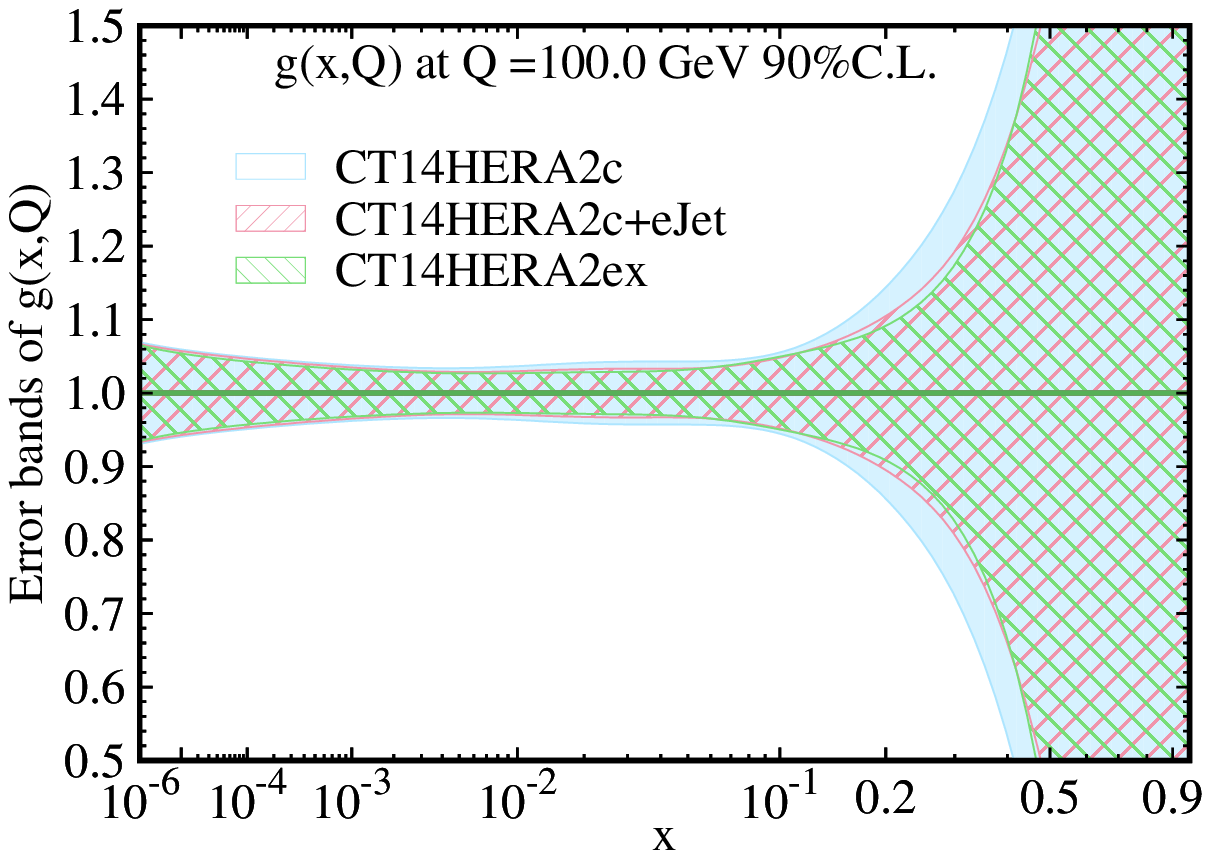}
\includegraphics[width=0.43\textwidth]{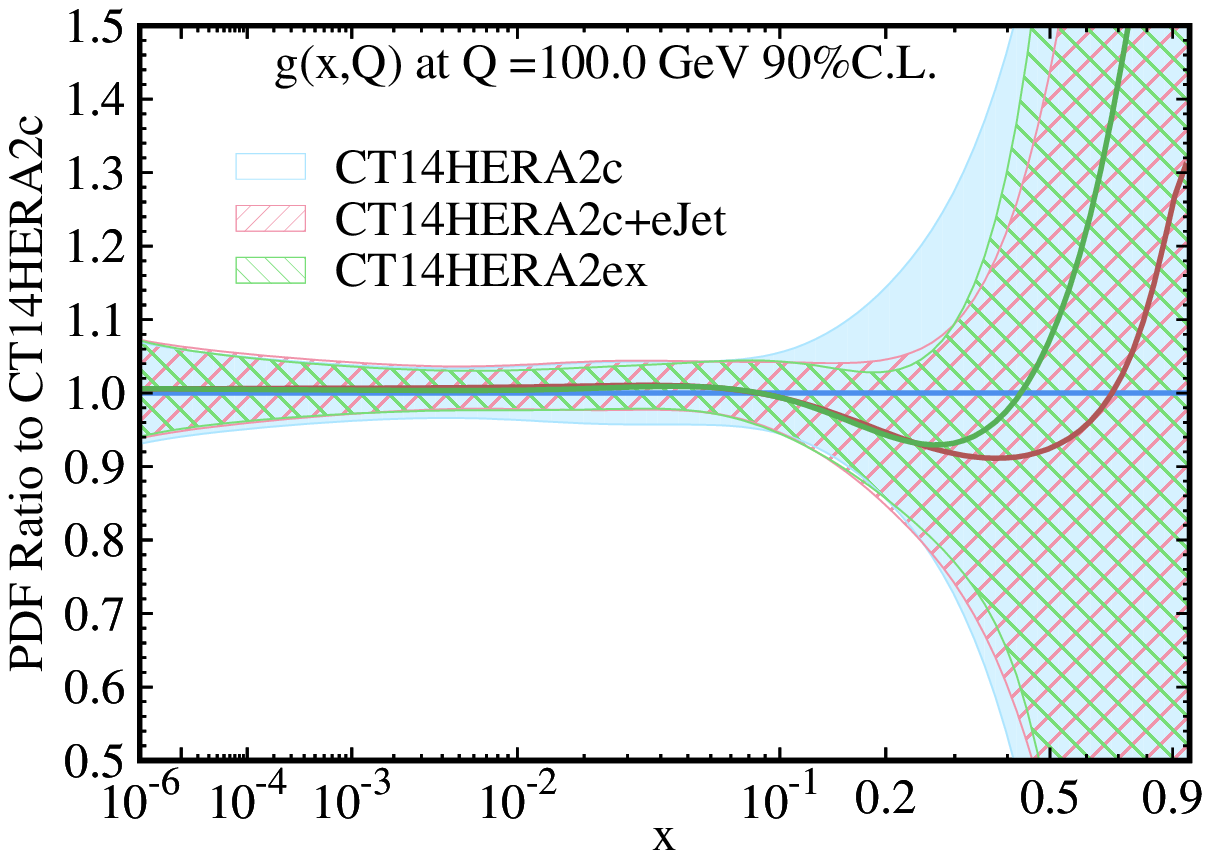}
\caption{Comparison of the gluon-PDF for three analyses: CT14HERA2c is the standard CT14HERA2 analysis but with
with all, except CDF, jet data removed, CT14HERA2c$+$eJet is the CT14HERA2c set updated by ePump to add back
the D\O~, ATLAS and CMS inclusive jet data, and CT14HERA2ex is the full CT14HERA2 analysis (but with the two gluon extreme sets excluded in computing the uncertainty bands). }\label{fig:ct14hera2J504}
\end{figure}

In practice, a more typical use of ePump would be to add new data to the original CT14 or CT14HERA2 PDF analysis, in which some jet data have already been included to constrain the $g$-PDF. Hence, the difference in the $g$-PDF between 
the original global fit and the updated ePump analysis would be expected to be smaller than what is observed in Fig.~\ref{fig:ct14hera2mjet}, unless the new data set had a strong tension with the old data set included in the CT14 global analysis. Though more detailed discussions about this topic will be presented in a separate paper, we shall illustrate this point by the following example.
The data set of the CT14HERA2 global analysis includes four sets of inclusive jet data, which are
CDF \cite{Aaltonen:2008eq}, D\O~ \cite{Abazov:2008ae},
ATLAS \cite{Aad:2011fc}, and CMS \cite{Chatrchyan:2012bja} inclusive jet data.
Let us consider a new global fit in which all of the CT14HERA2 non-jet data is included, but only the CDF jet data \cite{Aaltonen:2008eq} is included, which results in a
new error PDF set, called CT14HERA2c, with 54 error PDFs plus its central set.
In Fig.~\ref{fig:ct14hera2J504} we show the $g$-PDF and its error bands for CT14HERA2c, for CT14HERA2ex (excluding its last two extreme PDFs), and for CT14HERA2c updated by ePump to add back the other three (D\O~, ATLAS and CMS) jet data, whose result is labelled as CT14HERA2c$+$eJet. It is interesting to note that the CDF data alone already constrain the $g$-PDF in the relevant $x$ range.  
This is corroborated by considering the distance in parameter-space, both relative to the original and updated confidence-level boundaries.  For this case we obtain $d^0=0.42$ and $\tilde{d}^0=0.65$, respectively.  The first value shows that the additional
three jet data are consistent with the CDF data, while the second value shows that the CDF data alone is sufficiently strong that
the CT14HERA2c best-fit is well within the confidence-level boundaries obtained after including the full set of jet data. 
In addition, we see that the ePump result agrees very well with CT14HERA2 in the $g$-PDF for most of the range of $x$, with almost
perfect agreement up to about $x\sim0.3$.

\begin{figure}[t]
\includegraphics[width=0.43\textwidth]{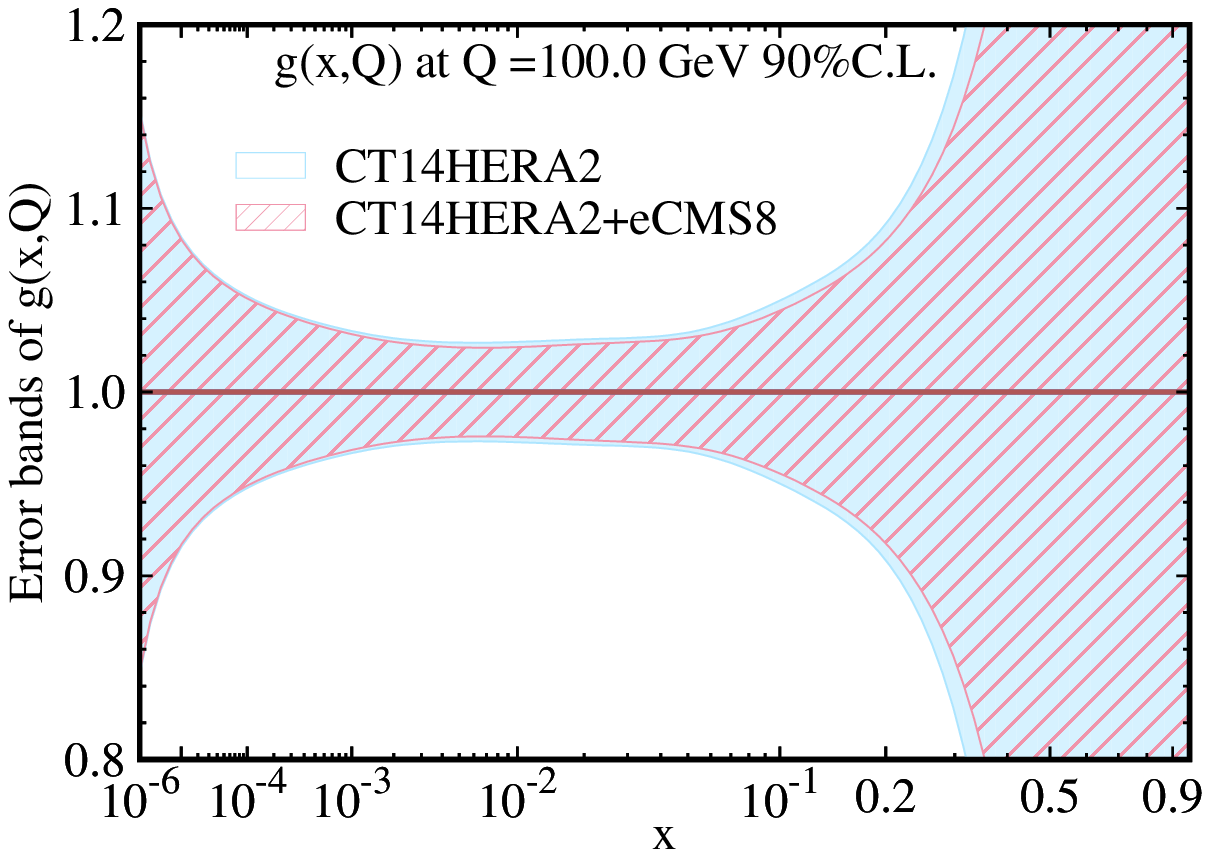}
\includegraphics[width=0.43\textwidth]{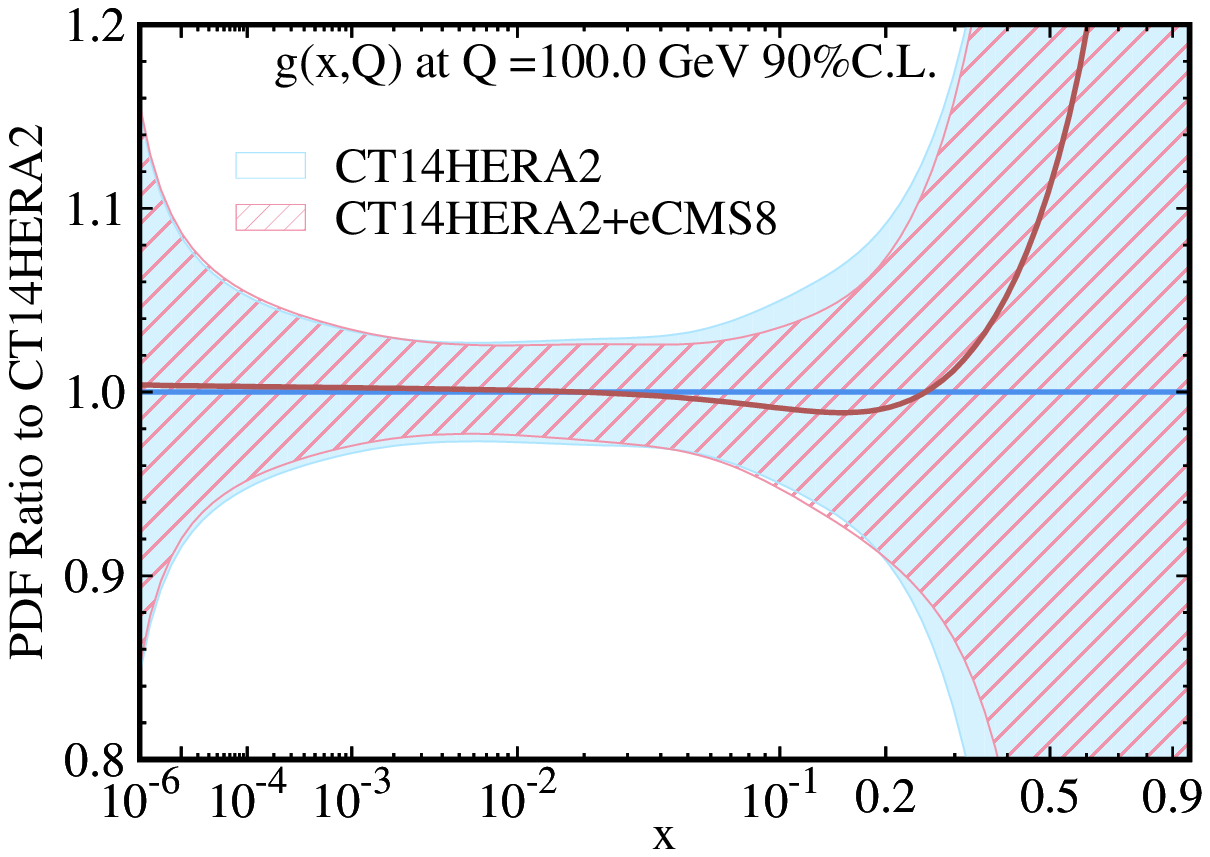}
\caption{
Comparison between the gluon-PDF at $Q=100$ GeV for CT14HERA2 NNLO and for CT14HERA2 updated by ePump using
 the CMS 8 TeV double-differential inclusive jet cross section measurements as a function of jet rapidity ($y$) and transverse momentum ($q_T$).
The left curves and error bands are normalized to their respective central fits, while the right curves and error bands are each normalized to the CT14HERA2NNLO central PDF.
}\label{fig:cms8jet}
\end{figure}

Finally, we present a simple example of the error PDF updating method using ePump by adding new data to a current standard PDF set.
We consider the effect of the
CMS inclusive jet data measured at the LHC at 8 TeV~\cite{Khachatryan:2016mlc},
when added to the CT14HERA2NNLO global analysis.  In other words, we use ePump to obtain an approximation
of the outcome of a full NNLO global analysis that includes all of the data included in the CT14 analysis plus the CMS double-differential inclusive jet cross section measurements as a function of jet rapidity ($y$) and transverse momentum ($p_T$). 
For this study we have included all of the rapidity bins for a total of 185 data points.
As in our PDF global analysis, we have considered all three different types of experimental error, including statistical, correlated and uncorrelated errors, in the ePump analysis. (Details of the error analysis are discussed in Appendix B.) For the theoretical prediction, we have used the FastNLO grids~\cite{Wobisch:2011ij,fastnlo-cms8jet}, including the corresponding $k$-factors, defined as the ratios of NNLO to NLO double-differential inclusive jet cross sections for each data point. 
For the initial PDF set we have used the CT14HERA2NNLO
central set (00) and the error PDF sets 01 through 56, including the final two extreme sets.
With this set of inputs, we can now use the ePump executable {\tt UpdatePDFs} to calculate the impact of the new data on the CT14HERA2NNLO PDFs,
using the method discussed in section~\ref{sec:UpdatePDFs}.
In Fig.~\ref{fig:cms8jet} we show the updated gluon-PDF evaluated at the scale $Q=100$ GeV.
As before, the left curves and error bands are normalized to their respective central fits in order to emphasize the changes to the uncertainty bands, while the right curves and error bands are each normalized to the CT14HERA2NNLO central PDF in order to emphasize the changes to the updated central fit.
These plots show that including these data makes the gluon-PDF become softer when $x$ is larger than around 0.2, and it reduces the 
uncertainty in the gluon PDF in the same region of $x$.

\begin{figure}[t]
\includegraphics[width=0.43\textwidth]{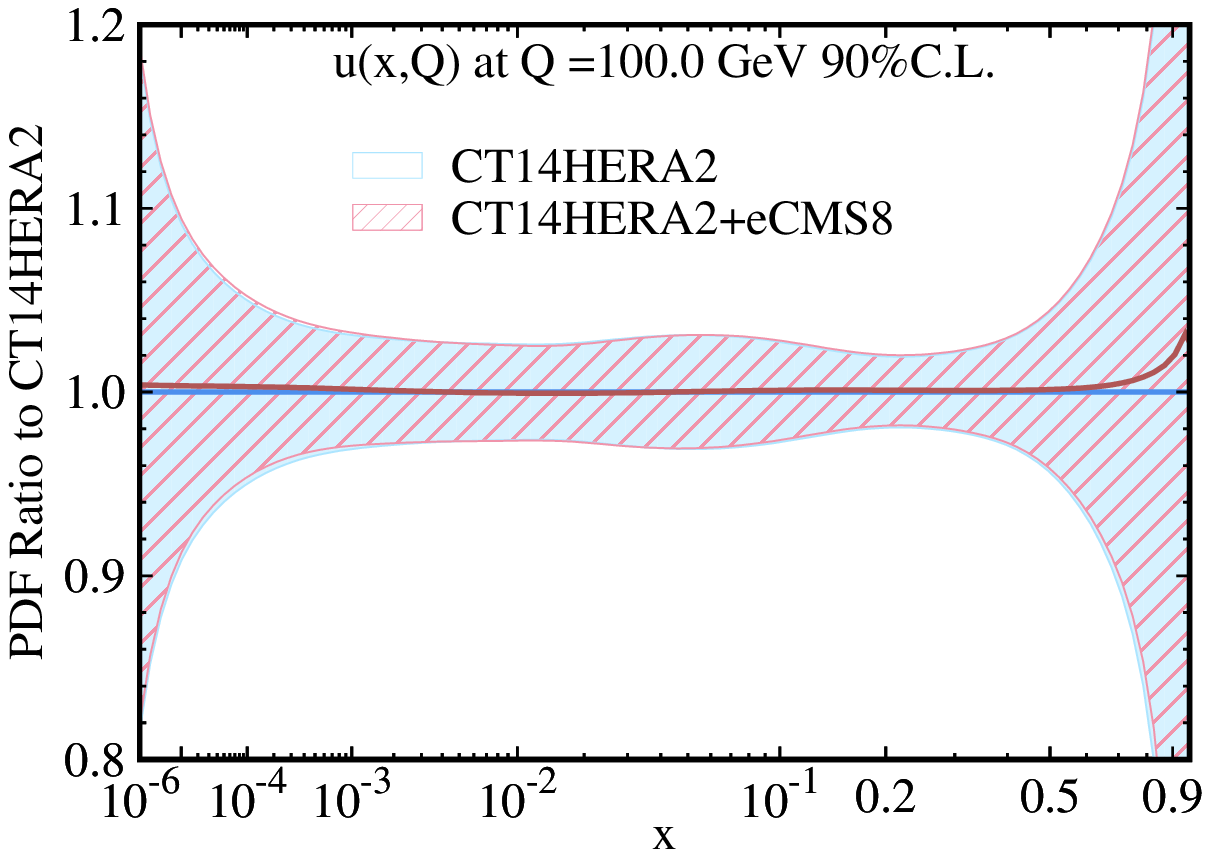}
\includegraphics[width=0.43\textwidth]{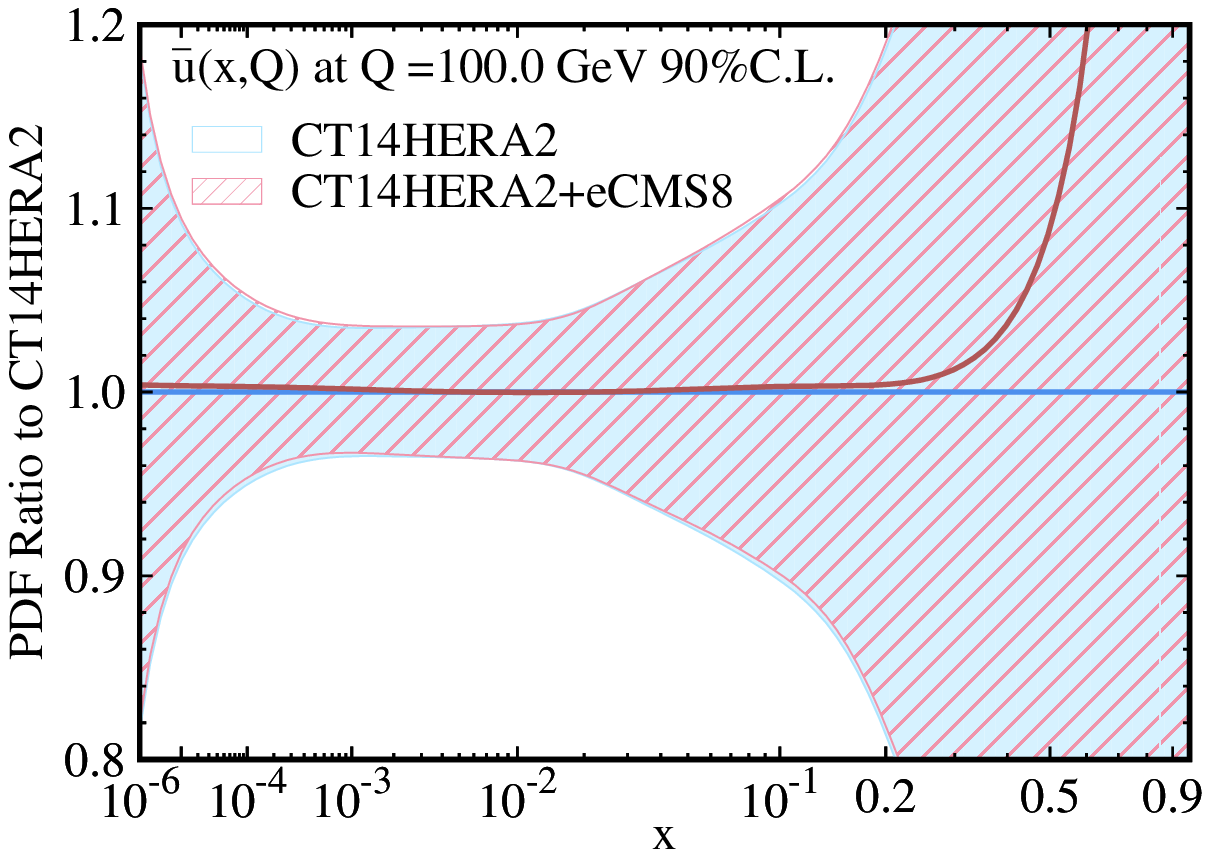}
\includegraphics[width=0.43\textwidth]{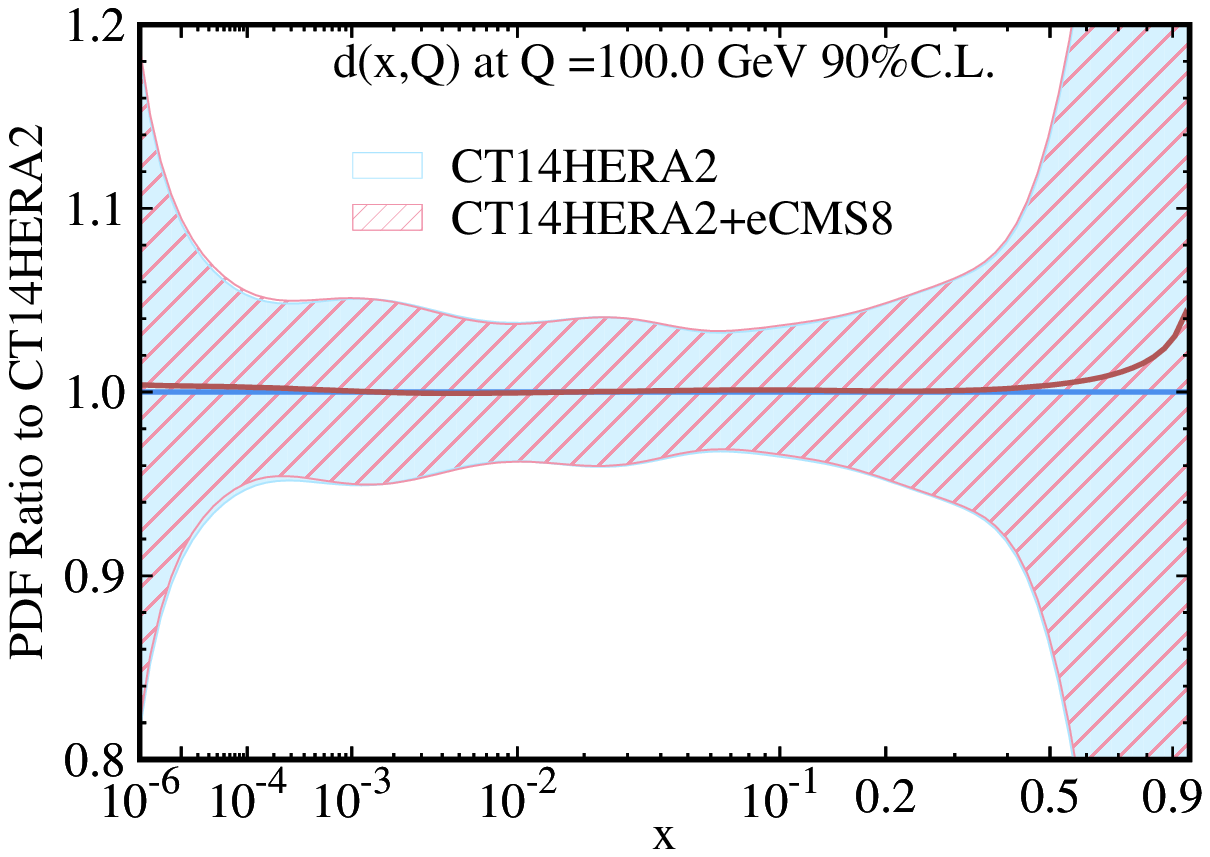}
\includegraphics[width=0.43\textwidth]{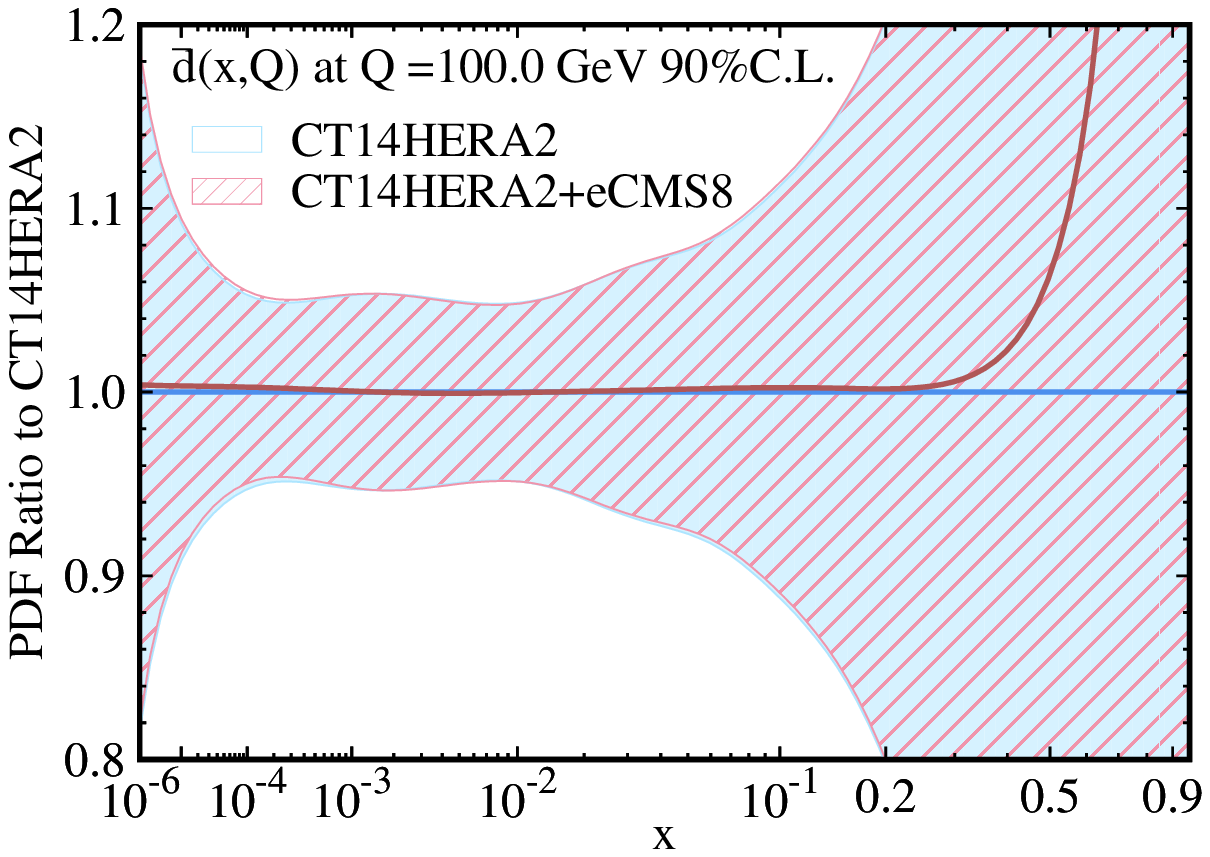}
\includegraphics[width=0.43\textwidth]{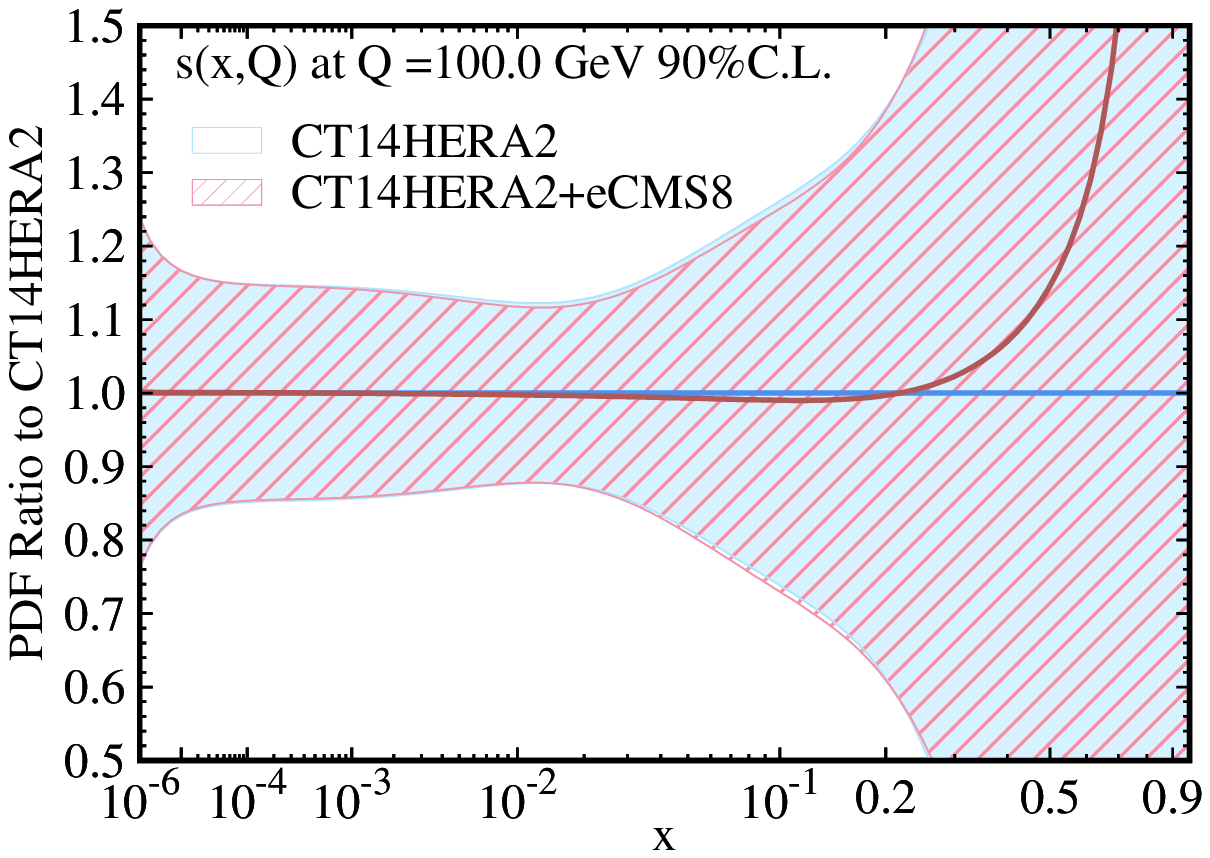}
\includegraphics[width=0.43\textwidth]{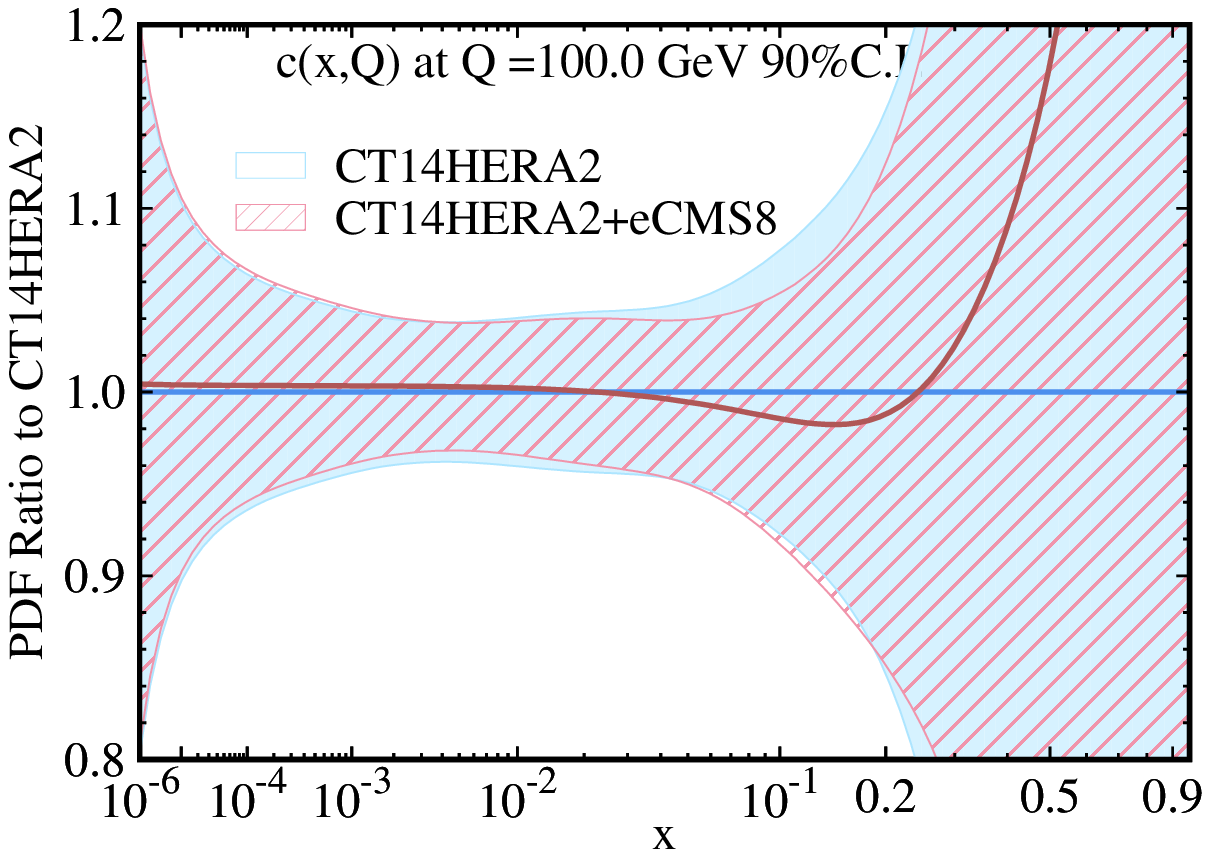}
\caption{Same as Fig.~\ref{fig:cms8jet}, but for the updated $u$, $d$, $\bar u$, $\bar d$, $s$ and $c$ PDFs.
}\label{fig:cms8jetrest}
\end{figure}

\begin{figure}[t]
\includegraphics[width=0.43\textwidth]{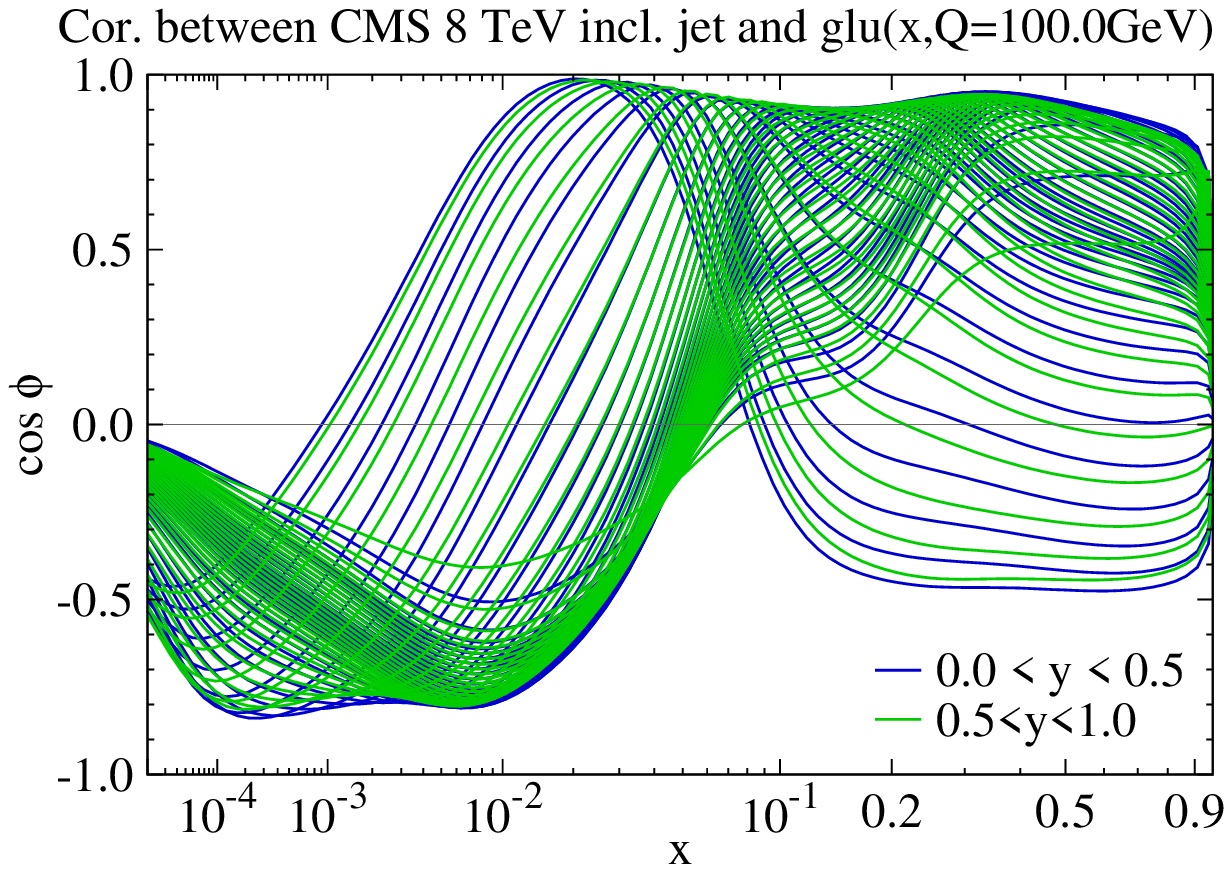}
\caption{Correlation cosine, cf. Eq.~(\ref{eq:corrcosine}),
of CMS 8 TeV double-differential inclusive jet cross section data points in the first two bins of rapidity versus various the
gluon PDF,
as a function of $x$ at $Q=100$ GeV.
}\label{fig:cosineCMS8}
\end{figure}

For completeness, we also compare in Fig.~\ref{fig:cms8jetrest} the changes in the other flavor
($u$, $d$, $\bar u$, $\bar d$, $s$ and $c$) PDFs after
updating the CT14HERA2 PDFs using ePump to include the
additional CMS 8 TeV jet data. Though these jet data are expected to have the largest effect on the $g$-PDF, it could also modify slightly the other flavor PDFs, as shown in the figure.  
In Ref.~\cite{Nadolsky:2008zw} it was proposed that the most relevant parton flavor and its range of $x$ values affected by this
new data set could be found by examining the correlation cosine, cf.~Eq.~(\ref{eq:corrcosine}), between
 each data point and the various PDF flavors. The result, displayed in
Fig.~\ref{fig:cosineCMS8} for the correlation cosines between the gluon PDF as a function of $x$ at $Q=100$ GeV with the CMS 8 TeV jet  data points in the first two bins of rapidity, shows that the CMS 8 TeV double-differential inclusive jet cross section data are
most sensitive to the $g$-PDF a $x$ between a few times $10^{-2}$ and 0.5.
 This result is consistent with the sensitivity analysis of Ref.~\cite{Wang:2018heo}, which identified
this data set as one of the more important new LHC data sets to constrain the PDFs.

The distance in parameter space between the original and updated best-fits for this analysis is $d^0=0.33$ and $\tilde{d}^0=0.42$,
relative to the original and updated 90\% confidence levels, respectively, indicating that the ePump result is most likely reliable.
We note that the change in the updated gluon PDF observed in Fig.~\ref{fig:cms8jet} for $x$ larger than about 0.5 
(and similarly for the other PDFs in Fig.~\ref{fig:cms8jetrest}) is not significant, due to the large uncertainty of the PDF 
in that region (not to mention that the PDFs themselves go rapidly to zero as $x$ goes to zero).
As mentioned in Sec.~\ref{sec:limitations}, we cannot be certain that the result of ePump would agree well with a full global analysis at very small or large $x$ values, because in those regions, the PDF uncertainties are so large compared to their central value 
that the linear approximation used to evaluate the PDFs in ePump is likely to fail.  

As discussed in the previous section, ePump can also directly update the predictions for any physical observables
after the inclusion of new data.  
For example, one might want to know how the inclusion of the CMS double-differential
inclusive jet cross section measurements at the LHC at 8 TeV~\cite{Khachatryan:2016mlc} in the global PDF fits would
modify the prediction for the total cross section of the Higgs boson produced via gluon-gluon fusion process at the LHC, at 8 and 13 TeV center-of-mass energies.
In Table~\ref{tbl:HiggsXStable}, two sets of predictions are compared by using the CT14HERA2  PDFs  and the CT14HERA2$+$eCMS8 PDFs, obtained by updating the CT14HERA2 with the CMS jet data using ePump.
The updated cross sections are the direct output of the ePump code, without recalculating it by running through all of the updated error PDFs.  As we can see from the table, the uncertainties decrease slightly after the update, while the change in the predicted central value is negligible compared to the uncertainties.

\begin{table}
\vspace{2ex}
\begin{center}
\begin{tabular}{c|c|c}
\hline \hline
  & CT14HERA2 & CT14HERA2$+$eCMS8\\
\hline
8 TeV &
 $18.60^{+1.65\%}_{-2.10\%}$
 &
 $18.57^{+1.59\%}_{-1.97\%}$

\\
\hline
13 TeV &
 $42.56^{+1.76\%}_{-2.12\%}$
 &
 $42.55^{+1.67\%}_{-1.91\%}$

\\
\hline
\hline
\end{tabular}
\end{center}
\caption{\label{tbl:HiggsXStable}
Higgs boson (with a 125 GeV mass) production cross sections (in pb)
for the gluon fusion channel at the LHC, at 8 and 13 TeV
center-of-mass energies, respectively, obtained using the CT14HERA2 and CT14HERA2$+$eCMS8
PDFs, with a common value of $\alpha_s(M_Z)=0.118$. The errors
given are due to the PDFs at the 68\% C.L.
}
\end{table}

\begin{figure}[t]
\includegraphics[width=0.50\textwidth]{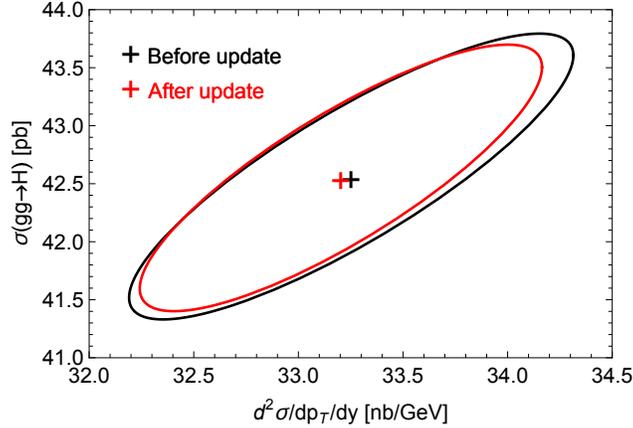}
\caption{The 90\% CL correlation ellipse of the following two observables:
the CMS 8 TeV double-differential inclusive jet cross section measured
for $y$ between 0 and 0.5 and $p_T$ between 74 GeV and 84 GeV and
the Higgs boson inclusive cross section $\sigma(gg \to H)$ at the 13 TeV LHC.
}\label{fig:ellipseone}
\end{figure}

\begin{figure}[t]
\includegraphics[width=0.50\textwidth]{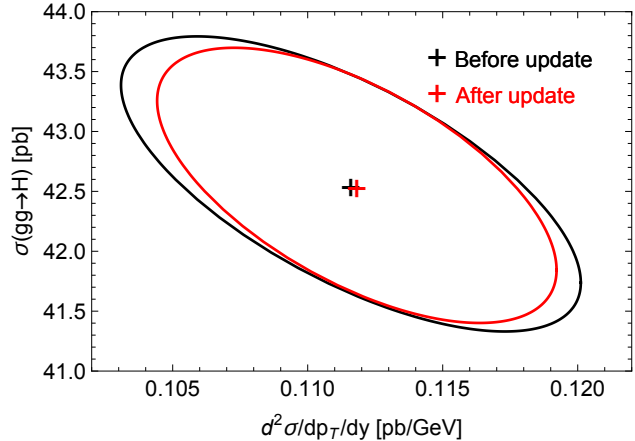}
\caption{Similar to Fig.~\ref{fig:ellipseone}, but
for $y$ between 0 and 0.5 and $p_T$ between 686 GeV and 737 GeV.
 }\label{fig:ellipsetwo}
\end{figure}

\begin{figure}[t]
\includegraphics[width=0.43\textwidth]{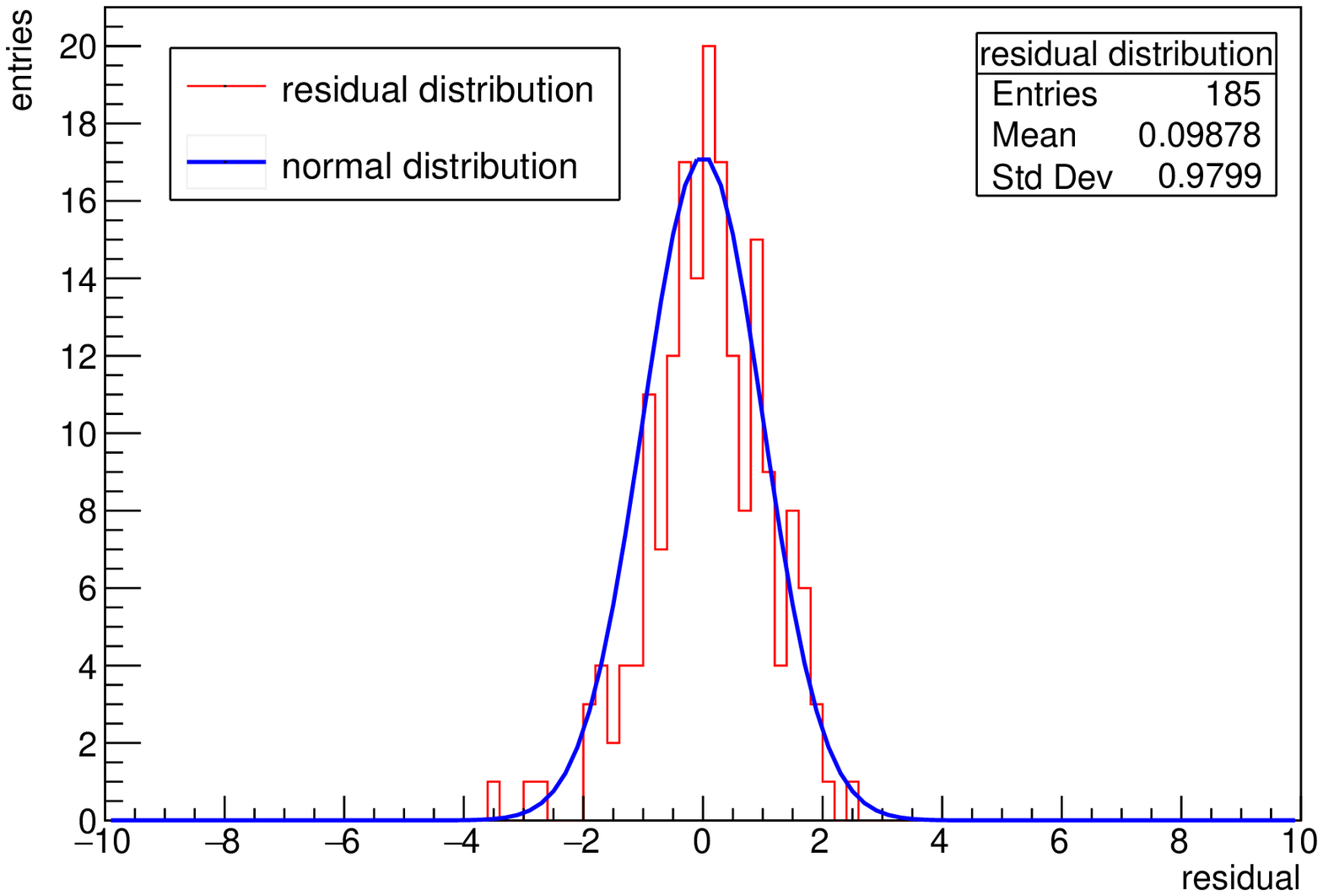}
\caption{ The distribution of residues for each data point of
the CMS 8 TeV double-differential inclusive jet cross section measurements.
 }\label{fig:chi2residue}
\end{figure}

In addition to updating PDFs and physical observables, ePump can also update correlation cosines in order to study
the degree of correlation between various experimental observables, before and after updating the PDFs
with the newly included data.
For example, let us examine the correlation ellipse of the following two observables:
the CMS 8 TeV double-differential inclusive jet cross section measured in the first $y$ and $p_T$ bin ($0\le y\le0.5$, 74 GeV $\le p_T\le$ 84 GeV), and 
the Higgs boson inclusive cross section $\sigma(gg \to H)$ at the 13 TeV LHC.  This is defined such that the predictions for the two observables will fall within the ellipse at the 90\% CL, when including only the uncertainties due to the PDFs (and using the Hessian approximations).
In Fig.~\ref{fig:ellipseone} we compare their correlation ellipses before and after updating the CT14HERA2 PDFs with the 
full CMS 8 TeV double-differential inclusive jet cross section. 
From the ePump output we can easily identify this CMS data point as the one with the largest correlation to the $\sigma(gg \to H)$ at 13 TeV (with $\cos\theta=0.85$ before and $\cos\theta_{\rm new}=0.83$ after the update).
Similarly, we can also identify the most anti-correlated (CMS 8 TeV) data point
to $\sigma(gg \to H)$ at the 13 TeV LHC, which is the bin with $0\le y\le0.5$ and 686 GeV $\le p_T\le$ 737 GeV (with $\cos\theta=-0.67$ before and $\cos\theta_{\rm new}=-0.61$ after the update).
In Fig.~\ref{fig:ellipsetwo} we show the correlation ellipses for this bin with $\sigma(gg \to H)$ at the 13 TeV LHC.
From both of these plots, we see that the PDF uncertainty in the Higgs cross section decreases, and that the 
magnitude of the correlation with the respective data points decreases after the update by ePump.

It is also interesting to examine the quality of the fit to this new CMS data after updating by ePump.
As proposed in Ref.~\cite{Gao:2013xoa}, this can be addressed by examining the distribution of
the residuals for each data point of the CMS data.  The residuals, which are automatically supplied by the output of ePump,
compare the difference between the theory prediction and the shifted data relative to the uncorrelated error
(with the full correlated systematic errors included via the data shift).
In an excellent
fit, the residuals follow a standard normal distribution, with a mean
of zero and a unit width. A non-zero mean observed in the
residual distribution would indicate a systematic
discrepancy affecting the whole data set; on the
other hand, a smaller or larger than normal width
may be due to unaccounted random effects or non-Gaussian errors
(see Appendix B.2 in Ref.~\cite{Pumplin:2002vw}).
The distribution of
the residuals for the CMS 8 TeV data, using the updated best fit, is
 plotted in Fig.~\ref{fig:chi2residue}.
It is evident that the frequencies of the residuals
agree well with a standard distribution; 
 the mean of the residual distribution is consistent
with zero and the width of the distribution is about one.

\section{Optimization of Hessian error PDFs}\label{sec:OptimizePDFs}

A scenario that arises frequently in experimental error analyses is the repeated simulation of events via Monte Carlo with the Hessian error PDFs,
in order to evaluate various cuts or experimental uncertainties and their interplay with the PDF uncertainties.  With 50 or more error PDFs, this
can be a time-consuming endeavor. Therefore, a smaller, reduced set of error PDFs that contain the majority of the PDF dependence of the observables under
consideration is often critical for the analysis.  In this section we present a method, entirely based within the Hessian approach, that can be used
to produce a new set of Hessian error PDFs that are optimized for a given set of observables.  In addition, the optimized set of error PDFs is
ordered in such a way that a reduced set of the error PDFs can be easily chosen to reproduce the PDF-induced variances in the observables to any desired precision.

Our optimization method is based on the observation that the Hessian parametrization of the $\chi^2$ function around its minimum is not unique. 
This idea was used in the data set diagonalization procedure by Pumplin~\cite{Pumplin:2009nm} to diagonalize the contribution of some subset of the data to the total $\chi^2$ function in order to assess its impact on the PDFs and its compatibility with the rest of the data.  In our present application, we use this additional freedom to find new eigenvector directions that have the maximal PDF sensitivity 
for a given set of observables, which may or may not be included in the original global analysis. The new eigenvectors contain exactly
the same information as the original eigenvectors, but are optimized so that a smaller set of error PDFs can be chosen for use
with the set of observables to any required PDF-sensitivity.
This procedure is easiest to understand when applied to a single
observable.  Thus, we quickly review the idea of extreme PDFs~\cite{Dulat:2013kqa}, and show how they can be constructed in the Hessian method, before
generalizing to an arbitrary number of observables.

\subsection{Extreme PDF Sets for a Single Observable}

For a single observable $X$ one can define two ``extreme'' PDF sets that give the maximum and minimum uncertainty values for
the observable at the CL corresponding to the tolerance $T$.  We can obtain extreme PDF sets by maximizing the function
\begin{eqnarray}
S({\bf z},\lambda)&=&w\Bigl(X({\bf z})-X(\mathrm{\bf 0})\Bigr)^2-\lambda\left(\Delta\chi^2({\bf z})/T^2- 1\right)\,,\label{eq:maximize1}
\end{eqnarray}
where $\lambda$ is a Lagrange multiplier and $w$ is an arbitrary positive constant, which we will set for convenience shortly.
The extreme sets can be calculated without any approximations on the behavior of $X({\bf z})$ or $\Delta \chi^2({\bf z})$ using the Lagrange multiplier method~\cite{Stump:2001gu}.
This method has been used to obtain the extreme PDF sets for describing the total
inclusive cross section of
$gg\rightarrow H$ at the LHC~\cite{Dulat:2013kqa}.

We can also obtain the extreme sets\footnote{In this presentation, we use a
single global tolerance $T$, but the derivation also follows for dynamical tolerances if we let $T_i$ be the average of $T^\pm_i$ and assume that the confidence level boundary is given by
$\sum_{i=1}^N(T/T_i)^2z_i^2= 1$.} using the quadratic approximation for the chi-square function,
Eq.~(\ref{eq:chisquare}) and the linear approximation for the observables,
\begin{eqnarray}
X({\bf z})&=&X(\mathrm{\bf 0})+\sum_{j=1}^N\Delta X^jz_j\ ,\label{eq:observablelin}
\end{eqnarray}
 to obtain
\begin{eqnarray}
S({\bf z},\lambda)&=&w\left(\sum_{j=1}^Nz_j\,\Delta {X}^j\right)^2-\lambda\left(\sum_{i=1}^Nz_i^2- 1\right)\nonumber\\
&=&\sum_{i,j=1}^Nz_i\,M_{ij}\,z_j-\lambda\left(\sum_{i=1}^Nz_i^2- 1\right)\,,\label{eq:maximize2}
\end{eqnarray}
where $M_{ij}=w\,\Delta{X}^i\,\Delta{X}^j$.  If we choose
\begin{equation}
w\,=\,\left(\Delta X\right)^{-2}\,=\,\left(\sum_{i=1}^N\left(\Delta{X}^i\right)^2\right)^{-1}\,,
\end{equation}
 then the matrix $M_{ij}$ has the nice property that ${\rm Tr}(M)=1$.
The maximization of $S$ is obtained by solving the eigenvector equation
\begin{eqnarray}
\sum_{j=1}^N M_{ij}z_j&=&\lambda z_i\,,
\end{eqnarray}
with the normalization condition $\sum_{i=1}^Nz_i^2= 1$.  Since $M_{ij}$ is an $N\times N$ real symmetric matrix, it has $N$ real linearly-independent eigenvectors, $U_i^{(r)}$ with real eigenvalues $\lambda^{(r)}$, $r=1$ to $N$, which satisfy
\begin{eqnarray}
\sum_{j=1}^N M_{ij}U^{(r)}_j&=&\lambda^{(r)} U_i^{(r)}\,,
\end{eqnarray}
and can be made orthonormal,
\begin{eqnarray}
\sum_{i=1}^NU_i^{(r)}U_i^{(s)}&=&\delta_{rs}\,.
\end{eqnarray}

By inspection one can see that the uncertainty in $X$ is saturated by a single eigenvector,
\begin{eqnarray}
U^{(1)}_i&=&\pm \frac{\Delta{X}^i}{\sqrt{\sum_{i=j}^N\left(\Delta{X}^j\right)^2}}\ ,
\end{eqnarray}
in agreement with Eq.~(\ref{eq:zextreme}) from Sec.~\ref{sec:ReviewHessian}.  The corresponding eigenvalue is $\lambda^{(1)}=1$.
The remaining eigenvectors, $U_i^{(r)}$ for $r=2$ to $N$, are annihilated by $M_{ij}$ and therefore have eigenvalues $\lambda^{(r)}=0$.  If we introduce new coordinates $(c_1,\dots,c_N)$
that are coefficients of the displacement along the eigenvectors of $M_{ij}$, we obtain
\begin{eqnarray}
z_i&=&\sum_{r=1}^Nc_rU_i^{(r)}\nonumber\\
\sum_{r=1}^Nc_r^2&=&\sum_{i=1}^Nz_i^2\\
S({\bf z},\lambda)&=&\sum_{r=1}^N\lambda^{(r)}c_r^2\ =\ c_1^2\ .\nonumber
\end{eqnarray}
Thus, the PDF uncertainty on $X$ is dependent only on displacements along eigenvector $U_i^{(1)}$ and is completely insensitive to displacements in the directions orthogonal to this.

This analysis of the extreme sets in the Hessian method suggests a re-diagonalization of the error PDFs.
We define new error PDFs as:
\begin{eqnarray}
f^{(\pm r)}(x,Q_0)&=&f^0(x,Q_0)  +\sum_{i=1}^NU^{(r)}_i\,\left( f^{\pm i}(x,Q_0)-f^0(x,Q_0)\right)
\ .\label{eq:extremePDF}
\end{eqnarray}
The PDFs $f^{(\pm1)}(x,Q)$ are the Hessian version of the extreme sets.  In the Hessian approximation, the dependence of the observable $X$ on the PDFs is
contained only in $f^{(\pm1)}(x,Q)$, whereas it is insensitive to variations along the remaining eigenvector directions, given by the re-diagonalized error PDFs $f^{(\pm r)}(x,Q_0)$ for $r=2$ to $N$.

For the same reasons that were discussed in Sec.~\ref{sec:UpdatePDFs}, we make the above linear approximation for the optimized PDFs with the constraint
$\sum_{i=1}^NU^{(r)}_i\ge0$.  This gives good, but not perfect, agreement between the original and optimized eigenvector bases for the asymmetric errors on the PDFs themselves.  However, it gives perfect agreement for the symmetric errors by construction.

\begin{figure}[t]
\includegraphics[width=0.70\textwidth]{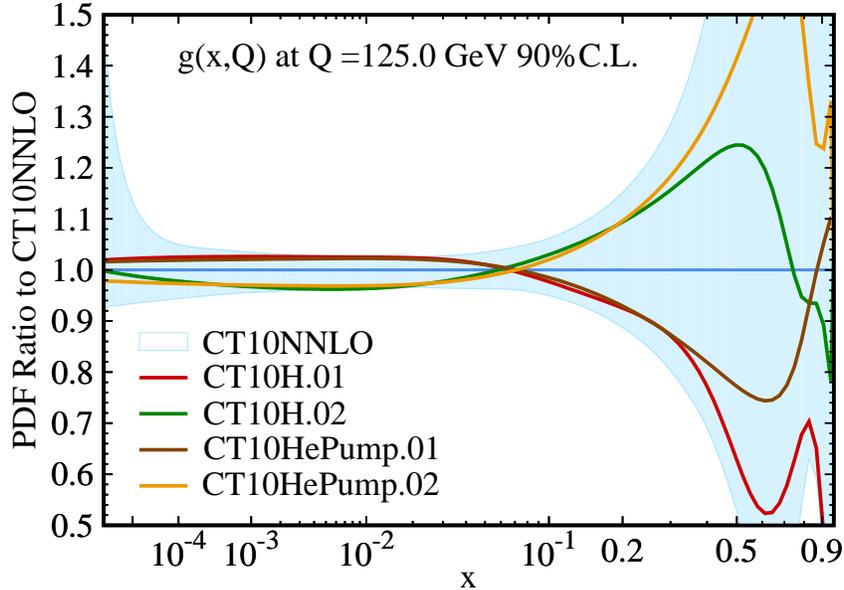}
\caption{The gluon extreme sets for the Higgs cross section through gluon fusion at the LHC with $\sqrt{s}=14$ TeV, as a ratio to the CT10NNLO central
prediction.  The green and red curves are the CT10H extreme sets, obtained using the Lagrange Multiplier method, while the gold and brown curves are
the Hessian extreme sets, obtained as discussed in the text.  The blue shaded region is the CT10NNLO gluon  uncertainty band at the 
90\% CL.}\label{fig:sym}
\end{figure}

In Fig.~\ref{fig:sym}, we plot the Hessian gluon extreme PDFs, $f^{(\pm1)}(x,Q)$, starting from the CT10NNLO and optimized for the $gg\rightarrow H$ cross section at the LHC at 14 TeV.  These are compared to the CT10H gluon PDFs from Ref.~\cite{Dulat:2013kqa}, obtained by the Lagrange Multiplier method for the same observable.  As can be seen from the figure, the two different versions of the extreme gluon sets agree well in the range $10^{-4}<x<0.3$  They only differ substantially at large $x$ or very small $x$ values, where
the PDFs are unconstrained and nonlinear dependence on the parameters becomes significant.

\subsection{Optimization of error PDFs for a set of observables}

The method of the last section can be generalized to a set of observables $X_\alpha$, where $\alpha=1$ to $N_X$.  The goal is to find the directions in
the parameter space, ${\bf z}$, that maximize the deviations of each of the observables from their best-fit values, while keeping $\Delta\chi^2({\bf z})\le T^2$
at the prescribed CL.  Thus, we maximize the function
\begin{eqnarray}
S({\bf z},\lambda)&=&\sum_{\alpha=1}^{N_X}w_\alpha\Bigl(X_\alpha({\bf z})-X_\alpha(\mathrm{\bf 0})\Bigr)^2-\lambda\left(\Delta\chi^2({\bf z})/T^2- 1\right)\,,\label{eq:maximize3}
\end{eqnarray}
where $\lambda$ is a Lagrange multiplier and the $w_\alpha$ are positive constants that weight the contribution of each of the observables to the sum.
In the Hessian approximation, we obtain
\begin{eqnarray}
S({\bf z},\lambda)&=&\sum_{\alpha=1}^{N_X}w_\alpha\left(\sum_{j=1}^Nz_j\,\Delta {X}_\alpha^j\right)^2-\lambda\left(\sum_{i=1}^Nz_i^2- 1\right)\nonumber\\
&=&\sum_{i,j=1}^Nz_i\,M_{ij}\,z_j-\lambda\left(\sum_{i=1}^Nz_i^2- 1\right)\,,\label{eq:maximize4}
\end{eqnarray}
where
\begin{equation}
M_{ij}\,=\,\sum_{\alpha=1}^{N_X}w_\alpha\Delta {X}_\alpha^i\Delta {X}_\alpha^j\,,
\end{equation}
and $\Delta{X}_{\alpha}^i=\left(X_\alpha(f^{+i})-X_\alpha(f^{-i})\right)/2$, as defined previously.
A natural choice for the constants are
\begin{equation}
w_\alpha\,=\,\left(\Delta X_\alpha\right)^{-2}\,=\,\left(\sum_{i=1}^N\left(\Delta{X}_\alpha^i\right)^2\right)^{-1}\,,
\end{equation}
so that each observable is weighted equally\footnote{This equal-weighting assumption can be easily modified, if different
observables require different precision on their PDF uncertainty for a given analysis.}, and the matrix satisfies
 ${\rm Tr}(M)=N_X$.

At this point, the solution of this problem is formally identical to that for a single observable.  To maximize $S$ we solve the eigenvector equation
\begin{eqnarray}
\sum_{j=1}^N M_{ij}z_j&=&\lambda z_i\,,
\end{eqnarray}
with the normalization condition $\sum_{i=1}^Nz_i^2= 1$.  We obtain $N_X$ orthonormal eigenvectors $U_i^{(r)}$ and corresponding non-negative real eigenvalues $\lambda^{(r)}$, $r=1$ to $N_X$. If $N_X<N$, then in general there will be $N_X$ non-zero eigenvalues and $(N-N_X)$ eigenvalues that are identically zero.

As before, we introduce new coordinates $(c_1,\dots,c_N)$
defined by
\begin{eqnarray}
z_i&=&\sum_{r=1}^Nc_rU_i^{(r)}
\end{eqnarray}
and satisfying
\begin{eqnarray}
\sum_{r=1}^Nc_r^2&=&\sum_{i=1}^Nz_i^2\\
S({\bf z},\lambda)&=&\sum_{r=1}^N\lambda^{(r)}c_r^2\ .\nonumber
\end{eqnarray}
The re-diagonalized error PDFs, optimized for this set of observables, are
\begin{eqnarray}
f^{(\pm r)}(x,Q_0)&=&f^0(x,Q_0)  +\sum_{i=1}^NU^{(r)}_i\,\left( f^{\pm i}(x,Q_0)-f^0(x,Q_0)\right)
\ ,\label{eq:extremePDF2}
\end{eqnarray}
where again we fix the sign by $\sum_{i=1}^NU^{(r)}_i\ge0$.

We can calculate the Hessian uncertainty on any observable using the re-diagonalized error PDFs.  Defining
\begin{eqnarray}
\Delta X_\alpha^{(r)}&=&\frac{X_\alpha(f^{(+r)})-X_\alpha(f^{(-r)})}{2}\,,\label{eq:dxr}
\end{eqnarray}
using the new error PDFs, then the Hessian uncertainty on the observable $X_\alpha$ using the Symmetric Master Equation is
\begin{eqnarray}
\Delta X_\alpha&=&\sqrt{\sum_{r=1}^N\left(\Delta X_\alpha^{(r)}\right)^2}\nonumber\\
&=&\sqrt{\sum_{r=1}^N\left(\sum_{i=1}^NU^{(r)}_i\Delta X_\alpha^{i}\right)\left(\sum_{j=1}^NU^{(r)}_j\Delta X_\alpha^{j}\right)}\nonumber\\
&=&\sqrt{\sum_{i,j=1}^N\Delta X_\alpha^{i}\Delta X_\alpha^{j}\left(\sum_{r=1}^NU^{(r)}_iU^{(r)}_j\right)}\nonumber\\
&=&\sqrt{\sum_{j=1}^N\left(\Delta X_\alpha^j\right)^2}\,,\label{eq:uncertaintyr}
\end{eqnarray}
where we have used the linear approximation assumed by the Hessian analysis.
Thus, the uncertainty calculated with the re-diagonalized error PDFs is identical to that calculated with the original error PDFs in this approximation.
In particular, the symmetric uncertainty on the PDFs themselves is exactly identical in the calculations using the two different error PDF sets, because the
re-diagonalized error PDFs are defined explicitly using a linear expansion in Eq.~(\ref{eq:extremePDF2}).

The advantage of using the re-diagonalized error PDFs, optimized for the set of observables, $X_\alpha$, is that we can now define a prescription for keeping a reduced set of error PDFs, by removing those that have negligible effect on the observables of interest. Let us order the optimized error PDFs by the size of their eigenvalues, so that $\lambda^{(1)}\ge\lambda^{(2)}\ge\dots\ge\lambda^{(N)}$.  Then, if we keep only the first $2n$ error PDFs (including both $\pm$ directions), the residual error in the variance calculated for observable $X_\alpha$ is
\begin{eqnarray}
\delta_\alpha^{(n)}&=&1-\frac{\sum_{r=1}^{n}\left(\Delta X_\alpha^{(r)}\right)^2}{\left(\Delta X_\alpha\right)^2}\nonumber\\
&=&1-w_\alpha\sum_{r=1}^{n}\left(\Delta X_\alpha^{(r)}\right)^2\,,\label{eq:deltan}
\end{eqnarray}
If we sum this residual error over all $N_X$ observables, and use the linear approximation in the parameter dependence of the observables, we obtain
\begin{eqnarray}
\sum_{\alpha=1}^{N_X}\delta_\alpha^{(n)}&=&N_X-\sum_{\alpha=1}^{N_X}w_\alpha\sum_{r=1}^{n}\left(\Delta X_\alpha^{(r)}\right)^2\nonumber\\
&=&N_X-\sum_{\alpha=1}^{N_X}w_\alpha\sum_{r=1}^{n}\left(\sum_{i=1}^{N}U_i^{(r)}\Delta X_\alpha^{i}\right)^2\nonumber\\
&=&N_X-\sum_{r=1}^{n}\sum_{i,j=1}^{N}U_i^{(r)}M_{ij}U_i^{(r)}\nonumber\\
&=&N_X-\sum_{r=1}^{n}\lambda^{(r)}\nonumber\\
&=&\sum_{r=n+1}^{N}\lambda^{(r)}\,.\label{eq:deltasum}
\end{eqnarray}
Thus, the sum of the eigenvalues of the discarded error PDFs gives the sum of the residual errors in the calculation of the variances for the set of observables.  This implies that if the number of observables is less than the number of eigenvector directions ($N_X<N$), then we need only keep the first $2N_X$ error PDFs to obtain the full dependence of the observables $X_\alpha$ on the PDF parameters, in the Hessian approximation.

Furthermore, in practice when performing studies on a set of experimentally-relevant observables, there will be correlations among the observables. Therefore, even if
$N_X>N$ there will be eigenvector directions that have negligible impact on the calculation of the $X_\alpha$.  Thus, by producing the optimized error PDFs for the set of observables, and noting the size of the eigenvalues obtained in the re-diagonalization procedure, one can determine how many error PDFs are necessary to retain in order to study the observables at a particular level of precision.

\begin{figure}[t]
\includegraphics[width=0.43\textwidth]{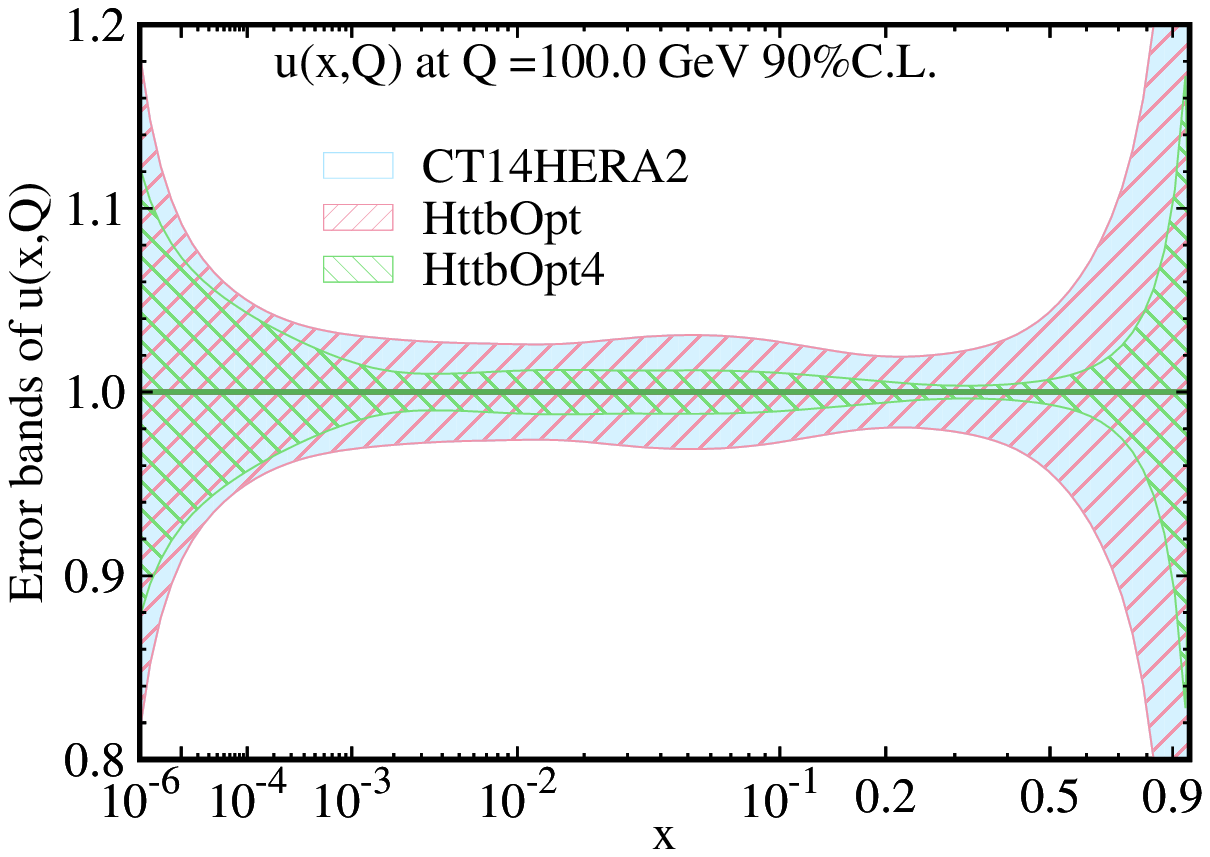}
\includegraphics[width=0.43\textwidth]{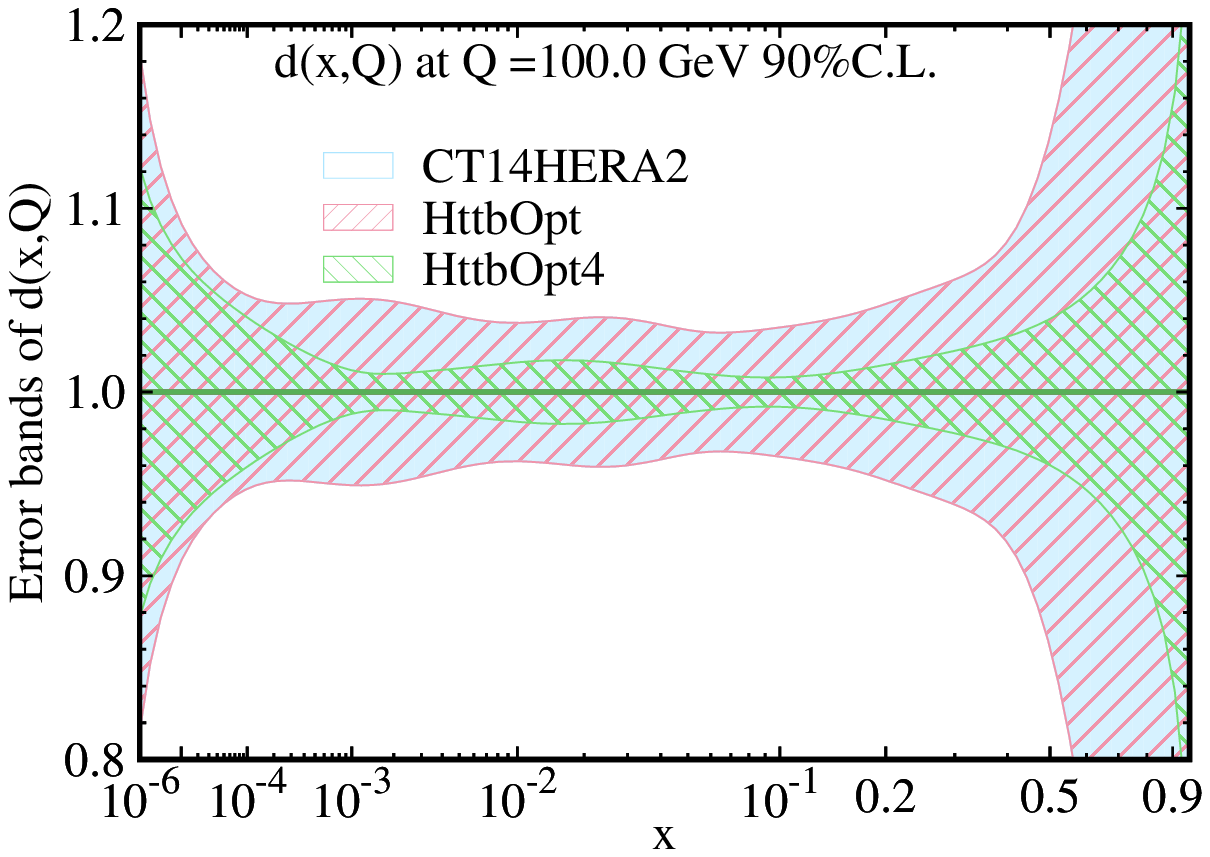}
\includegraphics[width=0.43\textwidth]{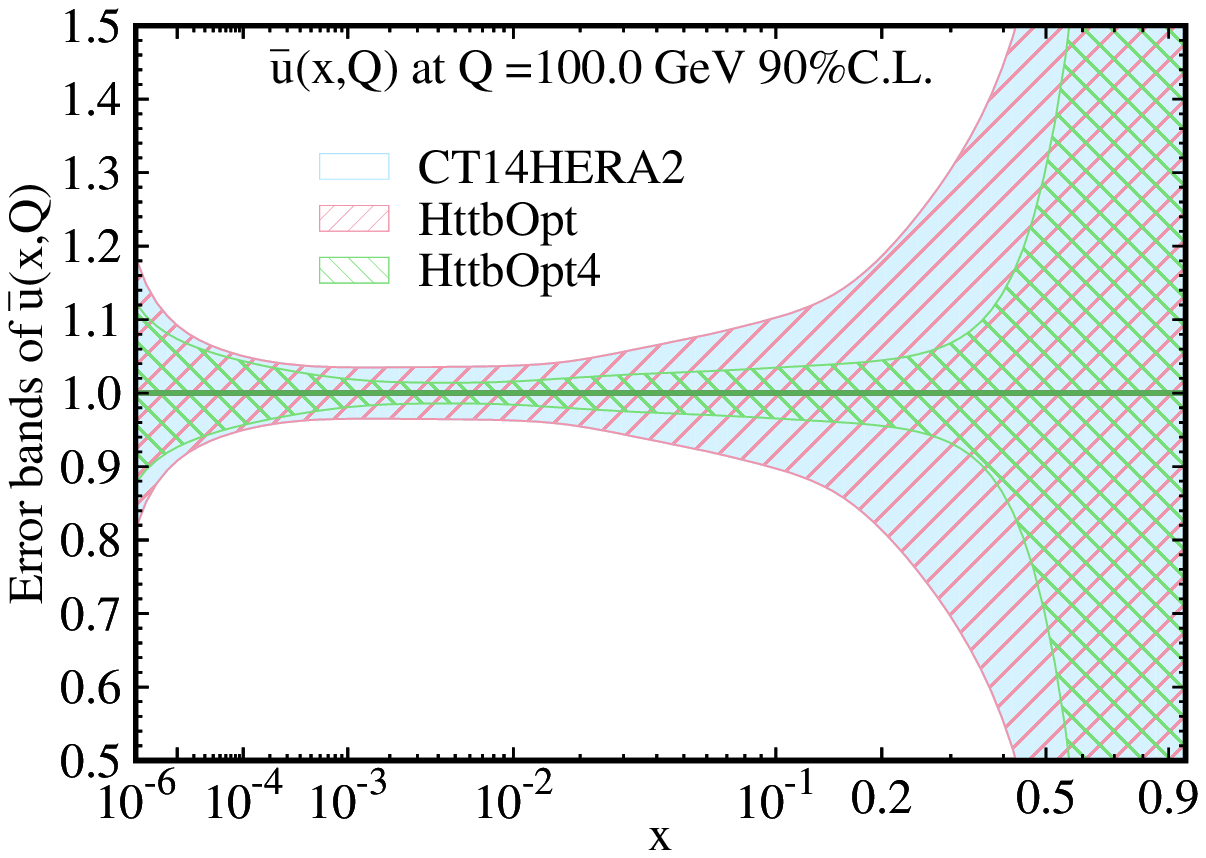}
\includegraphics[width=0.43\textwidth]{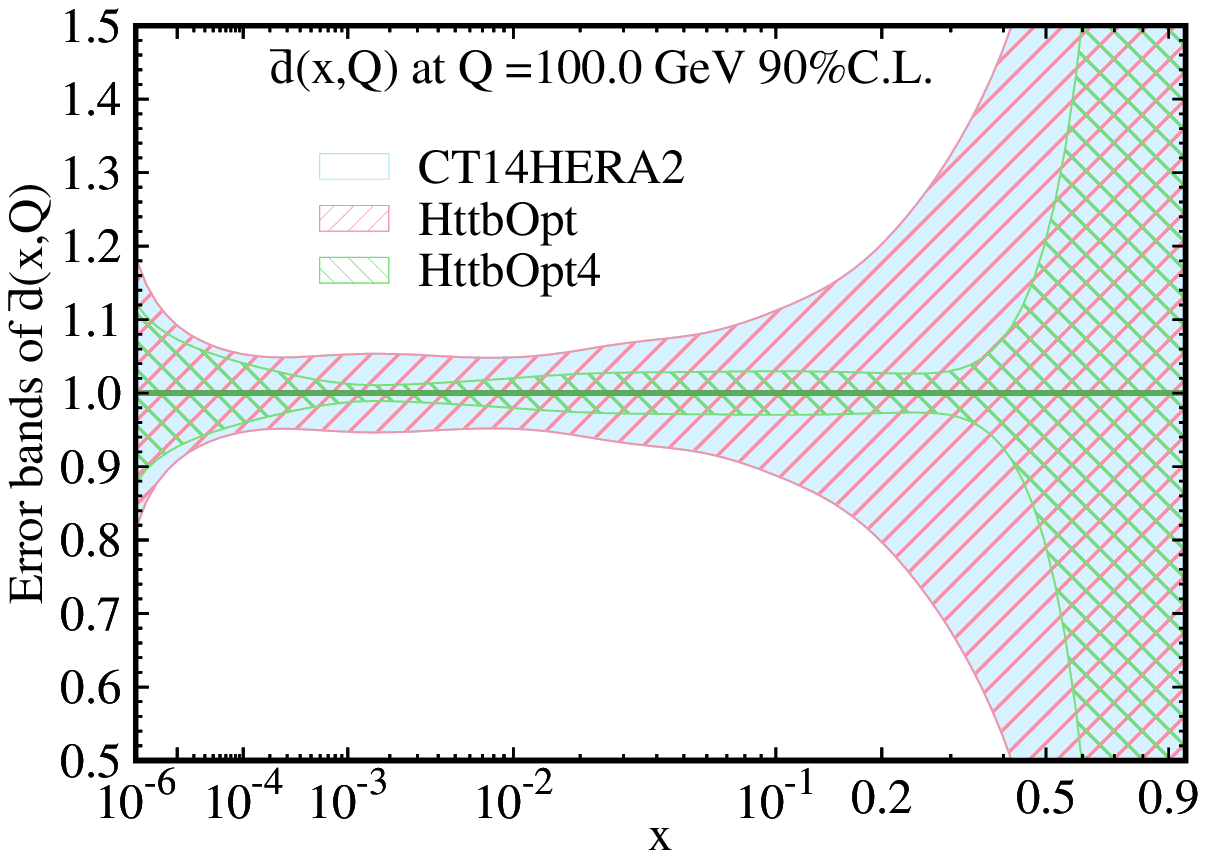}
\includegraphics[width=0.43\textwidth]{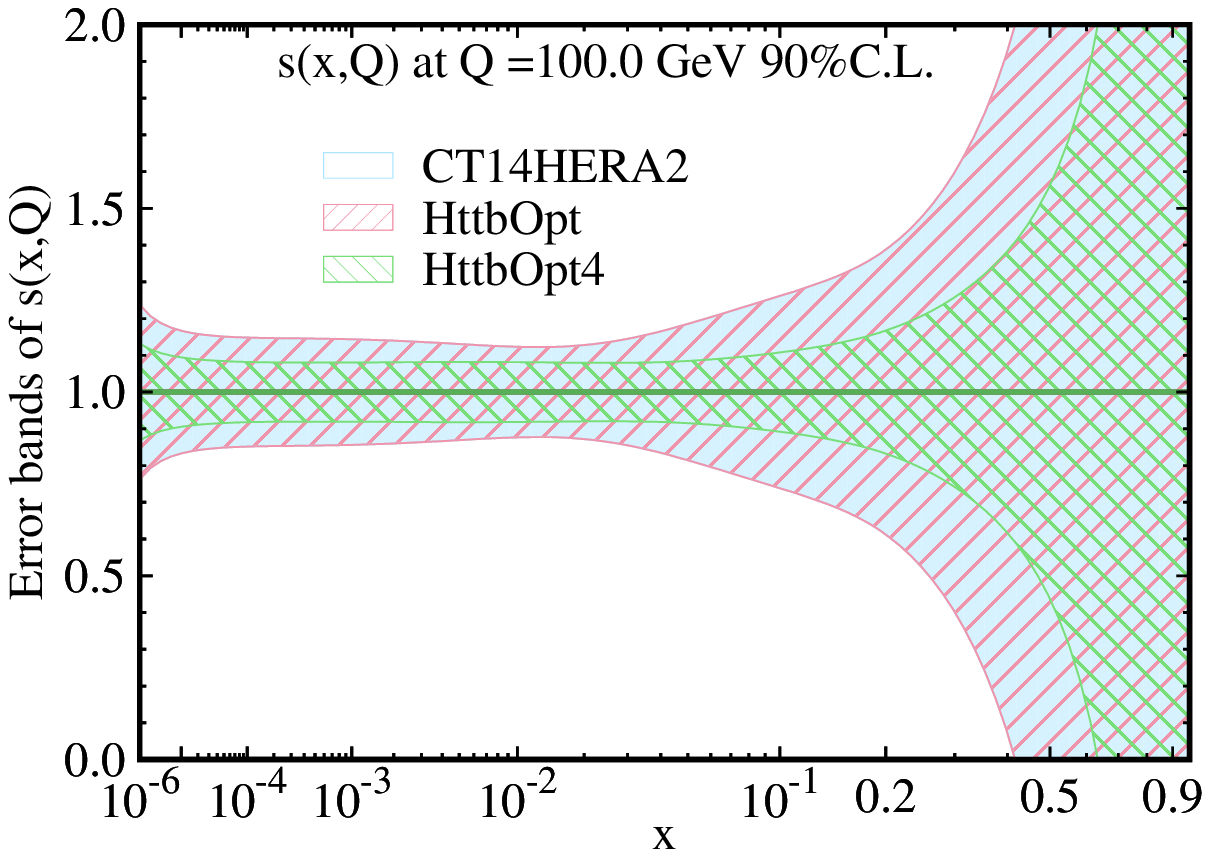}
\includegraphics[width=0.43\textwidth]{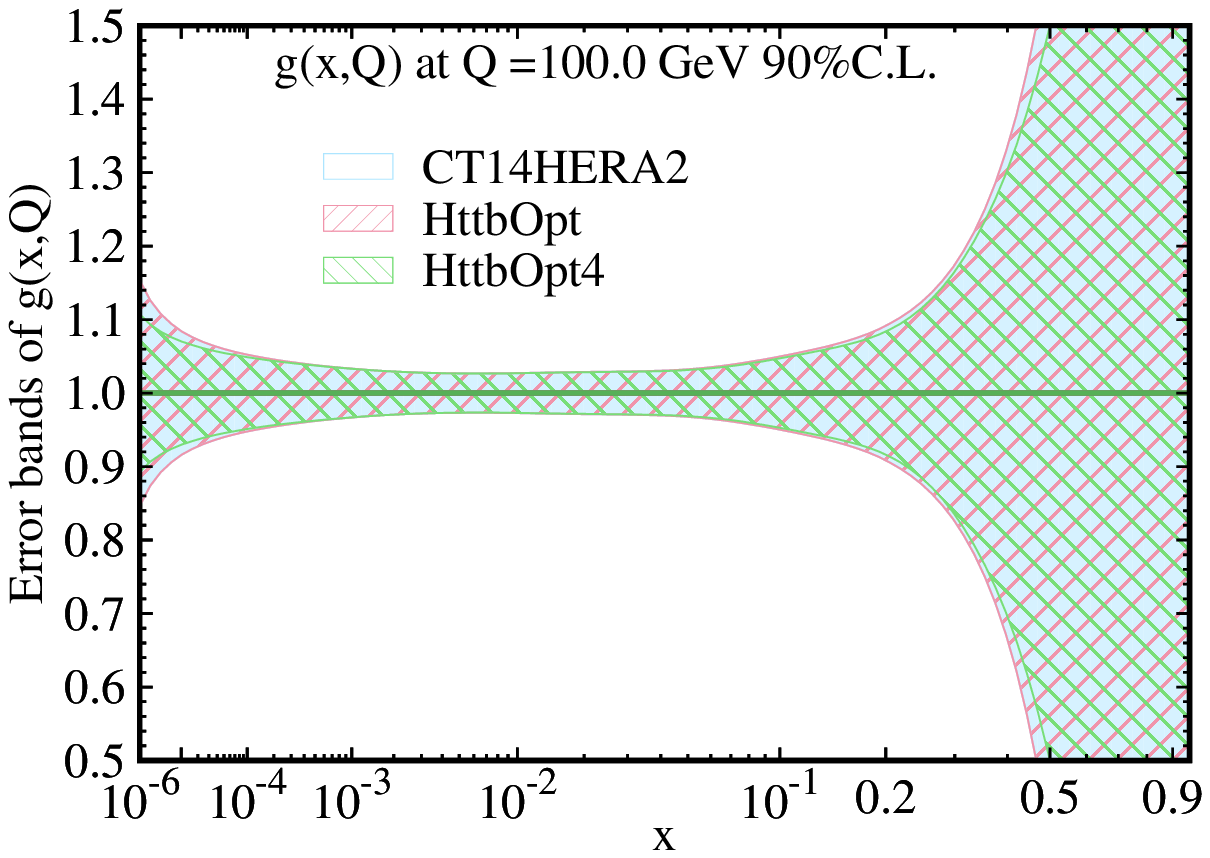}
\caption{Comparison of $u$, $d$, $\bar{u}$, $\bar{d}$, $s$, and $g$ PDF
uncertainties using original CT14HERA2 error PDFs (light blue), error PDFs optimized for Higgs and $t\bar{t}$ cross sections (red hatched), labelled as HttbOpt, and a reduced set of optimized error PDFs (green hatched), labelled as HttbOpt4.
}\label{fig:gghttb}
\end{figure}

\subsection{Examples of Optimizing using ePump}\label{sec:ExampleO}

As an example we have optimized the CT14HERA2 PDFs for
the following eight observables: the NNLO Higgs cross section through gluon fusion at the LHC
and the NNLO $t\bar t$ cross section at the LHC, each at center-of-mass energies 7, 8, 13, and 14 TeV.
The cross sections were calculated using the CT14HERA2 error PDFs, and
the masses of the top quark and Higgs boson are set to $m_t^{pole}=173.3$ GeV and
$m_H=125$ GeV, respectively, as done in Ref.~\cite{Dulat:2015mca}.
Namely, we work with $N=28$ error set pairs and $N_X=8$ observables.
The eigenvalues that we obtain are: 4.40, 3.53, $6.8\times10^{-2}$, $1.6\times10^{-3}$, $1.1\times10^{-4}$, $1.4\times10^{-5}$, $9.1\times10^{-7}$, $1.5\times10^{-7}$, and the remaining 20 eigenvalues are identically zero.  Two of the nonzero eigenvalues are large, corresponding to the two main cross sections, and the other nonzero eigenvalues are smaller, picking up the small changes
in the PDF dependence of the Higgs and $t\bar{t}$ cross sections as the energy is varied.  By choosing a reduced eigenvector set of the first 4 optimized eigenvector directions (corresponding to 8 error PDFs), one is guaranteed to cover the full PDF uncertainties of these eight observables in the Hessian approximation with a residual error of less than 0.01\% for any one of the observables. In Fig.~\ref{fig:gghttb} we have plotted the symmetric error bands at $Q=100$ GeV, relative to the CT14HERA2 central fit,
for the $u$, $d$, $\bar{u}$, $\bar{d}$, $s$, and gluon PDFs, calculated using the original CT14HERA2 error PDFs (light blue), the full set of optimized error PDFs (red hatched), labelled as HttbOpt, and the reduced set of the first 8 optimized
error PDFs (green hatched), labelled as HttbOpt4.  From these plots we can first verify that the symmetric uncertainties of the PDF themselves
are identical whether calculated with the original CT14HERA2 error PDFs or the full set of optimized error PDFs, as expected.  However, the error bands of the reduced set of optimized error PDFs are generally much smaller than those
calculated with the original CT14HERA2 error PDFs.  The only exception is for the gluon PDF error band, which is almost completely spanned by the reduced set of optimized PDFs for the range of $x$ between about $10^{-4}$ to 0.2.  This is consistent with our expectations that the 
Higgs and $t\bar{t}$ cross sections at the LHC are most strongly dependent on the gluon PDFs in that range of $x$.  
For the quark and antiquark PDFs, the reduced set spans a smaller region of the full uncertainty bands, indicating less dependence of the observables on these PDFs.  Note, however, that the error PDFs contain correlated shifts between all of the flavors of PDFs, which are not visible when plotting each of the PDF uncertainties individually.

For a more realistic application of optimized PDFs from ePump, we consider the rapidity distributions from $Z$ boson production at the LHCb experiment at 8 TeV center-of-mass energy.  The LHCb collaboration measured the cross section in 18 bins of $Z$ boson rapidity in the di-muon decay channel~\cite{Aaij:2015zlq}.  They also measured 17 bins of $Z$ boson rapidity in the $e^+e^-$ decay channel~\cite{Aaij:2015vua}.  Therefore, for this exercise we consider the optimization of the CT14HERA2 PDFs for these 35 observables.  Note that this is more than the 28 pairs of eigenvectors in the CT14HERA2 set.  For the calculation of the 35 observables that goes into the 
ePump analysis, we used a NNLO calculation obtained from the FEWZ~\cite{Melnikov:2006kv,Gavin:2010az,Gavin:2012sy,Li:2012wna} and ApplGrid~\cite{applgrid} codes.
For this set of observables, the first 6 eigenvalues are 28.64, 4.61, 1.29, 0.30, 0.12, 0.03, with the remaining eigenvalues decreasing rapidly after that.  From this we see that the first eigenvector pair contains most of the dependence on the PDFs of the 35 observables.
The rest of the eigenvector pairs are necessary to provide the sensitivity to the different flavor and $x$-range probed by the different observables in the set of 35.
In order to decide how many eigenvector pairs are necessary for this particular analysis, it is useful to consider the residual errors on the variance of each observable $\alpha$, as a function of the number of eigenvector pairs $n$ that are retained in the reduced set.  These values, $\delta_\alpha^{(n)}$, which were defined in Eq.~(\ref{eq:deltan}), are included in the output of ePump.  The residual errors for each of the observables, using up to 6 eigenvector pairs is shown in Table \ref{tbl:LHCbOpt}. The maximal residual error for each
number of pairs used is enclosed in a box.
From this table, we find $\delta_\alpha^{(4)}<0.05$ for all observables $\alpha$, so that 4 eigenvector pairs are sufficient to ensure that all residual errors are less than 5\%.  Similarly, we find $\delta_\alpha^{(5)}<0.004$ and $\delta_\alpha^{(6)}<0.001$ for all observables $\alpha$, so that a reduced set of 5 or 6 eigenvector pairs is sufficient to ensure that the residual error
on the variance for any of the 35 observables is less than 0.4\% or 0.1\%, respectively.

\begin{table}
\vspace{2ex}
\begin{center}
\begin{tabular}{r|c|c|c|c|c|c|c|c}
\hline \hline
$\alpha$\ &&$y_Z$ & $\delta_\alpha^{(1)}$& $\delta_\alpha^{(2)}$& $\delta_\alpha^{(3)}$& $\delta_\alpha^{(4)}$& $\delta_\alpha^{(5)}$& $\delta_\alpha^{(6)}$\\
\hline 
  1&   \ &\    2.000 -- 2.125 \ &\     0.2924 \ &\  0.1165 \ &\  0.0209 \ &\  0.0099 \ &\  0.0027 \ &\ 0.0005  \\
  2&   \ &\    2.125 -- 2.250 \ &\     0.2371 \ &\  0.0803 \ &\  0.0067 \ &\  0.0018 \ &\  0.0002 \ &\ 0.0002  \\
  3 &   \ &\    2.250 -- 2.375 \ &\     0.1787 \ &\  0.0449 \ &\  0.0013 \ &\  0.0007 \ &\  0.0007 \ &\ 0.0002  \\
  4 &    \ &\    2.375 -- 2.500 \ &\     0.1322 \ &\  0.0223 \ &\  0.0034 \ &\  0.0030 \ &\  0.0014 \ &\ 0.0001  \\
  5 &    \ &\    2.500 -- 2.625 \ &\     0.0975 \ &\  0.0116 \ &\  0.0082 \ &\  0.0043 \ &\  0.0009 \ &\ 0.0000  \\
  6 &    \ &\    2.625 -- 2.750 \ &\     0.0757 \ &\  0.0128 \ &\  0.0123 \ &\  0.0038 \ &\  0.0002 \ &\ 0.0000  \\
  7 &    \ &\    2.750 -- 2.875 \ &\     0.0638 \ &\  0.0224 \ &\  0.0131 \ &\  0.0022 \ &\  0.0001 \ &\ 0.0001  \\
  8 &     \ &\    2.875 -- 3.000 \ &\     0.0582 \ &\  0.0357 \ &\  0.0102 \ &\  0.0010 \ &\  0.0006 \ &\ 0.0001  \\
  9 &\   $Z\rightarrow\mu^+\mu^-$  \ &\    3.000 -- 3.125 \ &\     0.0562 \ &\  0.0484 \ &\  0.0061 \ &\  0.0011 \ &\  0.0009 \ &\ 0.0000  \\
10 &     \ &\   3.125 -- 3.250 \ &\    0.0576 \ &\  0.0572 \ &\  0.0033 \ &\  0.0023 \ &\  0.0007 \ &\ 0.0001  \\
11 &     \ &\   3.250 -- 3.375  \ &\   0.0621 \ &\  0.0584 \ &\  0.0038 \ &\  0.0035 \ &\  0.0002 \ &\ 0.0001  \\
12 &     \ &\   3.375 -- 3.500  \ &\   0.0752 \ &\  0.0509 \ &\  0.0084 \ &\  0.0033 \ &\  0.0002 \ &\ 0.0001  \\
13 &     \ &\   3.500 -- 3.625  \ &\   0.1033 \ &\  0.0362 \ &\  0.0139 \ &\  0.0018 \ &\  0.0007 \ &\ 0.0001  \\
14 &     \ &\   3.625 -- 3.750  \ &\   0.1559 \ &\  0.0207 \ &\  0.0169 \ &\  0.0009 \ &\  0.0008 \ &\ 0.0001  \\
15 &     \ &\   3.750 -- 3.875  \ &\   0.2442 \ &\  0.0165 \ &\  0.0140 \ &\  0.0032 \ &\  0.0002 \ &\ 0.0001  \\
16 &     \ &\   3.875 -- 4.000  \ &\   0.3670 \ &\  0.0360 \ &\  0.0087 \ &\  0.0071 \ &\  0.0006 \ &\ 0.0000  \\
17 &     \ &\   4.000 -- 4.250  \ &\   0.5447 \ &\  0.0897 \ &\  0.0116 \ &\  0.0023 \ &\  0.0016 \ &\ 0.0001  \\
18 &     \ &\   4.250 -- 4.500  \ &\   \fbox{0.7396} \ &\  \fbox{0.2061} \ &\  \fbox{0.1232} \ &\  \fbox{0.0408} \ &\  \fbox{0.0037} \ &\ 0.0001  \\
    \hline
 19 &\ &\  2.000 -- 2.125   \ &\   0.3020 \ &\  0.1259 \ &\  0.0219 \ &\  0.0113 \ &\  0.0029 \ &\ \fbox{0.0009}  \\
 20   &&\  2.125 -- 2.250  \ &\   0.2337 \ &\  0.0800 \ &\  0.0071 \ &\  0.0019 \ &\  0.0001 \ &\ 0.0001  \\
 21   &&\ 2.250 -- 2.375   \ &\   0.1821 \ &\  0.0468 \ &\  0.0015 \ &\  0.0008 \ &\  0.0008 \ &\ 0.0003  \\
 22  &&\  2.375 -- 2.500   \ &\   0.1327 \ &\  0.0227 \ &\  0.0034 \ &\  0.0030 \ &\  0.0015 \ &\ 0.0001  \\
 23  &&\  2.500 -- 2.625   \ &\   0.0975 \ &\  0.0117 \ &\  0.0083 \ &\  0.0043 \ &\  0.0009 \ &\ 0.0001  \\
 24   &&\  2.625 -- 2.750   \ &\   0.0764 \ &\  0.0130 \ &\  0.0125 \ &\  0.0038 \ &\  0.0002 \ &\ 0.0000  \\
 25   &&\  2.750 -- 2.875   \ &\   0.0631 \ &\  0.0224 \ &\  0.0131 \ &\  0.0024 \ &\  0.0002 \ &\ 0.0001  \\
 26   &&\  2.875 -- 3.000   \ &\   0.0579 \ &\  0.0355 \ &\  0.0104 \ &\  0.0009 \ &\  0.0006 \ &\ 0.0001  \\
 27   &$\ Z\rightarrow e^+e^-$\ &\    3.000 -- 3.125   \ &\   0.0564 \ &\  0.0486 \ &\  0.0062 \ &\  0.0011 \ &\  0.0009 \ &\ 0.0000  \\
 28   &&\   3.125 -- 3.250  \ &\   0.0574 \ &\  0.0570 \ &\  0.0034 \ &\  0.0025 \ &\  0.0007 \ &\ 0.0001  \\
 29   &&\   3.250 -- 3.375   \ &\  0.0619 \ &\  0.0584 \ &\  0.0041 \ &\  0.0037 \ &\  0.0003 \ &\ 0.0002  \\
 30   &&\   3.375 -- 3.500   \ &\  0.0751 \ &\  0.0510 \ &\  0.0085 \ &\  0.0033 \ &\  0.0002 \ &\ 0.0001  \\
 31   &&\   3.500 -- 3.625   \ &\  0.1033 \ &\  0.0362 \ &\  0.0134 \ &\  0.0018 \ &\  0.0008 \ &\ 0.0001  \\
 32   &&\   3.625 -- 3.750   \ &\  0.1562 \ &\  0.0208 \ &\  0.0170 \ &\  0.0009 \ &\  0.0009 \ &\ 0.0001  \\
 33   &&\   3.750 -- 3.875   \ &\  0.2422 \ &\  0.0154 \ &\  0.0134 \ &\  0.0034 \ &\  0.0006 \ &\ 0.0004  \\
 34   &&\   3.875 -- 4.000   \ &\  0.3698 \ &\  0.0383 \ &\  0.0087 \ &\  0.0072 \ &\  0.0008 \ &\ 0.0001  \\
 35   &&\   4.000 -- 4.250   \ &\  0.5474 \ &\  0.0940 \ &\  0.0121 \ &\  0.0033 \ &\  0.0024 \ &\ 0.0005  \\
\hline
\hline
\end{tabular}
\end{center}
\vspace{-2ex}
\caption{\label{tbl:LHCbOpt}
Residual error in the PDF-induced variance, $\delta_\alpha^{(n)}$, for each of the observables ($\alpha=1,35$) as a function of the number $n$ of optimized eigenvectors retained in the 
reduced Hessian PDF set.
}
\end{table}

\begin{figure}[t]
\includegraphics[width=0.43\textwidth]{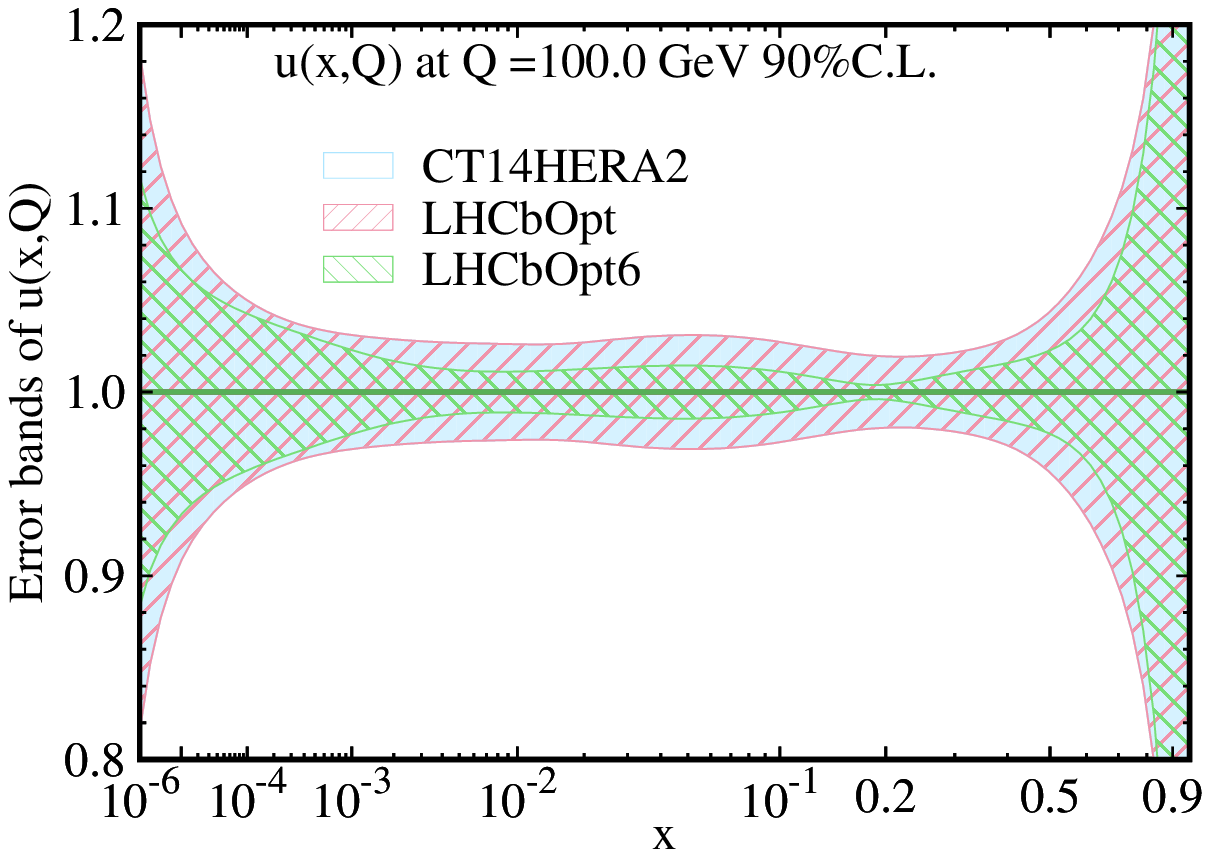}
\includegraphics[width=0.43\textwidth]{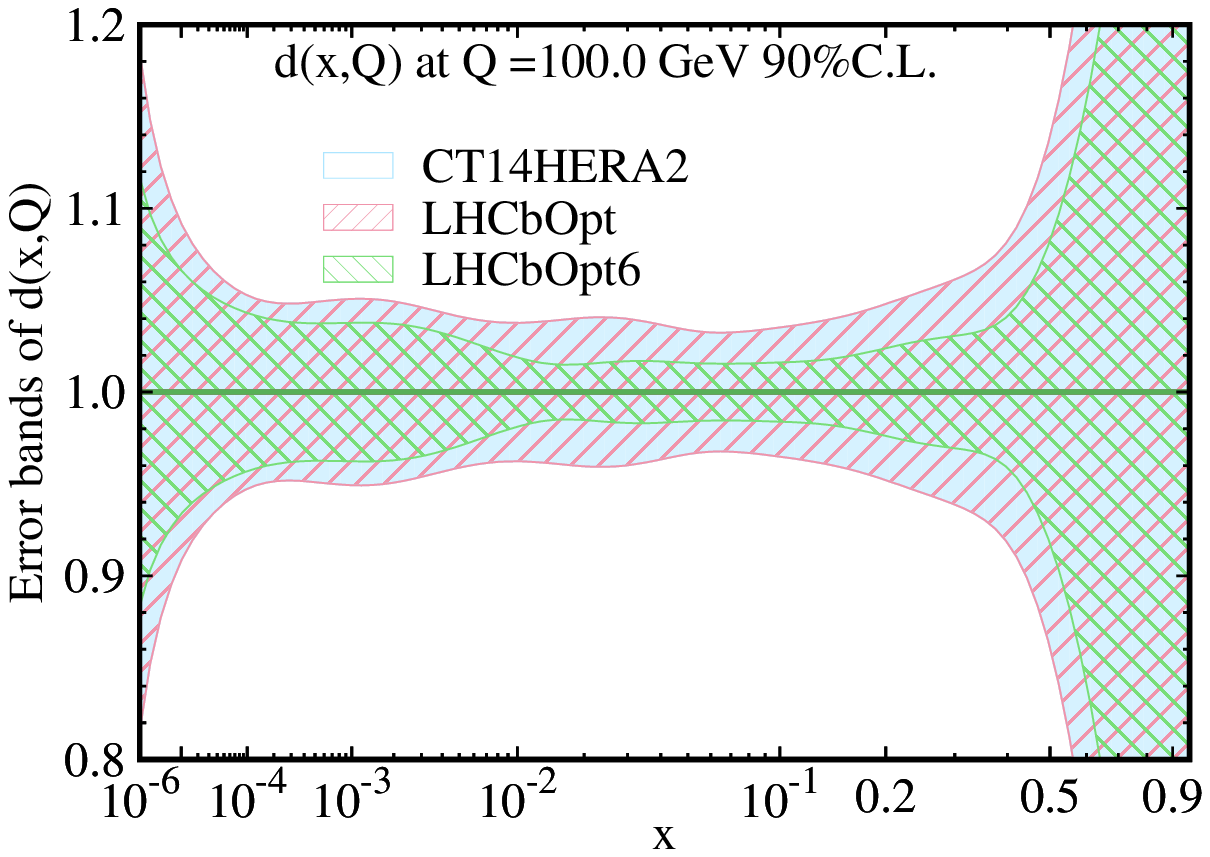}
\includegraphics[width=0.43\textwidth]{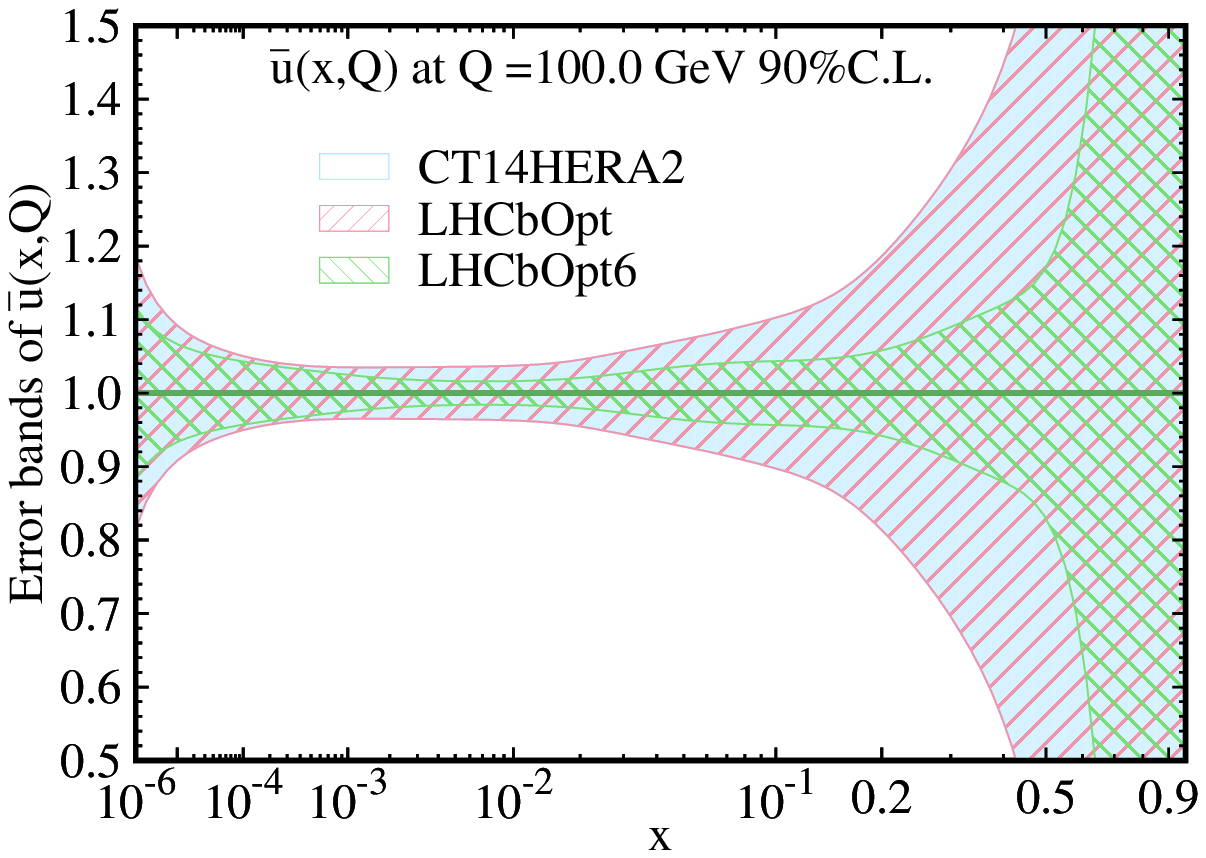}
\includegraphics[width=0.43\textwidth]{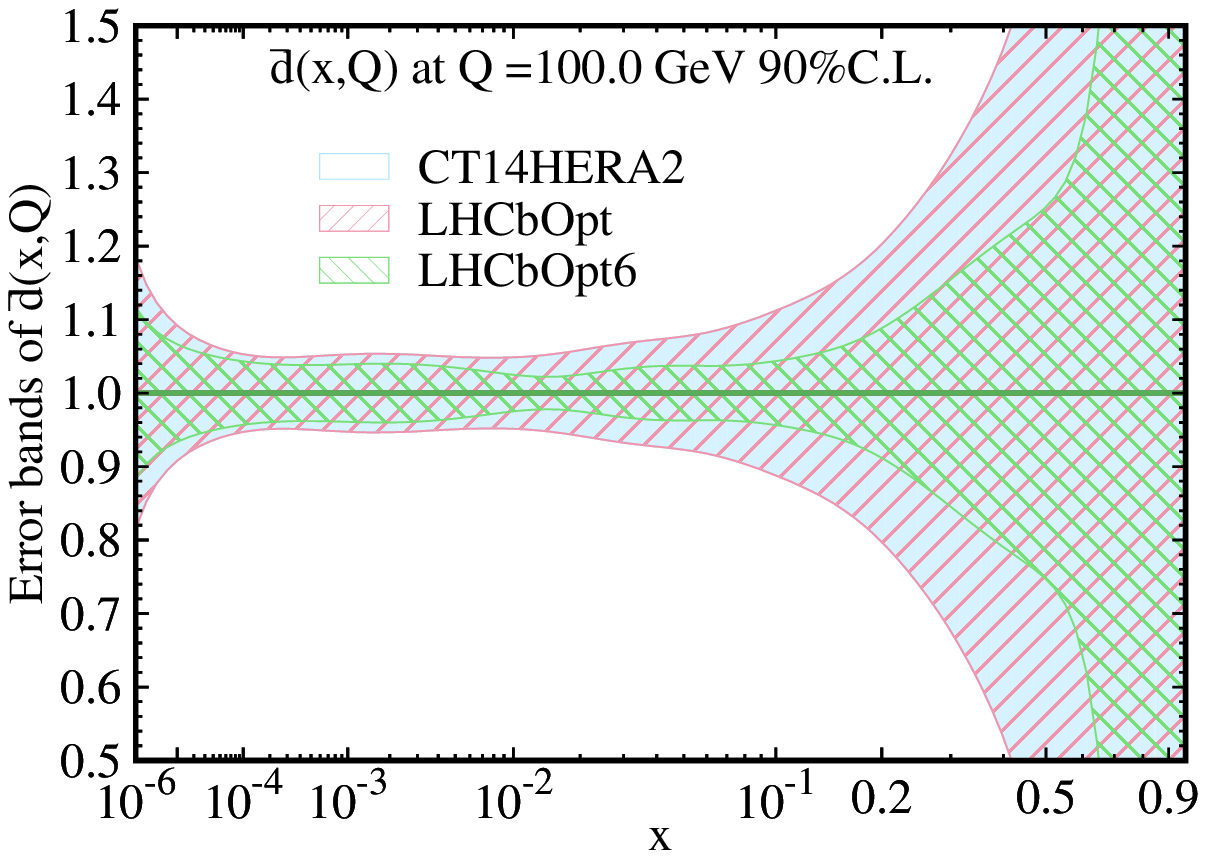}
\includegraphics[width=0.43\textwidth]{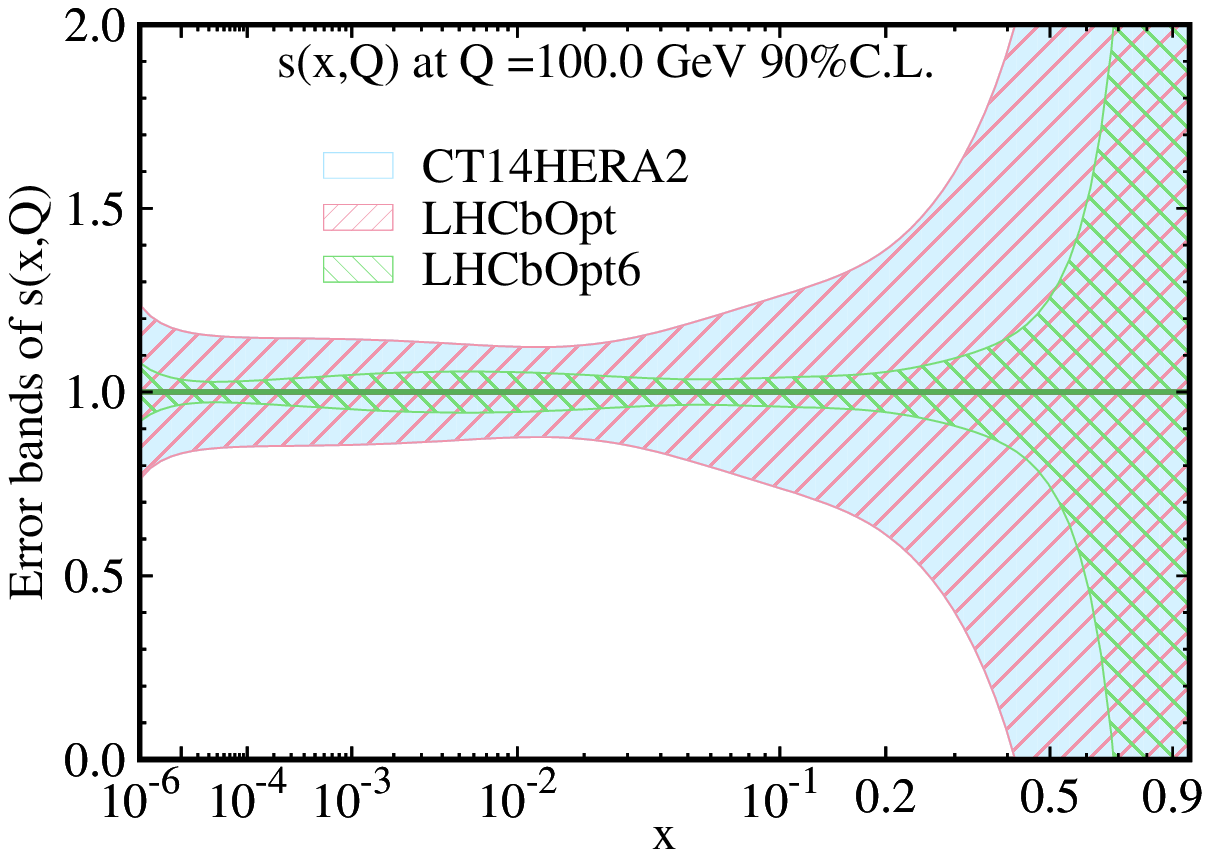}
\includegraphics[width=0.43\textwidth]{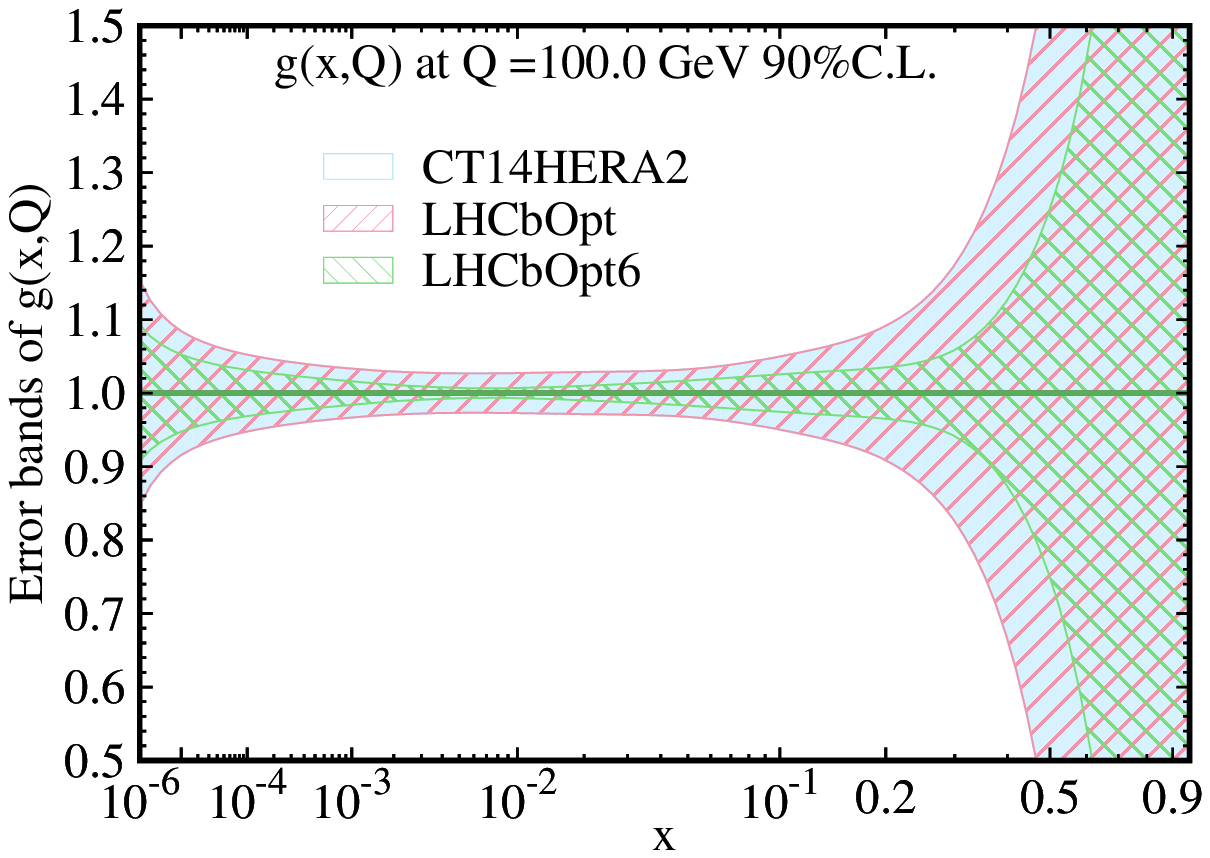}
\caption{Comparison of $u$, $d$, $\bar{u}$, $\bar{d}$, $s$, and $g$ PDF
uncertainties using original CT14HERA2 error PDFs (light blue), error PDFs optimized for LHCb $Z$ rapidity distributions
at 8 TeV (red hatched), labelled as LHCbOpt, and reduced set of optimized error PDFs (green hatched), labelled as LHCbOpt6.
}\label{fig:LHCb}
\end{figure}

Given these results, a reasonable choice might be to use a reduced set of 5 or 6 optimized eigenvector pairs (corresponding to 10 or 12 optimized error PDFs), depending on the precision required.
In Fig.~\ref{fig:LHCb} we have plotted the symmetric error bands at $Q=100.0$ GeV, relative to the CT14HERA2 best fit,
for the $u$, $d$, $\bar{u}$, $\bar{d}$, $s$, and gluon PDFs, calculated using the original CT14HERA2 error PDFs (light blue), the full set of optimized error PDFs (red hatched), labelled as LHCbOpt, and the reduced set of the first 12 optimized
error PDFs (green hatched), labelled as LHCbOpt6.

\section{ePump Code}\label{sec:ePump}

The {\tt ePump} code is written in C++ and consists of two main executables, {\tt UpdatePDFs} and {\tt OptimizePDFs}.  We discuss some general features of the code here, and direct the interested reader to the website \url{http://hep.pa.msu.edu/epump/} to obtain the code and to find specific details of its usage.

The {\tt UpdatePDFs} executable updates the best-fit and Hessian eigenvector PDFs, as well as predictions for observables, given one or more new data sets of observables, using the approximations outlined in Sec.~\ref{sec:Updating}.  The PDF set to be updated can be in either LHAPDF format (``.dat'' files) or CTEQ format (``.pds'' files), and the updated central and error PDFs are output in the same format
as the input files ({\it i.e.}, either ``.dat'' or ``.pds'' files).   Either a global tolerance value may be used, or dynamical tolerances may be used, in which case a file of tolerance-squared values, defined as $(T_i^\pm)^2$ for each $\pm$ error PDF, in a ``.tol'' file must be included with the PDF set.  For each data set of observables to be used to update the PDFs, one must supply a ``.data'' file and a ``.theory'' file.
The ``.data'' file contains the experimental values and errors for each of the data points in a particular data set.  The errors can be included in several different formats, including a table
of uncorrelated statistical and systematic errors and correlated systematic errors, as detailed in Appendix~\ref{sec:app2}. This is essentially identical to the format of the data tables used in the standard CT global analysis. The ``.theory'' file contains a list of the theoretical predictions for the observables in the data set
for each of the PDFs in the original best-fit and Hessian error PDF set.  The location of the PDFs, the ``.data'' and ``.theory" files, as well as other basic information to be used
by {\tt UpdatePDFs} is contained in a master ``.in'' file.    Individual weights can be given to the contributions of each data set to the $\chi^2$ function, as well as an overall global weight to the contribution of all new data sets, in order to easily probe the impact of the PDFs to each of the new experimental data.  The code takes only a few seconds to run on an early-2013 MacBook Pro, so it is quick and easy to try different combinations of data sets to compare their impact on the PDFs.

In addition to the updated best-fit and error PDFs, the output (``.out'' file ) of the {\tt UpdatePDFs} executable includes information on the updated $\chi^2$ contribution from each of the data sets, including updated reduced $\chi^2$ and residuals for each of the data points, and best-fit values for the nuisance parameters~\cite{Gao:2013xoa,Dulat:2015mca}.   Updated predictions and PDF uncertainties for each of the observables, as well as updated correlation cosines for pairs of observables are also included in the ``.out'' file.  These updated predictions can also include auxiliary observables, not used in the update, by including additional ``.theory' files.

The {\tt OptimizePDFs} executable optimizes the Hessian eigenvector PDFs for one or more sets of observables, using the method described in Sec.~\ref{sec:OptimizePDFs}.  The PDFs can be in LHAPDF or CTEQ formats (with the optimized error PDFs output in
the same format as the original PDFs), and the theoretical predictions used for optimizing the PDFs are given in ``.theory''
files, identical to those used by {\tt UpdatePDFs}.  The locations of the PDFs and ``.theory'' files and supplementary information is given in a single ``.in'' file, and useful information
on the optimized PDF set is output into the ``.out'' file.  In particular, the ``.out'' file gives the eigenvalues for each of the optimized eigenvector direction, as well
as the residual errors to the variance of each of the observables, when using a reduced set of the optimized Hessian eigenvector PDFs.  With this information it is then easy to decide how large the reduced set must for the desired precision of a particular analysis.

The complete {\tt ePump} package, as well as more detailed instructions for installing and file formatting can be found at the website \url{http://hep.pa.msu.edu/epump/}.

\section{Conclusions}\label{sec:Conclusions}

Already, a large amount of precision data has been collected at the LHC, and many advanced theoretical calculations 
have been produced in order to compare to the data to better determine the parameters of the Standard Model and to search for possible new physics signals. One of the key inputs to this important task is the parton distribution functions, which must be determined as precisely and with as much information of their uncertainties as possible. However, in the typical QCD global analysis of PDFs, the
long computing time that is necessary to obtain preliminary results can slow down the improvement of the PDFs, even when just including a few LHC jet, high mass Drell-Yan, $W$, $Z$, and top quark data sets in the fits at the NNLO accuracy in QCD interaction.

In this work, we have presented a theoretical method which allows us to obtain preliminary results of the updated PDFs in just a few seconds of CPU time. We have extended the method of Paukkunen and Zurita to update the Hessian PDFs, of which CTEQ-TEA PDFs are typical examples. Some specific advances, as compared to the previous work~\cite{Paukkunen:2014zia,Camarda:2015zba},  are given as follows.
As briefly reviewed in Secs.~II A and C, the CT error PDF sets were determined with a total tolerance $T$ (around 100 at the 90\% CL), which also includes the so-called Tier-2 penalties. In this work, we have consistently included the effect of the Tier-2 penalties associated with each error PDF set along its eigenvector direction in the CTEQ-TEA fitting parameter space.  This allows us to obtain both the updated best-fit PDF and the updated eigenvector PDFs (after making a reasonable assumption about the the behavior of the Tier-2 penalties after the update).  In addition to updating the central fit and its associated error PDFs, we have also showed how to directly update the predictions on experimental observables and their PDF uncertainties without requiring the use of the updated PDFs to re-do the theory predictions in Sec. ~II D. 

We have implemented these functions in a numerical software package, called ePump.  We showed some examples and verifications of
its functionality in Sec.~II F, and discussed where the approximations used in the method may break down in Sec.~II E.  The Hessian method relies on a quadratic approximation for the parameter dependence of the $\chi^2$ minimization function and a linear approximation for the parameter dependence of the observables, including the PDFs and any other observables whose prediction will be impacted by the new data. The Hessian updating also relies on the linear approximation for the parameter dependence of the observables that describe the new data used in the updating.  We presented some ways to probe when these approximations might be pushed to
their limits.  In practice, the most important limitation of the Hessian updating method is that it is tied to all of the systematic assumptions 
used in the original PDF error set.  For instance, if the non-perturbative functional form of the original PDFs is not sufficiently flexible to
describe the new data, then the updating method will likely fail.

Recently, a public program, PDFSense~\cite{Wang:2018heo}, was developed as a tool to identify the relative importance of new
experimental data sets in reducing the uncertainties of the PDFs.  The ePump code provides complementary information to this by
directly calculating the changes to the PDFs and their uncertainties when the new data is included.  In the process we have 
corroborated that the largest impact on the PDFs do come from the experiments suggested by the PDFSense program.

The second part of this paper was devoted to converting a given set of Hessian error PDFs to a new (and smaller) set of Hessian error PDFs that are optimized for a given set of observables.
The motivation for this effort comes from the fact that it is 
frequently required in experimental error analyses to repeat Monte Carlo simulation
of events with the Hessian error PDFs, in order to evaluate various cuts
or experimental uncertainties and their interplay with the PDF uncertainties. With 50 or
more error PDFs, this can be a time-consuming endeavor. Therefore, a smaller, reduced set
of error PDFs, which provides the same information on the PDF dependence of the observables under
consideration is often critical for the analysis. 
In Sec.~III, we presented a novel method, entirely
based within the Hessian approach, that can be used to produce the new set of Hessian error
PDFs that are optimized for a given set of observables, and contain the full uncertainty information of the original
error set. This optimized set of
error PDFs is ordered in such a way that a reduced set of the error PDFs can be easily
chosen to reproduce the PDF-induced variances in the observables to any desired precision.
This functionality has also been implemented in the ePump code, and we discussed a couple applications
of it in Sec.~III C.

The ePump code is a simple and efficient package for performing the two tasks of updating and optimizing the Hessian PDFs.
With regards to updating PDFs to include the effects of new data, we emphasize that
the ePump code is not meant as a replacement for a true global analysis, which very often requires the introduction of new parametrization forms for the parton distributions at the initial evolution scale (around 1.3 GeV in the CT fits).
In addition, the final judgment for the applicability of the quadratic and linear approximations used in ePump will always be 
the full global analysis.
Nevertheless, ePump provides a quick way to estimate the impact of a given new data set (inclusive jets, for example)
on the updated error PDFs, which in turn affects other observables (the Higgs boson production cross section at the LHC, in
the example of Sec.~II F). 
With regards to optimizing PDFs, the ePump code can produce an optimized set for a particular set of experimental observables,
from which the user can then choose a reduced set for their required precision.  Again the advantage is to shorten the CPU time 
needed for tedious Monte Carlo simulations, while retaining as much of the dependence of the complete set of error PDFs as 
determined by the user.

\begin{acknowledgments}
We thank our CTEQ-TEA colleagues and Raymond Brock for support and discussions.
Special thanks to Tie-Jiun Hou and Zhite Yu who helped to test the ePump code and provided insightful suggestions.
This work was supported by the U.S. National Science Foundation under Grant No. PHY-1719914.
C.-P. Yuan is also grateful for the support from
the Wu-Ki Tung endowed chair in particle physics.
\end{acknowledgments}

\clearpage

\appendix
\section{Inclusion of diagonal quadratic terms in $X({\bf z})$
\label{sec:app1}}

Although the Hessian eigenvector PDFs do not provide enough information to numerically calculate the full set of quadratic terms in the Taylor expansion
of the operators $X({\bf z})$ as given by Eq.~(\ref{eq:observable}), they can be used to obtain the diagonal quadratic terms.  The results for the linear and diagonal quadratic coefficients, while using dynamical tolerances, was given in Eqs.~(\ref{eq:dxalphaDyn}) and (\ref{eq:RalphaDyn}).  In this appendix we discuss the inclusion of these quadratic terms in the Hessian updating.

The new best-fit parameters, ${\bf z}^0$, of the $\chi^2$ function, Eq.~(\ref{eq:chi2new}), are a solution of the equation 
\begin{eqnarray}
0&=&\frac{1}{2}
\frac{\partial\Delta\chi^2({\bf z})_{\rm new}}{\partial z_i}\Bigg|_{{\bf z}={\bf z}^0}
\ =\ T^2z^0_i\,+\,\sum_{\alpha,\beta=1}^{N_X}\left[\left(X_\alpha({\bf z})-X_{\alpha}^E\right)C^{-1}_{\alpha\beta}\frac{\partial X_\beta}{\partial z_i}\right]_{{\bf z}={\bf z}^0}\,.\label{eq:chi2min}
\end{eqnarray}
If the quadratic terms are not too large, this can be solved iteratively as 
\begin{eqnarray}
z_i^{0(n+1)}&=&z_i^{0(n)}+\sum_{j=1}^N(\delta+M^{(n)})^{-1}_{ij}\left(A^{j(n)}-z_j^{0(n)}\right)\, ,\label{eq:bestfititer}
\end{eqnarray}
where 
\begin{eqnarray}
A^{i(n)}&=&
\frac{1}{T^2}\sum_{\alpha,\beta=1}^{N_X}\left[\left(X^E_\alpha-X_\alpha(\mathrm{\bf z})\right)\,C^{-1}_{\alpha\beta}\,\frac{\partial X_\beta}{\partial z_i}\right]_{{\bf z}={\bf z}^{0(n)}}\,,\nonumber\\
M^{ij(n)} &=&\frac{1}{T^2}\sum_{\alpha,\beta=1}^{N_X}\left[\frac{\partial X_\alpha}{\partial z_i}\,C^{-1}_{\alpha\beta}\,\frac{\partial X_\beta}{\partial z_j}-\left(X^E_\alpha-X_\alpha(\mathrm{\bf z})\right)\,C^{-1}_{\alpha\beta}\,\frac{\partial^2 X_\beta}{\partial z_i\partial z_j}\right]_{{\bf z}={\bf z}^{0(n)}}
\ ,\label{eq:AandMiter}
\end{eqnarray}
and $z_i^{0(0)}=0$.  

Note that if the quadratic terms are exactly zero, this converges in the first iteration to the result found previously, with $z_i^0\equiv z_i^{0(1)}$.  In most applications, we have found it to converge in 5 to 10 iterations, starting from the original best-fit parameters or from the best-fit parameters from the updated linear solution.  In these cases the difference in the speed of ePump is hardly
noticeable compared to the linear version.  On the other hand, if the $\chi^2$ function deviates too much from the quadratic approximation and the initial trial best-fit $z_i^{0(0)}$ is too far from the exact best-fit solution, then the iterative procedure may not converge.  In that case it would be necessary to invoke a more sophisticated algorithm to find the minimum, which is beyond the scope of the ePump project.
Furthermore, it is not possible to include the off-diagonal quadratic terms in $X({\bf z})$ using the error PDFs, and we have not considered 
comparable corrections to the contributions to $\chi^2$ from the original data.
Therefore, we advocate using the implementation of ePump with linear terms in $X({\bf z})$ only, as described in the main text, while including the quadratic terms only to check for consistency of the results, as discussed in Sec.~\ref{sec:limitations}.

\section{Inverse Covariance Matrix from Uncorrelated and Correlated Systematic Uncertainties
\label{sec:app2}}

In many modern analyses, the experimental errors are given in terms of uncorrelated experimental errors $s_\alpha$ for each data point
(labeled by $\alpha$) and correlated systematic errors $\beta_{\alpha a}$ with $a=1,N_\lambda$.  The discussion here follows that of Ref.~\cite{Gao:2013xoa}. The assumption is that the measured value and the true value are related by $X_\alpha^E=X_\alpha^{\rm(true)}+s_\alpha\delta_\alpha
+\sum_a\beta_{\alpha a}\lambda_a$, where the nuisance parameters, $\delta_\alpha$ and $\lambda_a$, are Gaussian random numbers with standard deviation of one.  The numbers $\delta_\alpha$
are uncorrelated between data points, while the $\lambda_a$ are independent of $\alpha$ ({\it i.e.,} $\lambda_a$ is the same for each data point).
If we replace the true value $X_\alpha^{\rm(true)}$ by the theoretical prediction $X_\alpha({\bf z})$ in the above equation,
then the $\chi^2$ function is
\begin{eqnarray}
\chi^2&=&\sum_{\alpha=1}^{N_X}\delta_\alpha^2+\sum_{a=1}^{N_\lambda}\lambda_a^2\nonumber\\
&=&\sum_{\alpha=1}^{N_X}\frac{\left(X_\alpha({\bf z})-X_\alpha^E+\sum_{a=1}^{N_\lambda}\beta_{\alpha a}\lambda_a\right)^2}{s_\alpha^2}+\sum_{a=1}^{N_\lambda}\lambda_a^2\ ,\label{eq:correrrors}
\end{eqnarray}
and we obtain the best fit by minimizing with respect to the fitting parameters $\{{\bf z}\}$ and the parameters $\lambda_a$.

Minimizing with respect to $\lambda_a$ and replacing back into the $\chi^2$ function gives
\begin{eqnarray}
\chi^2&=&\sum_{\alpha,\beta=1}^{N_X}\left(X_{\alpha}({\bf z})-X_\alpha^E\right)C^{-1}_{\alpha\beta}\left(X_{\beta}({\bf z})-X_\beta^E\right)\ ,
\end{eqnarray}
where
\begin{eqnarray}
C^{-1}_{\alpha\beta}&=&\frac{\delta_{\alpha\beta}}{s_\alpha^2}-\sum_{a,b=1}^{N_\lambda}\frac{\beta_{\alpha a}}{s_\alpha^2}{\cal A}^{-1}_{ab}\frac{\beta_{\beta b}}{s_\beta^2}\ ,
\end{eqnarray}
with
\begin{eqnarray}
{\cal A}_{ab}&=&\delta_{ab}+\sum_{\alpha=1}^{N_X}\frac{\beta_{\alpha a}\beta_{\alpha b}}{s_\alpha^2}\ .
\end{eqnarray}
Note that we can write the (un-inverted) covariance matrix as
\begin{eqnarray}
C_{\alpha\beta}&=&s_\alpha^2\delta_{\alpha\beta}+\sum_{a=1}^{N_\lambda}\beta_{\alpha a}\beta_{\beta a}\ .
\end{eqnarray}
However, the former expression for $C^{-1}_{\alpha\beta}$ is typically less computationally intense than to directly invert $C_{\alpha\beta}$ if $N_\lambda<N_X$.

The residual for each data point $\alpha$ is given by the $\delta_\alpha$ at the best fit value of the parameters. That is,
\begin{eqnarray}
r_\alpha\ =\ \bar{\delta}_\alpha&=&\frac{X_\alpha^E-\sum_{a=1}^{N_\lambda}\beta_{\alpha a}\lambda_a-X_\alpha({\bf z}^0)}{s_\alpha}\nonumber\\
&=&-\sum_{\beta=1}^{N_X}s_\alpha C_{\alpha\beta}^{-1}\left(X_\beta({\bf z}^0)-X^E_\beta\right)\ .
\end{eqnarray}
Comparison of the residuals to a normalized Gaussian distribution gives information about the self-consistency of the data points and the uncorrelated
errors.
Similarly, the best-fit values of $\lambda_a$ give information about the self-consistency of the correlated systematic errors.
They are
\begin{eqnarray}
\bar{\lambda}_a&=&-\sum_{b=1}^{N_\lambda}{\cal A}^{-1}_{ab}\sum_{\alpha=1}^{N_X}\frac{\beta_{\alpha b}\left(X_\alpha({\bf z}^0)-X_\alpha^E\right)}{s_\alpha^2}\ .
\end{eqnarray}

In practice the correlated systematic errors $\beta_{\alpha a}$ are given as percentages of $X_\alpha^{\rm(true)}$.  In ePump, we treat these errors as multiplicative errors and calculate the $\beta_{\alpha a}$ using the original best-fit theory predictions.
Note that the difference between using the original best-fit predictions and the updated best-fit predictions in the computation of
$C^{-1}_{\alpha\beta}$ is greater than quadratic order in the parameter expansion, and so is beyond the Hessian approximation that we are using.


\begin{thebibliography}{99}


%\cite{Dulat:2015mca}
\bibitem{Dulat:2015mca} 
  S.~Dulat {\it et al.},
  %``New parton distribution functions from a global analysis of quantum chromodynamics,''
  Phys.\ Rev.\ D {\bf 93}, no. 3, 033006 (2016)
  doi:10.1103/PhysRevD.93.033006
  [arXiv:1506.07443 [hep-ph]].
  %%CITATION = doi:10.1103/PhysRevD.93.033006;%%
  %540 citations counted in INSPIRE as of 08 Apr 2018

%\cite{Harland-Lang:2014zoa}
\bibitem{Harland-Lang:2014zoa} 
  L.~A.~Harland-Lang, A.~D.~Martin, P.~Motylinski and R.~S.~Thorne,
  %``Parton distributions in the LHC era: MMHT 2014 PDFs,''
  Eur.\ Phys.\ J.\ C {\bf 75}, no. 5, 204 (2015)
  doi:10.1140/epjc/s10052-015-3397-6
  [arXiv:1412.3989 [hep-ph]].
  %%CITATION = doi:10.1140/epjc/s10052-015-3397-6;%%
  %535 citations counted in INSPIRE as of 08 Apr 2018

%\cite{Ball:2014uwa}
\bibitem{Ball:2014uwa} 
  R.~D.~Ball {\it et al.} [NNPDF Collaboration],
  %``Parton distributions for the LHC Run II,''
  JHEP {\bf 1504}, 040 (2015)
  doi:10.1007/JHEP04(2015)040
  [arXiv:1410.8849 [hep-ph]].
  %%CITATION = doi:10.1007/JHEP04(2015)040;%%
  %1073 citations counted in INSPIRE as of 08 Apr 2018
  
  %\cite{Abramowicz:2015mha}
\bibitem{Abramowicz:2015mha} 
  H.~Abramowicz {\it et al.} [H1 and ZEUS Collaborations],
  %``Combination of measurements of inclusive deep inelastic ${e^{\pm }p}$ scattering cross sections and QCD analysis of HERA data,''
  Eur.\ Phys.\ J.\ C {\bf 75}, no. 12, 580 (2015)
  doi:10.1140/epjc/s10052-015-3710-4
  [arXiv:1506.06042 [hep-ex]].
  %%CITATION = doi:10.1140/epjc/s10052-015-3710-4;%%
  %265 citations counted in INSPIRE as of 08 Apr 2018

%\cite{Alekhin:2013nda}
\bibitem{Alekhin:2013nda} 
  S.~Alekhin, J.~Blumlein and S.~Moch,
  %``The ABM parton distributions tuned to LHC data,''
  Phys.\ Rev.\ D {\bf 89}, no. 5, 054028 (2014)
  doi:10.1103/PhysRevD.89.054028
  [arXiv:1310.3059 [hep-ph]].
  %%CITATION = doi:10.1103/PhysRevD.89.054028;%%
  %220 citations counted in INSPIRE as of 08 Apr 2018

%\cite{Accardi:2016qay}
\bibitem{Accardi:2016qay} 
  A.~Accardi, L.~T.~Brady, W.~Melnitchouk, J.~F.~Owens and N.~Sato,
  %``Constraints on large-$x$ parton distributions from new weak boson production and deep-inelastic scattering data,''
  Phys.\ Rev.\ D {\bf 93}, no. 11, 114017 (2016)
  doi:10.1103/PhysRevD.93.114017
  [arXiv:1602.03154 [hep-ph]].
  %%CITATION = doi:10.1103/PhysRevD.93.114017;%%
  %81 citations counted in INSPIRE as of 08 Apr 2018


%\cite{Giele:1998gw}
\bibitem{Giele:1998gw} 
  W.~T.~Giele and S.~Keller,
  %``Implications of hadron collider observables on parton distribution function uncertainties,''
  Phys.\ Rev.\ D {\bf 58}, 094023 (1998)
  doi:10.1103/PhysRevD.58.094023
  [hep-ph/9803393].
  %%CITATION = doi:10.1103/PhysRevD.58.094023;%%
  %142 citations counted in INSPIRE as of 08 Apr 2018

%\cite{Giele:2001mr}
\bibitem{Giele:2001mr} 
  W.~T.~Giele, S.~A.~Keller and D.~A.~Kosower,
  %``Parton distribution function uncertainties,''
  hep-ph/0104052.
  %%CITATION = HEP-PH/0104052;%%
  %113 citations counted in INSPIRE as of 08 Apr 2018

\bibitem{Pumplin:2001ct}
  J.~Pumplin, D.~Stump, R.~Brock, D.~Casey, J.~Huston, J.~Kalk, H.~L.~Lai and W.~K.~Tung,
  %``Uncertainties of predictions from parton distribution functions. 2. The Hessian method,''
  Phys.\ Rev.\ D {\bf 65}, 014013 (2001)
  [hep-ph/0101032].
  %%CITATION = HEP-PH/0101032;%%
  %205 citations counted in INSPIRE as of 07 Dec 2013

%\cite{Ball:2010gb}
\bibitem{Ball:2010gb} 
  R.~D.~Ball {\it et al.} [NNPDF Collaboration],
  %``Reweighting NNPDFs: the W lepton asymmetry,''
  Nucl.\ Phys.\ B {\bf 849}, 112 (2011)
  Erratum: [Nucl.\ Phys.\ B {\bf 854}, 926 (2012)]
  Erratum: [Nucl.\ Phys.\ B {\bf 855}, 927 (2012)]
  doi:10.1016/j.nuclphysb.2011.03.017, 10.1016/j.nuclphysb.2011.10.024, 10.1016/j.nuclphysb.2011.09.011
  [arXiv:1012.0836 [hep-ph]].
  %%CITATION = doi:10.1016/j.nuclphysb.2011.03.017, 10.1016/j.nuclphysb.2011.10.024, 10.1016/j.nuclphysb.2011.09.011;%%
  %118 citations counted in INSPIRE as of 08 Apr 2018

%\cite{Ball:2011gg}
\bibitem{Ball:2011gg} 
  R.~D.~Ball {\it et al.},
  %``Reweighting and Unweighting of Parton Distributions and the LHC W lepton asymmetry data,''
  Nucl.\ Phys.\ B {\bf 855}, 608 (2012)
  doi:10.1016/j.nuclphysb.2011.10.018
  [arXiv:1108.1758 [hep-ph]].
  %%CITATION = doi:10.1016/j.nuclphysb.2011.10.018;%%
  %119 citations counted in INSPIRE as of 08 Apr 2018


%\cite{Hou:2016sho}
\bibitem{Hou:2016sho} 
  T.~J.~Hou {\it et al.},
  %``Reconstruction of Monte Carlo replicas from Hessian parton distributions,''
  JHEP {\bf 1703}, 099 (2017)
  doi:10.1007/JHEP03(2017)099
  [arXiv:1607.06066 [hep-ph]].
  %%CITATION = doi:10.1007/JHEP03(2017)099;%%
  %12 citations counted in INSPIRE as of 08 Apr 2018

%\cite{Paukkunen:2014zia}
\bibitem{Paukkunen:2014zia} 
  H.~Paukkunen and P.~Zurita,
  %``PDF reweighting in the Hessian matrix approach,''
  JHEP {\bf 1412}, 100 (2014)
  doi:10.1007/JHEP12(2014)100
  [arXiv:1402.6623 [hep-ph]].
  %%CITATION = doi:10.1007/JHEP12(2014)100;%%
  %45 citations counted in INSPIRE as of 08 Apr 2018
  
  %\cite{Paukkunen:2013grz}
\bibitem{Paukkunen:2013grz} 
  H.~Paukkunen and C.~A.~Salgado,
  %``Agreement of Neutrino Deep Inelastic Scattering Data with Global Fits of Parton Distributions,''
  Phys.\ Rev.\ Lett.\  {\bf 110}, no. 21, 212301 (2013)
  doi:10.1103/PhysRevLett.110.212301
  [arXiv:1302.2001 [hep-ph]].
  %%CITATION = doi:10.1103/PhysRevLett.110.212301;%%
  %29 citations counted in INSPIRE as of 10 Sep 2018

%\cite{Camarda:2015zba}
\bibitem{Camarda:2015zba} 
  S.~Camarda {\it et al.} [HERAFitter developers' Team],
  %``QCD analysis of $W$- and $Z$-boson production at Tevatron,''
  Eur.\ Phys.\ J.\ C {\bf 75}, no. 9, 458 (2015)
  doi:10.1140/epjc/s10052-015-3655-7
  [arXiv:1503.05221 [hep-ph]].
  %%CITATION = doi:10.1140/epjc/s10052-015-3655-7;%%
  %26 citations counted in INSPIRE as of 18 Jun 2018



\bibitem{lhapdf}
\url{https:/lhapdf.hepforge.org/}.

%\cite{Gao:2013bia}
\bibitem{Gao:2013bia} 
  J.~Gao and P.~Nadolsky,
  %``A meta-analysis of parton distribution functions,''
  JHEP {\bf 1407}, 035 (2014)
  doi:10.1007/JHEP07(2014)035
  [arXiv:1401.0013 [hep-ph]].
  %%CITATION = doi:10.1007/JHEP07(2014)035;%%
  %75 citations counted in INSPIRE as of 17 Sep 2018

%\cite{Carrazza:2015aoa}
\bibitem{Carrazza:2015aoa} 
  S.~Carrazza, S.~Forte, Z.~Kassabov, J.~I.~Latorre and J.~Rojo,
  %``An Unbiased Hessian Representation for Monte Carlo PDFs,''
  Eur.\ Phys.\ J.\ C {\bf 75}, no. 8, 369 (2015)
  doi:10.1140/epjc/s10052-015-3590-7
  [arXiv:1505.06736 [hep-ph]].
  %%CITATION = doi:10.1140/epjc/s10052-015-3590-7;%%
  %56 citations counted in INSPIRE as of 09 Apr 2018

%\cite{Carrazza:2016htc}
\bibitem{Carrazza:2016htc}
  S.~Carrazza, S.~Forte, Z.~Kassabov and J.~Rojo,
  %``Specialized minimal PDFs for optimized LHC calculations,''
  Eur.\ Phys.\ J.\ C {\bf 76}, no. 4, 205 (2016)
  doi:10.1140/epjc/s10052-016-4042-8
  [arXiv:1602.00005 [hep-ph]].
  %%CITATION = doi:10.1140/epjc/s10052-016-4042-8;%%
  %10 citations counted in INSPIRE as of 21 Apr 2017



%\cite{Pumplin:2009nm}
\bibitem{Pumplin:2009nm}
  J.~Pumplin,
  %``Data set diagonalization in a global fit,''
  Phys.\ Rev.\ D {\bf 80}, 034002 (2009)
  doi:10.1103/PhysRevD.80.034002
  [arXiv:0904.2425 [hep-ph]].
  %%CITATION = doi:10.1103/PhysRevD.80.034002;%%
  %13 citations counted in INSPIRE as of 21 Apr 2017


%\cite{Nadolsky:2001yg}
\bibitem{Nadolsky:2001yg} 
  P.~M.~Nadolsky and Z.~Sullivan,
  %``PDF uncertainties in WH production at Tevatron,''
  eConf C {\bf 010630}, P510 (2001)
  [hep-ph/0110378].
  %%CITATION = HEP-PH/0110378;%%
  %48 citations counted in INSPIRE as of 09 Apr 2018

%\cite{Lai:2010vv}
\bibitem{Lai:2010vv} 
  H.~L.~Lai, M.~Guzzi, J.~Huston, Z.~Li, P.~M.~Nadolsky, J.~Pumplin and C.-P.~Yuan,
  %``New parton distributions for collider physics,''
  Phys.\ Rev.\ D {\bf 82}, 074024 (2010)
  doi:10.1103/PhysRevD.82.074024
  [arXiv:1007.2241 [hep-ph]].
  %%CITATION = doi:10.1103/PhysRevD.82.074024;%%
  %2379 citations counted in INSPIRE as of 09 Apr 2018


%\cite{Dulat:2013hea}
\bibitem{Dulat:2013hea} 
  S.~Dulat, T.~J.~Hou, J.~Gao, J.~Huston, J.~Pumplin, C.~Schmidt, D.~Stump and C.-P.~Yuan,
  %``Intrinsic Charm Parton Distribution Functions from CTEQ-TEA Global Analysis,''
  Phys.\ Rev.\ D {\bf 89}, no. 7, 073004 (2014)
  doi:10.1103/PhysRevD.89.073004
  [arXiv:1309.0025 [hep-ph]].
  %%CITATION = doi:10.1103/PhysRevD.89.073004;%%
  %50 citations counted in INSPIRE as of 09 Apr 2018

\bibitem{Pumplin:2002vw} % CTEQ6
  J.~Pumplin, D.~R.~Stump, J.~Huston, H.~L.~Lai, P.~M.~Nadolsky and W.~K.~Tung,
  %``New generation of parton distributions with uncertainties from global QCD analysis,''
  JHEP {\bf 0207}, 012 (2002).
  [hep-ph/0201195].

%\bibitem{Stump:2003yu}  % CTEQ6.1
%  D.~Stump, J.~Huston, J.~Pumplin, W.~-K.~Tung, H.~L.~Lai, S.~Kuhlmann and J.~F.~Owens,
%  %``Inclusive jet production, parton distributions, and the search for new physics,''
%  JHEP {\bf 0310}, 046 (2003)
%  [hep-ph/0303013].

%\bibitem{Tung:2006tb} % CTEQ6.5
%  W.~K.~Tung, H.~L.~Lai, A.~Belyaev, J.~Pumplin, D.~Stump and C.~-P.~Yuan,
%  %``Heavy Quark Mass Effects in Deep Inelastic Scattering and Global QCD Analysis,''
%  JHEP {\bf 0702}, 053 (2007)
%  [hep-ph/0611254].

\bibitem{Nadolsky:2008zw} % CTEQ6.6
  P.~M.~Nadolsky, H.~-L.~Lai, Q.~-H.~Cao, J.~Huston, J.~Pumplin, D.~Stump, W.~-K.~Tung and C.~-P.~Yuan,
  %``Implications of CTEQ global analysis for collider observables,''
  Phys.\ Rev.\ D {\bf 78}, 013004 (2008)
  [arXiv:0802.0007 [hep-ph]].

%\bibitem{Pumplin:2009nk} % CT09
%  J.~Pumplin, J.~Huston, H.~L.~Lai, P.~M.~Nadolsky, W.~-K.~Tung and C.~-P.~Yuan,
%  %``Collider Inclusive Jet Data and the Gluon Distribution,''
%  Phys.\ Rev.\ D {\bf 80}, 014019 (2009)
%  [arXiv:0904.2424 [hep-ph]].


%\cite{Gao:2013xoa}
\bibitem{Gao:2013xoa} %The CT10 Global Analysis of QCD
  J.~Gao {\it et al.},
  %``CT10 next-to-next-to-leading order global analysis of QCD,''
  Phys.\ Rev.\ D {\bf 89}, no. 3, 033009 (2014)
  doi:10.1103/PhysRevD.89.033009
  [arXiv:1302.6246 [hep-ph]].
  %%CITATION = doi:10.1103/PhysRevD.89.033009;%%
  %464 citations counted in INSPIRE as of 04 Mar 2018


\bibitem{paper-2} T.-J.~Hou, C.~Schmidt, C.-P.~Yuan, Z.~Yu, (In preparation) 2018.

\bibitem {Aaltonen:2008eq}  T. Aaltonen {\em et.~al.} (CDF Collaboration), Phys. Rev. D78, 052006 (2008).
\bibitem {Abazov:2008ae}  V. Abazov {\em et.~al.} (D\O~ Collaboration), Phys.Rev.Lett. 101, 062001 (2008), 0802.2400.
\bibitem {Aad:2011fc}  G. Aad {\em et.~al.} (ATLAS Collaboration), Phys.Rev. D86, 014022 (2012), 1112.6297.
\bibitem {Chatrchyan:2012bja}  S. Chatrchyan et al. (CMS Collaboration), Phys.Rev. D87, 112002 (2013), 1212.6660.

%%\cite{Chatrchyan:2014gia}
%\bibitem{Chatrchyan:2014gia}
%  S.~Chatrchyan {\it et al.} [CMS Collaboration],
%  %``Measurement of the ratio of inclusive jet cross sections using the anti-$k_T$ algorithm with radius parameters R=0.5 and 0.7 in pp collisions at $\sqrt{s}=7$  TeV,''
%  Phys.\ Rev.\ D {\bf 90}, no. 7, 072006 (2014)
%  doi:10.1103/PhysRevD.90.072006
%  [arXiv:1406.0324 [hep-ex]].
%  %%CITATION = doi:10.1103/PhysRevD.90.072006;%%
%  %34 citations counted in INSPIRE as of 08 M

\bibitem{Khachatryan:2016mlc}
  V.~Khachatryan {\it et al.} [CMS Collaboration],
  %``Measurement and QCD analysis of double-differential inclusive jet cross sections in pp collisions at $ \sqrt{s}=8 $ TeV and cross section ratios to 2.76 and 7 TeV,''
  JHEP {\bf 1703}, 156 (2017)
  doi:10.1007/JHEP03(2017)156
  [arXiv:1609.05331 [hep-ex]].
  %%CITATION = doi:10.1007/JHEP03(2017)156;%%
  %34 ci

%\cite{Wobisch:2011ij}
\bibitem{Wobisch:2011ij} 
  M.~Wobisch {\it et al.} [fastNLO Collaboration],
  %``Theory-Data Comparisons for Jet Measurements in Hadron-Induced Processes,''
  arXiv:1109.1310 [hep-ph].
  %%CITATION = ARXIV:1109.1310;%%
  %69 citations counted in INSPIRE as of 18 Jun 2018

\bibitem{fastnlo-cms8jet}
D.~Britzger, K.~Rabbertz, G.~Sieber, F.~Stober, M.~Wobisch,
http://fastnlo.hepforge.org.


%\cite{Wang:2018heo}
\bibitem{Wang:2018heo} 
  B.~T.~Wang, T.~J.~Hobbs, S.~Doyle, J.~Gao, T.~J.~Hou, P.~M.~Nadolsky and F.~I.~Olness,
  %``Visualizing the sensitivity of hadronic experiments to nucleon structure,''
  arXiv:1803.02777 [hep-ph].
  %%CITATION = ARXIV:1803.02777;%%



%\cite{Dulat:2013kqa}
\bibitem{Dulat:2013kqa} 
  S.~Dulat {\it et al.},
  %``Higgs Boson Cross Section from CTEQ-TEA Global Analysis,''
  Phys.\ Rev.\ D {\bf 89}, no. 11, 113002 (2014)
  doi:10.1103/PhysRevD.89.113002
  [arXiv:1310.7601 [hep-ph]].
  %%CITATION = doi:10.1103/PhysRevD.89.113002;%%
  %12 citations counted in INSPIRE as of 09 Apr 2018

%\cite{Stump:2001gu}
\bibitem{Stump:2001gu} 
  D.~Stump, J.~Pumplin, R.~Brock, D.~Casey, J.~Huston, J.~Kalk, H.~L.~Lai and W.~K.~Tung,
  %``Uncertainties of predictions from parton distribution functions. 1. The Lagrange multiplier method,''
  Phys.\ Rev.\ D {\bf 65}, 014012 (2001)
  doi:10.1103/PhysRevD.65.014012
  [hep-ph/0101051].
  %%CITATION = doi:10.1103/PhysRevD.65.014012;%%
  %248 citations counted in INSPIRE as of 09 Apr 2018

%\cite{Aaij:2015zlq}
\bibitem{Aaij:2015zlq} 
  R.~Aaij {\it et al.} [LHCb Collaboration],
  %``Measurement of forward W and Z boson production in $pp$ collisions at $ \sqrt{s}=8 $ TeV,''
  JHEP {\bf 1601}, 155 (2016)
  doi:10.1007/JHEP01(2016)155
  [arXiv:1511.08039 [hep-ex]].
  %%CITATION = doi:10.1007/JHEP01(2016)155;%%
  %65 citations counted in INSPIRE as of 09 Apr 2018

%\cite{Aaij:2015vua}
\bibitem{Aaij:2015vua} 
  R.~Aaij {\it et al.} [LHCb Collaboration],
  %``Measurement of forward $\rm Z\rightarrow e^+e^-$ production at $\sqrt{s}=8$ TeV,''
  JHEP {\bf 1505}, 109 (2015)
  doi:10.1007/JHEP05(2015)109
  [arXiv:1503.00963 [hep-ex]].
  %%CITATION = doi:10.1007/JHEP05(2015)109;%%
  %39 citations counted in INSPIRE as of 09 Apr 2018
  
  %\cite{Melnikov:2006kv}
\bibitem{Melnikov:2006kv} 
  K.~Melnikov and F.~Petriello,
  %``Electroweak gauge boson production at hadron colliders through O(alpha(s)**2),''
  Phys.\ Rev.\ D {\bf 74}, 114017 (2006)
  doi:10.1103/PhysRevD.74.114017
  [hep-ph/0609070].
  %%CITATION = doi:10.1103/PhysRevD.74.114017;%%
  %825 citations counted in INSPIRE as of 18 Jun 2018
  
  %\cite{Gavin:2010az}
\bibitem{Gavin:2010az} 
  R.~Gavin, Y.~Li, F.~Petriello and S.~Quackenbush,
  %``FEWZ 2.0: A code for hadronic Z production at next-to-next-to-leading order,''
  Comput.\ Phys.\ Commun.\  {\bf 182}, 2388 (2011)
  doi:10.1016/j.cpc.2011.06.008
  [arXiv:1011.3540 [hep-ph]].
  %%CITATION = doi:10.1016/j.cpc.2011.06.008;%%
  %629 citations counted in INSPIRE as of 18 Jun 2018
  
  %\cite{Gavin:2012sy}
\bibitem{Gavin:2012sy} 
  R.~Gavin, Y.~Li, F.~Petriello and S.~Quackenbush,
  %``W Physics at the LHC with FEWZ 2.1,''
  Comput.\ Phys.\ Commun.\  {\bf 184}, 208 (2013)
  doi:10.1016/j.cpc.2012.09.005
  [arXiv:1201.5896 [hep-ph]].
  %%CITATION = doi:10.1016/j.cpc.2012.09.005;%%
  %234 citations counted in INSPIRE as of 18 Jun 2018
  
  %\cite{Li:2012wna}
\bibitem{Li:2012wna} 
  Y.~Li and F.~Petriello,
  %``Combining QCD and electroweak corrections to dilepton production in FEWZ,''
  Phys.\ Rev.\ D {\bf 86}, 094034 (2012)
  doi:10.1103/PhysRevD.86.094034
  [arXiv:1208.5967 [hep-ph]].
  %%CITATION = doi:10.1103/PhysRevD.86.094034;%%
  %287 citations counted in INSPIRE as of 18 Jun 2018
  
  %\cite{Carli:2010rw}
\bibitem{applgrid} 
  T.~Carli, D.~Clements, A.~Cooper-Sarkar, C.~Gwenlan, G.~P.~Salam, F.~Siegert, P.~Starovoitov and M.~Sutton,
  %``A posteriori inclusion of parton density functions in NLO QCD final-state calculations at hadron colliders: The APPLGRID Project,''
  Eur.\ Phys.\ J.\ C {\bf 66}, 503 (2010)
  doi:10.1140/epjc/s10052-010-1255-0
  [arXiv:0911.2985 [hep-ph]].
  %%CITATION = doi:10.1140/epjc/s10052-010-1255-0;%%
  %183 citations counted in INSPIRE as of 18 Jun 2018
  

\end{thebibliography}
\end{document}